\documentclass[11pt]{article}
\usepackage{amsmath}
\usepackage{amssymb}
\usepackage{amsfonts}
\usepackage{amsbsy}
\usepackage{amsthm}
\usepackage{booktabs}
\usepackage{epsfig}
\usepackage{graphicx}
\usepackage{rotating}
\usepackage[margin=1in]{geometry}
\usepackage{layout}
\usepackage{enumerate}
\usepackage{natbib} 
\usepackage{url} 

\usepackage{appendix}
    
    \usepackage{nicefrac}
    
\usepackage{float}
\usepackage{verbatimbox,stackengine}
\usepackage{xcolor}
\usepackage[list=true]{subcaption}
\usepackage{makecell}

\usepackage{multirow}

\usepackage[ruled,vlined]{algorithm2e}
\usepackage{algorithmic}

\usepackage[implicit=false,
            colorlinks = false]{hyperref}
\usepackage{comment}
\usepackage{todonotes}
\usepackage{soul}

\usepackage[printonlyused,withpage]{acronym}

\usepackage{import}

\definecolor{ao}{rgb}{0.0, 0.5, 0.0}

\begin{document}
		\def\spacingset#1{\renewcommand{\baselinestretch}%
			{#1}\small\normalsize} \spacingset{1}
     \title{\bf Evaluating diversion and treatment policies for opioid use disorder}
     \author{Veronica M. White$^{1,2}$ \and Laura A. Albert $^1$ \thanks{CONTACT: V. M. W.  Email:     vwhite2@fsu.edu \quad L. A. A. Email:     laura@engr.wisc.edu *corresponding author}}
\date{$^1$Industrial and Systems Engineering, University of Madison-Wisconsin, Madison, WI, USA
$^2$Industrial and Manufacturing Engineering, Florida A\&M University, FAMU-FSU College of Engineering, Tallahassee, FL, USA }
	\date{}
	\maketitle
\begin{center}
    \textbf{Disclaimer:} This is the accepted manuscript version of the article published in \textit{IISE Transactions on Healthcare Systems Engineering}. This manuscript may differ slightly from the final version, which is available at \href{https://www.tandfonline.com/eprint/HBSQVZPEMBXTTHNSF3FH/full?target=10.1080/24725579.2025.2523273}{https://doi.org/10.1080/24725579.2025.2523273}.
    \end{center}
		

\begin{abstract}
\noindent
{The \ac{US} opioid crisis contributed to 81,806 fatalities in 2022}. It has strained hospitals, treatment facilities, and law enforcement agencies due to the enormous resources and procedures needed to respond to the crisis. As a result, many individuals who use opioids never receive or finish the treatment they need and instead have many interactions with hospitals or the criminal justice system. 
This paper introduces a discrete event simulation model that evaluates three opioid use disorder treatment policies: arrest diversion, re-entry case management, and overdose diversion. Publicly available data from 2011 to 2019 in Dane County, Wisconsin, was used to forecast opioid-related outcomes through 2032. 
Through analyzing a variety of policy-mix implementations, the study offers a versatile framework for evaluating policies at various implementation levels. The results demonstrate that treatment policies that create new pathways and programming by utilizing treatment services and successfully divert at least 20\% of eligible individuals can lead to more opioid-resilient communities. The benefits increase when more policies are enacted and/or offered to more individuals{, with the largest impact from {overdose diversion, followed by re-entry case management, and the smallest impact from arrest diversion.}} 
The {statistically significant 10-year} cumulative total {reduction in societal costs from {2023 through} 2032 ranges from {\$39 M (USD) to \$584 M (USD)}, excluding implementation costs of policies.} To reverse the opioid crisis within a community, treatment policies may need to be combined with other strategies, such as harm reduction, supply reduction, and use prevention.
\end{abstract}
 
	
	
\spacingset{1} 
\pdfoutput=1  
	
	
\defcitealias{nida_2020}{NIDA, Jun. 2020}
\defcitealias{Victor_2021}{Victor et al., 2021}
\defcitealias{abuse_how_nodate}{NIDA, Apr. 2020}
\defcitealias{comprehensive_opioid_stimulant_and_substance_abuse_program_cossap_law_2020}{COSSAP, 2020}
\defcitealias{NIDA2016}{NIDA, Nov. 2016}
\defcitealias{samhsa_national_2016}{SAMHSA, multiple: 2016 - 2019    } 
\defcitealias{noauthor_paari_nodate    }{PAARI, 2021}
\defcitealias{substance_abuse_and_mental_health_services_administration_samhsa_2019_2019}{{SAMHSA (2019)}}

\newcommand*{\InitAgeMu}{2.08}
\newcommand*{\InitAgeSig}{0.76}
\newcommand*{\PrevAgeMu}{3.74}
\newcommand*{\PrevAgeSig}{0.49}
\newcommand*{\PopLow}{27299}
\newcommand*{\PopMedian}{34224}
\newcommand*{\PopHigh}{43261}
\newcommand*{\StartCJLow}{15}
\newcommand*{\StartCJMedian}{25}
\newcommand*{\StartCJHigh}{50}
\newcommand*{\StartHELow}{5}
\newcommand*{\StartHEMedian}{11}
\newcommand*{\StartHEHigh}{15}
\newcommand*{\StartTreatLow}{300}
\newcommand*{\StartTreatMedian}{450}
\newcommand*{\StartTreatHigh}{500}
\newcommand*{\StartIALow}{0.2}
\newcommand*{\StartIAMedian}{0.4}
\newcommand*{\StartIAHigh}{0.8}
\newcommand*{\ArrivalLam}{10.87}
\newcommand*{\ODDeathMu}{17.78}
\newcommand*{\ODDeathSigma}{4.22}
\newcommand*{\HEMu}{9.07}
\newcommand*{\HESigma}{3.01}
\newcommand*{\ArrestMu}{10.04}
\newcommand*{\ArrestSigma}{2.35}
\newcommand*{\TreatMu}{7.48}
\newcommand*{\TreatSigma}{1.03}
\newcommand*{\ProbHEtoDeath}{0.0218}
\newcommand*{\ProbHEtoArrest}{0.01}
\newcommand*{\ProbHEtoTreat}{0.2227}
\newcommand*{\InactiveMu}{4.82}
\newcommand*{\InactiveSigma}{2.20}
\newcommand*{\NatDeathLam}{Age-based}
\newcommand*{\HEServiceMu}{0.82}
\newcommand*{\HEServiceSigma}{0.48}
\newcommand*{\ArrestServiceMu}{2.16}
\newcommand*{\ArrestServiceSigma}{1.47}
\newcommand*{\TreatServiceMu}{4.78}
\newcommand*{\TreatServiceSigma}{1.18}
\newcommand*{\InactiveArrestServiceMu}{3.29}
\newcommand*{\InactiveArrestServiceSigma}{1.61}
\newcommand*{\InactiveTreatServiceMu}{4.52}
\newcommand*{\InactiveTreatServiceSigma}{1.09}
\newcommand*{\InactiveHEServiceMu}{1.95}
\newcommand*{\InactiveHEServiceSigma}{1.40}
\newcommand*{\InactiveActiveServiceMu}{6.29}
\newcommand*{\InactiveActiveServiceSigma}{2.51}
\section{Introduction} \label{s:intro}
There were 81,806 opioid-related deaths in 2022 in the \ac{US}, a significant increase from 10,000 deaths in 1999 \citep{national_center_for_health_statistics_cdc_2022}. The rapid increase in opioid-related deaths has become known as the ``Opioid Crisis.'' It is characterized by \emph{three waves}: prescription opioids in the 1990s, heroin in the 2010s, and synthetic opioids since 2013 \citep{pardo_future_2019}. Most of the literature in industrial engineering focuses on limiting the available supply of opioids through stricter laws and network interdiction modeling and decisions \citep{anzoom_review_2021}. However, the opioid crisis requires a wider array of interventions \citep{pardo_future_2019}. 

{\Ac{SUD} is defined as the uncontrollable chronic use of a substance (i.e., legal or illegal drugs) that causes clinically significant distress or impairment. \Ac{OUD} is a type of \ac{SUD} where the substance used is an opioid (e.g., tramadol, heroin, oxycodone, fentanyl). Not all individuals who use opioids are classified as having \ac{OUD} and, instead, may use opioids occasionally or have regular controlled use that does not affect their well-being, function, or social relationships. However, all individuals who use opioids are at risk of opioid-related death and potential arrest in the majority of the \ac{US}.} In fact, individuals {who use illicit drugs} often have many interactions with police and the \ac{CJS}. These individuals are more likely to be arrested for various crimes related to their drug use, such as drug possession or sale, driving under the influence of drugs, burglary, and prostitution \citep{national_institute_on_drug_abuse_principles_2018}, which creates a cycle and often leads to time in prison. This cycle is costly for individuals and society \citep{the_council_of_economic_advisers_role_2019}. 
It follows that {illicit drug use} heavily impacts the \ac{CJS}. The \ac{US} Department of Justice estimates that more than half of male and two-thirds of female state prisoners and sentenced jail inmates have a \ac{SUD}, with 16.6\% and 18.9\% of state prisoners and sentenced jail inmates, respectfully, using heroin or other opiates \citep{bronson_drug_2017}.
Following release from prison, individuals are estimated to be between seven to twelve times more likely {than the general population to die of an overdose} \citep{binswanger_release_2007,bird_male_2003}.

\Ac{MAT}{, also known as opioid agonist therapy, }is considered the best practice for treating \ac{OUD} \citep{national_institute_on_drug_abuse_effective_2016,national_center_for_biotechnology_information_introduction_2018}. \ac{MAT} {is a type of \ac{OUD} treatment} that consists of combining behavioral counseling {(e.g., in-patient)} with an \ac{OUD} medication such as buprenorphine, extended-release naltrexone, or methadone. Early evidence suggests that {\ac{MAT}} can help inmates stay engaged in treatment longer, reduce death rates after they are released from prison, and reduce recidivism rates \citep{moore_effectiveness_2019,malta_opioid-related_2019}. However, utilization of {any type of \ac{OUD} treatment} within correctional facilities and criminal justice institutions is poor, if it exists at all \citep{csete_criminal_2019,reichert_probation_2019,aronowitz_screaming_2016}. Additionally, connecting individuals to continuing treatment after release adds further challenges such as transportation to appointments, insurance issues, and long wait times when establishing care \citep{white_operations_2022, brinkley-rubinstein_benefits_2019, hamilton_substance_2019}.

 Several initiatives have aimed to break the cycle of recidivism and OUD through \textit{diversion programming}, which redirects individuals to treatment and counseling instead of the \ac{CJS}. One diversion program is \textit{arrest diversion}, also known as officer intervention, where officers refer individuals suspected of drug-related crimes to mental health treatment, behavioral health counseling, or case management \citepalias{comprehensive_opioid_stimulant_and_substance_abuse_program_cossap_law_2020}. Arrest diversion programs have shown promising effects on recidivism outcomes (PAARI, \citeyear{kopak_initial_2019,white_impact_2021,the_police_assisted_addiction__recovery_initiative_paari_paari_2021}) but face various barriers, making it difficult to evaluate whether the program was executed as planned \citep{joudrey_law_2021}.
Similarly, after someone serves their sentence and re-enters society, \textit{re-entry case management} can connect individuals to relevant resources, such as \ac{OUD} treatment via a case manager or parole officer.
Another treatment policy is \textit{overdose diversion}, which works similarly to re-entry case management except for the medical system. 
 
This paper aids decision-making for community leaders, organizations, and stakeholders against the opioid crisis by testing three policies that leverage OUD treatment in a \ac{DES} framework. The three policies include arrest diversion, overdose diversion, and re-entry case management.  The paper introduces a \ac{DES} model that captures the demand system of persons with OUD and uses publicly available data. A case study based on Dane County, WI, illustrates how policy decisions can affect outcomes through a detailed analysis. We test varying levels of program implementation of the three OUD treatment policies that provide broader insight into the benefits of different OUD treatment programming at the community level.

This paper is organized as follows. In Section \ref{s:Background}, we provide a literature review of modeling OUD policies. We introduce the \ac{DES} model in Section \ref{s:model}. In Section \ref{s:casestudy}, we introduce the case study used for testing the model along with the associated data and input parameters used. We describe the calibration and validation procedure in Section \ref{sec:validation}. In Section \ref{s:results}, we discuss the results. The paper concludes in Section \ref{s:conclusion}.

\section{Materials and Methods}  
\subsection{Literature Review}  
\label{s:Background}
Opioids have both a licit supply chain, through the prescription of opioids by medical professionals and drug companies, and an illicit supply chain, involving the re-distribution of prescription opioids and the manufacturing and distribution of fentanyl and heroin. To disrupt individuals from being introduced to opioids and potentially misusing opioids from licit supply chains, many medical organizations have created protocols to reduce the number of unused opioids and implement opioid {tapering} for patients who have become opioid dependent. One study uses time-series and geospatial modeling to understand the effects of drug prescribing limit policies in South Carolina \citep{fakhrabad_impact_2023}. Another uses a fixed effects model and a difference-in-difference model to gain insight into how changes in the complex opioid supply chain affect licit opioid dispensing and the racial disparities of these effects \citep{attari_hiding_2023}. A recent review of illicit supply chain networks discusses the need for invasive and non-invasive disruption of illicit supply chains \citep{anzoom_review_2021}. An invasive disruption requires interaction with {individuals distributing or manufacturing} an illicit good. In contrast, a noninvasive disruption does not involve illicit network stakeholders. Our paper explores additional noninvasive disruptions not mentioned in \citet{anzoom_review_2021}, such as policies and programs that influence the opioid supply chain through the reduction in demand for opioids.

Some argue for applying mathematical modeling and simulation to these treatment policies using available data {to predict policy performance} before implementation \citep{bansback_how_2021}. {Simulation has proven useful for planning, implementation, and preparedness for various public sector issues, including public health crises. For example, during the COVID-19 pandemic, simulation informed policy decisions such as minimum staffing needs, social distancing requirements, and delivery of testing and vaccines \citep{currie_how_2020, cao_covid-19_2022}. Additionally, simulation has been applied to various applications in the healthcare system, from disease screening to health and care system operations \citep{zhang_application_2018}.}
The need for a simulation modeling approach within the context of the opioid crisis is highlighted in literature reviews (e.g., \citet{sharareh_evidence_2019}, \citet{hoang_systematic_2016}). Specifically, \citet{sharareh_evidence_2019}  review opioid-related mathematical and conceptual models, concluding that more mathematical and analytical modeling is needed to tackle the complexity of the opioid crisis. Many simulations seek to evaluate if a specific program or treatment will reduce adverse outcomes (e.g., crime, fatal overdoses) and evaluate the program's cost-effectiveness.  Literature reviews by \citet{beaulieu_economic_2021} and \citet{barbosa_economic_2020} find a lack of simulations that model the trade-offs of implementing various interventions. We propose using a \ac{DES} to test multiple OUD treatment policies. We chose DES as the type of simulation due to the discrete nature of an individual's experience (e.g., an overdose, starting treatment, an arrest). In contrast to the existing papers in the literature, this paper models at the individual level while testing the effects of separate policies. We also follow the recommendations for opioid-related simulations by \citet{cerda_systematic_2021}, who conduct a literature review of parameters used in opioid-related simulations and conclude that many simulations do not meet standard transparency and reproducibility criteria.

This paper is most similar to four existing OUD treatment policy simulation models, including \citet{zarkin_lifetime_2015}, \citet{homer_dynamic_2021}, \citet{ortiz_modeling_2017}, and \citet{bernard_health_2020}. Our arrest diversion policy is similar to that considered by \citet{zarkin_lifetime_2015}. \citet{zarkin_lifetime_2015} test the implementation of re-arrest diversion programming for state prisoners nationally, whereas this paper tests the effects of an arrest diversion program in a county. We use data sources and test multiple OUD treatment policies similar to \citet{homer_dynamic_2021}. However, our policy interventions are more specific to OUD treatment and incorporate the \ac{CJS}. Our modeling approach is similar to \citet{ortiz_modeling_2017} in that we use \ac{DES} to evaluate drug policies. However, our results are more specific to the opioid crisis, whereas \citet{ortiz_modeling_2017} focus on those arrested for crack/cocaine purchases. Lastly, the scope of our outcomes is most similar to \citet{bernard_health_2020}, who use a microsimulation to model individual interactions and the result of implementing an arrest diversion program in King County, Washington. We model multiple policies that direct individuals to OUD treatment, whereas they more explicitly model different aspects of the \ac{CJS} and also model Human Immunodeficiency Virus and Hepatitis C Virus disease transmission. 

\subsection{Simulation Model}\label{s:model}
We introduce a \ac{DES} to test three OUD treatment policies{:} (1) an arrest diversion intervention, where individuals arrested for an opioid-related crime are instead referred to \ac{OUD} treatment (2) an overdose diversion intervention, where individuals who use opioids {encounter inpatient hospitalization or \ac{ED} admittance} are then connected to treatment following {hospital or \ac{ED} discharge}, and (3) re-entry case management where individuals who use opioids are paired with a case manager when re-entering the community following time in jail or prison. We compare the policies within the DES model against a base model with limited OUD treatment policies. In the model, \ac{OUD} treatment broadly includes all types of \ac{OUD} treatment. 

\begin{figure}[ht]
    \centering
\includegraphics[width=\textwidth,height=\textheight,keepaspectratio]{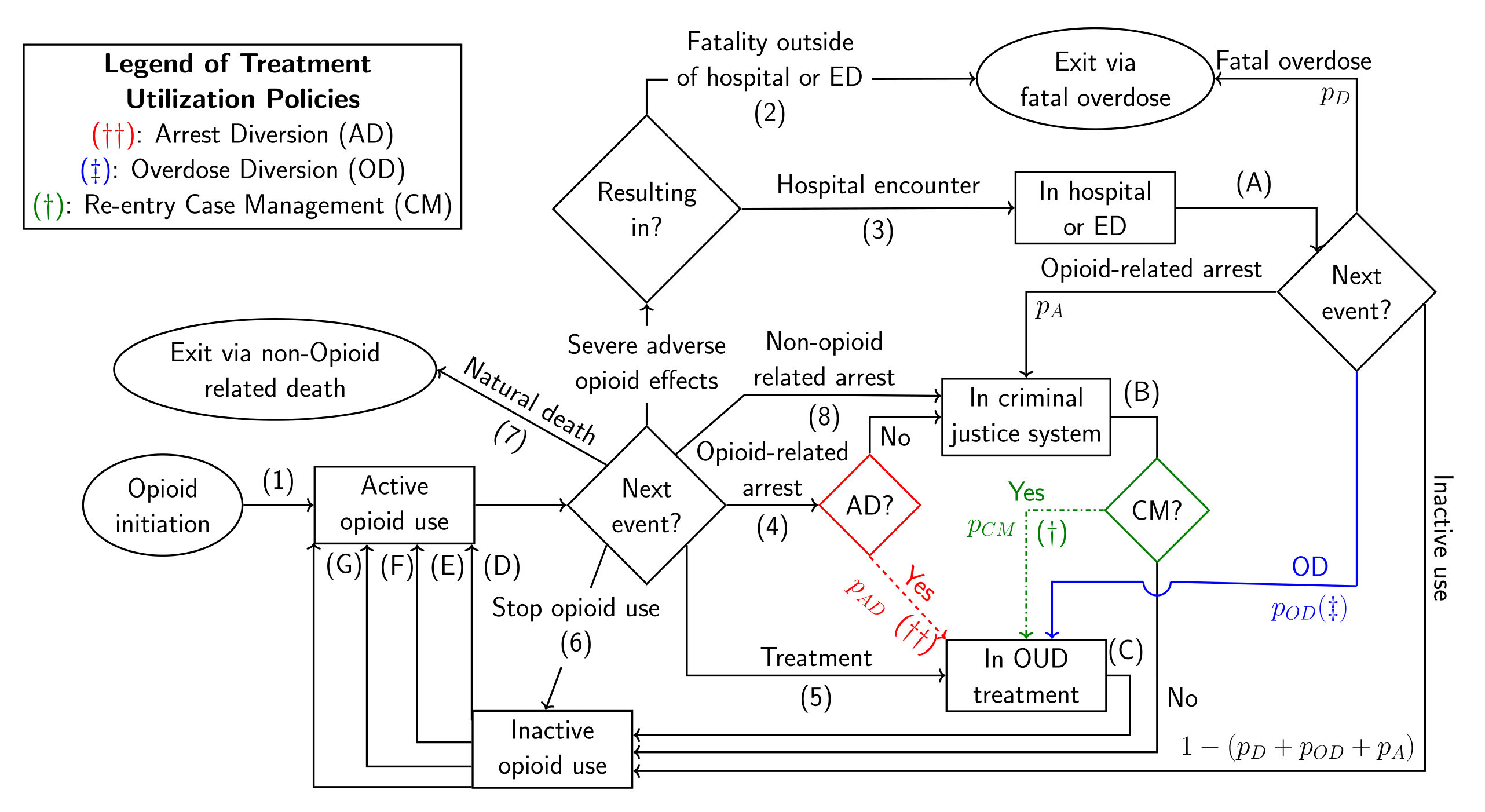}
        \caption{\ac{DES} model of individuals who have used opioids}
        \label{fig:DES}
\end{figure}

Figure \ref{fig:DES} shows the \ac{DES} model used for evaluating all policies. The arcs with numbers determine the individual's next event, whereas arcs with letters generate service times.  {The base simulation has six states---illustrated as rectangles in Figure \ref{fig:DES}---that identify where an individual is in the simulation. Possible states are active opioid use, inactive opioid use, in CJS, in hospital or ED, in OUD treatment, and opioid- and non-opioid-related death. Individuals move between states or exit the simulation through the following possible events: start opioid use, stop opioid use, start OUD treatment, stop OUD treatment, non-opioid-related arrest, opioid-related arrest, re-entry to society, hospital encounter, exit hospital or ED, non-opioid related death, and opioid-related death.}

Individuals enter the simulation via distribution (1), representing an individual\textquotesingle s first use or initiation of opioids. Upon entering the simulation, the individual is classified as actively using opioids for a period of time before the next event occurs. At the time of an individual's first start opioid use, {seven} event times are generated that follow distributions (2)-{(8)}. The {seven} possible next events are natural death, opioid-related death, opioid-related arrest, {non-opioid-related arrest,} opioid-related hospital encounter, stop opioid use, and the start of \ac{OUD} treatment. The event with the earliest time is scheduled, and the others are discarded, except for the natural death time {and non-opioid-related arrest}. {Note, we model an active use state rather than an \ac{OUD} state. Therefore, individuals in the active use state could have anything from prescription use to casual misuse to severe \ac{OUD}. No matter an individual's dependence on opioids, they are still at risk of opioid-related death, having an opioid-related hospital encounter, and entering the CJS for opioid-related arrests such as drug paraphernalia, illicit sale, and possession. For OUD treatments, individuals who are referred to treatment or start treatment may or may not have OUD. This is realistic since individuals who arrive at a hospital or have an opioid-related arrest could have varying levels of use and treatment needs}.

Suppose the next event is a natural death. In that case, the individual has a duration in the active opioid use state following distribution (7) and subsequently exits the simulation due to a non-opioid-related death. Similarly, suppose the next event is an opioid-related death. In that case, the individual has a duration in the active use state following distribution (2) and subsequently exits the simulation due to an opioid-related death. 

If the next event is to stop opioid use, the individual remains in the active use state for a duration that follows distribution (6). The individual then moves into the inactive opioid use state with a duration that follows distribution (G). Note that we model the duration of the inactive use state by selecting the distribution (D)-(G) corresponding to the previous state. We expect individuals' opioid use behavior to differ depending on whether they stop opioid use independently versus being released from the \ac{CJS}, hospital, or \ac{OUD} treatment. After the inactive use state, the individual re-enters the active opioid use state. Five new next event times are generated that follow distributions (2)-(6), where the earliest time among these times{, non-opioid-related arrest time,} and the individual's natural death time are selected as the next event.

If the next event is to start \ac{OUD} treatment, following a duration in the active use state that follows distribution (5), the individual is in treatment for a duration with distribution (C). Following treatment, they automatically move to the inactive opioid use state with a duration following distribution (E). Finally, after the inactive use state, they move to the active opioid use state, where the next event is determined using the same process described above. Similarly, if the next event is an opioid-related arrest following a duration in the active use state, which follows distribution (4), the next non-opioid related arrest time is discarded and the individual remains in the \ac{CJS} for a duration following distribution (B). Following time in the \ac{CJS}, a new non-opioid related arrest time for the individual is generated, and they automatically move to the inactive opioid use state with a duration following distribution (D). {Suppose the next event is a non-opioid-related arrest. In that case, the individual has a duration in the active opioid use state following distribution (8) and remains in the CJS for a duration following distribution (B). Once an individual exits the CJS, a new non-opioid-related arrest time is generated, and they automatically move to the inactive opioid use state with a duration following distribution (D).} From the inactive opioid use state, {if the individual time of starting opioid use is before the end of the simulation,} the individual will at that time move back to the active opioid use state, where the next event {is again assigned using the process described above. }

An opioid-related hospital encounter is one in which an individual is admitted to a hospital {or \ac{ED} }and is identified as either having adverse opioid-related effects or is found to be misusing an opioid \citep{wisconsin_department_of_health_services_opioid_2017}. Suppose the next event is a hospital encounter following the time in the active use state that follows distribution (3). In that case, the individual remains in the hospital for a duration following distribution (A). A hospital encounter may result in death with probability, $p_D$, an arrest with probability, $p_A$, a start of treatment with probability, $p_{OD}$, or inactive opioid use otherwise. Individuals with a hospital encounter and who do not enter the \ac{CJS} or treatment services have inactive opioid use duration following distribution (F).

The base model does not have arcs from jail/prison to treatment{. J}ails and prisons typically have poor quality \ac{OUD} treatment within facilities, if they have any, and do not set up treatment with outside services once an individual leaves the \ac{CJS} \citep{csete_criminal_2019,reichert_probation_2019}. However, these pathways still exist in the base model. For example, an individual released from the \ac{CJS} can start opioid use again, following time in the inactive use state, and subsequently goes to treatment within a short period. {Individuals can additionally move to \ac{CJS} from OUD treatment, the hospital, or inactive use if their non-opioid-related arrest time occurs during their time in that state. }  

Lastly, the only way to exit the simulation is through death via a fatal opioid overdose or other natural death. This is because the continuum of care often involves multiple treatment episodes, with relapses occurring weeks, months, or years after opioid abstinence \citep{hser_long-term_2015}.

\subsubsection{Tested OUD Treatment Policies}
We test three OUD treatment policies by adjusting the probabilities of moving along arcs $\textcolor{red}{(\dagger \dagger)}$, \textcolor{blue}{$(\ddagger)$}, and $\textcolor{ao}{(\dagger)}$ in Figure \ref{fig:DES}. We describe each of these treatment policies in the following paragraphs.

The first policy we test is the addition of an \textit{arrest diversion} (AD) pathway \citep{zgierska_pre-arrest_2021}. Existing arrest diversion programs typically divert individuals with low-level, non-violent offenses to SUD treatment, where not all referred individuals are eligible to participate due to program restrictions \citep{zhang_relationship_2022}. Additionally, not all of those eligible for arrest diversion programming agree to participate{,} and even fewer successfully adhere to prescribed OUD treatment \citep{white_impact_2021}. 

In our model, arrest diversion re-directs an individual to OUD treatment when arrested for an opioid-related offense by adjusting $p_{AD}$ as shown in Figure \ref{fig:DES} by the red two daggers $\textcolor{red}{(\dagger \dagger)}$ arc. The base model is at 0\% arrest diversion. Therefore, we test the alternative policies of implementing arrest diversion where $p_{AD}$ equals 20\%, 40\%, 60\%, 80\%, or 100\%. These percentages represent the percentage of individuals who are successfully diverted to treatment following an opioid-related arrest. When an individual is about to enter the \ac{CJS} in the simulation, a random $U(0,1)$ is generated from a separate random number stream. If the random number is below the given threshold, the individual follows this new arc to the OUD treatment state. Otherwise, the individual moves to the \ac{CJS} state as in the base model. When an individual is diverted to the OUD treatment state via arrest diversion, they are assumed to have been referred to an arrest diversion program and will complete any required OUD treatment. Testing various levels of arrest diversion implementation can indicate what program adherence and enrollment levels lead to the best opioid-related county outcomes over the simulated time period. 

Another pathway we test is an \textit{overdose diversion} (OD) program \citep{wisconsin_voices_for_recovery_ed2recovery_2021}. Individuals who experience opioid-related hospitalizations are more likely to have a subsequent fatal hospitalization in the following year, making hospital encounters a crucial time for potential treatment intervention \citep{king_designing_2021}. An overdose diversion program may be structured such that when a person presents in the emergency room as experiencing an overdose, a recovery coach is called and responds to the hospital. The individual is then offered the opportunity to connect with the recovery coach before discharge from the emergency room. One program obtained a 90\% success rate in connecting individuals to treatment following a hospitalization \citep{wisconsin_voices_for_recovery_ed2recovery_2021}.

 Overdose diversion can be modeled by adjusting $p_{OD}$ in Figure \ref{fig:DES} along the blue double dagger \textcolor{blue}{$(\ddagger)$} arc. The base model has an overdose diversion of around 22.3\% to account for subsequent in-patient OUD treatment that is typical in medical care for opioid-related hospitalizations \citep{singh_national_2020}.  Similar to {the AD policy}, we can test various rates of implementing an overdose diversion program by adjusting $p_{OD}$ from 22.3\% in the base model to 30\%, 45\%, 60\%, 75\%, or 90\% in alternative policy levels. Individuals who do not move to the treatment state, fatality state, or CJS state move to the inactive use state with a time in the inactive state following distribution (F). These tested probabilities are different from the other two policies for two reasons. First, the base model $p_{OD}$ has a non-zero starting value, as described above \citep{singh_national_2020}. Second, it is impossible to divert 100\% of individuals from the hospital and ED since some individuals have a fatal encounter with opioids and{,} therefore, cannot be diverted to treatment following {hospital or ED discharge}.

Third, individuals can be diverted to OUD treatment via a \textit{re-entry case management} (CM) policy. Re-entry case management involves the use of a case manager or parole officer that directs an individual to useful services, such as OUD treatment, upon re-entry to society (i.e., leaving prison or jail) \citep{fahmy_examining_2022, eastern_district_of_wisconsin_reentry_2017}.

In the DES model, re-entry case management directs an individual to OUD treatment following time in the \ac{CJS}, indicated by $p_{CM}$ and green dagger $\textcolor{ao}{(\dagger)}$ arc in Figure \ref{fig:DES}. Similar to arrest diversion, we test thresholds of implementing the re-entry case management policy where 20\%, 40\%, 60\%, 80\%, and 100\% are successfully diverted to OUD treatment instead of the inactive use state, following time in the \ac{CJS} for an arrest{, regardless if they were arrested for an opioid- or non-opioid-related crime}. Individuals who do not move to the treatment state move to the inactive use state with a service time following distribution (D).

\subsubsection{Model Outputs}
The main outcomes of the model are the number of opioid- and non-opioid-related deaths, opioid- and non-opioid-related non-diverted arrests, OUD treatments, and opioid-related hospital encounters in a given year. The model also estimates the amount of time in and the number of individuals in each state at a given time, which can be used to predict demand for that given resource. Demand for hospital beds, \ac{CJS} usage, and \ac{OUD} treatment are not typically publicly available or known and are critical to understanding potential policy impacts on existing systems. Additionally, we estimate the total yearly societal costs of the opioid crisis for each OUD treatment policy scenario, ignoring the initial implementation costs. Our {main analysis reports each scenario's 10-year cumulative outcomes starting in 2023 through the end of 2032. We then compare the difference in outcomes} against the base model to provide additional understanding of the trade-offs involved in referring more individuals to \ac{OUD} treatment.

\subsection{Case study: Dane County, Wisconsin}\label{s:casestudy}
The DES model is evaluated using a case study based in Dane County, which is in south-central Wisconsin and contains the metropolitan area of Madison. Only persons aged 12+ are included. Dane County had an estimated 12+ population of 473,436 in July 2019, making it a mid-size county in the U.S. \citep{us_census_bureau_quickfacts_2019}. Dane County is a natural choice for a case study since it implemented an arrest diversion policy named the Madison Addiction Recovery Initiative (MARI) that began in August of 2017 \citep{madison_police_department_madison_2020} as well as an overdose diversion policy, named ED2Recovery, in 2017 \citep{wisconsin_voices_for_recovery_ed2recovery_2021}. This study addresses analytic questions that the MARI program could not directly study. For example, the MARI team only had data on MARI participants and could not address broader Dane County outcomes \citep{white_impact_2021, nyland_law_2024}. Additionally, since the program was not implemented in isolation, it is difficult to assess {Dane County-level trends} due to MARI or ED2Recovery. Therefore, the DES simulation model provides insight into the effects of the three different OUD treatment programs.

Table \ref{tab:inputs} reports model inputs used in the case study. Table \ref{tab:inputs} includes distributions, model inputs, {data sources used, and parameter estimation type}. The distributions include: exponential (Exp) with parameter $\lambda$, triangular (Tri) with parameters minimum, median, and maximum, and \ac{LN} with parameters $\mu$, $\sigma$, a location parameter of 0 {, and a truncation parameter of infinity} unless otherwise stated. A \ac{LN} distribution is used to estimate times in OUD treatment, \ac{CJS}, and the hospital, since individuals are most likely to spend a short amount of time in jail, OUD treatment, and the hospital with a low likelihood of remaining in each of these states for an extended period of time. {Due to the ``new wave" of opioid deaths starting in 2019 caused by the introduction of fentanyl in the drug supply, we split arc (2) into pre-2019 (i.e., through the end of 2018) and post-2019 (i.e., 2019 and on) to capture the increased deadliness of opioids due to fentanyl \citep{national_center_for_health_statistics_cdc_2022}. }

\spacingset{1}
{\centering

\begin{table}[htbp]
    \centering \small
    \caption{Summary of discrete event simulation model inputs }   
    \resizebox{\textwidth}{!}{
    \begin{tabular}{|c|c|c|l|c|} \hline
        \multirow{2}{*}{{Dist.}} & \multirow{2}{*}{Parameter}  &  \multirow{2}{*}{Distribution for Base Model} &\multirow{2}{*}{Data Source} & \multirow{1}{*}{{Parameter}} \\ &&&&{Est. Type}\\\hline\hline
        \multirow{2}{*}{NA} & \multirow{2}{5cm}{\centering Age (years) at opioid initiation} 
        &  \multirow{1}{*}{\centering {Truncated }LN({\InitAgeMu, \InitAgeSig})}  &  \multirow{2}{4.45cm}{{\citetalias{substance_abuse_and_mental_health_services_administration_samhsa_2019_2019}}} & \multirow{2}{*}{{D}}\\  &&with Loc = 12 {and {T}runc=105}&& \\\hline
        \multirow{2}{*}{NA} & \multirow{2}{5cm}{\centering Age (years) in starting population} & \multirow{1}{*}{\centering {Truncated }LN({\PrevAgeMu, \PrevAgeSig})}  &  \multirow{2}{4.45cm}{{\citetalias{substance_abuse_and_mental_health_services_administration_samhsa_2019_2019}}} & \multirow{2}{*}{{D}}\\  &&with Loc = 12 {and {T}runc=105}
        &&\\\hline
        \multirow{2}{*}{NA} & \multirow{2}{*}{Starting population size} 
        &  \multirow{2}{*}{\centering Tri(\PopLow, \PopMedian, \PopHigh)} &  \multirow{2}{4.45cm}{\citetalias{substance_abuse_and_mental_health_services_administration_samhsa_2019_2019}} & \multirow{2}{*}{{D}}\\   &&&&\\\hline 
        \multirow{2}{*}{NA} & \multirow{2}{*}{Starting state} 
        &  \multirow{2}{*}{\centering Multinomial, See Section \ref{sec:InitiatingModel}}  & \multirow{2}{4.45cm}{Expert Opinion} & \multirow{2}{*}{{C}}\\   &&&&\\\hline \hline
        \multirow{2}{*}{(1)} & \multirow{2}{*}{ Time of next arrival (days)} 
        &  \multirow{2}{*}{\centering Exp(\ArrivalLam)} &  \multirow{2}{4.45cm}{Multiple years 2015-2019 \citetalias{substance_abuse_and_mental_health_services_administration_samhsa_2019_2019}} & \multirow{2}{*}{{D}}\\   &&&&\\\hline
        \multirow{3}{1.5cm}{\centering (2) {pre-2019}} & \multirow{2}{5cm}{\centering Time in active state given next event is an opioid-related death { before 2019} (days)}   
        &  \multirow{3}{5cm}{\centering Right-skewed non-parametric version of LN({17.60, 4.19})}  & \multirow{3}{4.5cm}{\cite{national_center_for_health_statistics_cdc_2022} 2016-{2018} }& \multirow{2}{*}{{DM}}\\  &&&&\\  &&&&\\ \hline
        \multirow{3}{1.5cm}{ \centering {(2) post-2019}} & \multirow{2}{5cm}{\centering { Time in active state given next event is an opioid-related death 2019 and later (days)}}   
        &  \multirow{3}{5cm}{\centering {Right-skewed non-parametric version of LN(17.13, 4.14})}  & \multirow{3}{4.5cm}{{\cite{national_center_for_health_statistics_cdc_2022} 2019-2020 } }& \multirow{2}{*}{{DM}}\\  &&&&\\  &&&&\\ \hline
        \multirow{3}{*}{(3)} & \multirow{3}{5cm}{\centering Time in active state given next event is a hospital encounter (days)}  
        & \multirow{3}{5cm}{\centering Right-skewed non-parametric version of LN(\HEMu, \HESigma)} & \multirow{3}{4.5cm}{\cite{wisconsin_department_of_health_services_wish_2017} 2018-2019}& \multirow{2}{*}{{DM}}\\  &&&&\\  &&&&\\ \hline
        \multirow{3}{*}{(4)} & \multirow{3}{5cm}{\centering Time in active state given next event is an opioid-related arrest (days)}  
        &  \multirow{3}{5cm}{\centering Right-skewed non-parametric version of LN(\ArrestMu, \ArrestSigma)}& \multirow{3}{4.5cm}{\cite{wisconsin_department_of_justice_uniform_2016} 2016-2019}& \multirow{2}{*}{{DM}}\\  &&&&\\  &&&&\\ \hline
        \multirow{3}{*}{(5)} & \multirow{3}{5cm}{\centering Time in active state given next event is OUD treatment (days)} 
        &  \multirow{3}{5cm}{\centering Right-skewed non-parametric version of LN(\TreatMu, \TreatSigma)}  & \multirow{3}{4.5cm}{\cite{wisconsin_department_of_health_services_opioids_2019} 2016-2017} & \multirow{2}{*}{{DM}}\\  &&&&\\  &&&&\\ \hline
        \multirow{2}{*}{(6)} & \multirow{2}{5cm}{\centering Time in active state given next event is inactive use (days)} 
        & \multirow{2}{5cm}{\centering Right-skewed non-parametric version of LN(\InactiveMu, \InactiveSigma)}   & \multirow{2}{4.45cm}{\cite{rivera_risk_2018, bauer_contributions_2019}}& \multirow{2}{*}{{DM}}\\  &&&&\\ \hline
        \multirow{2}{*}{(7)} & \multirow{2}{5cm}{\centering Time until individual\textquotesingle s non-opioid related death (days)} 
        & \multirow{2}{*}{\centering {Empirical, See Section \ref{App:Dist7}}}  & \multirow{2}{4.45cm}{{\cite{arias_united_2013}}} & \multirow{2}{*}{{D}}\\  &&&&\\ \hline
        \multirow{2}{*}{{(8)}} & \multirow{2}{5cm}{\centering {Time until individual\textquotesingle s non-opioid related arrest (days)}} 
        & \multirow{2}{5cm}{\centering {Right-skewed non-parametric version of LN(7.88, 2.38)}}  & \multirow{2}{4.45cm}{{\cite{wisconsin_department_of_justice_uniform_2016} 2016-2019}} & \multirow{2}{*}{{DM}}\\  &&&&\\ \hline \hline
        \multirow{2}{*}{(A)} & \multirow{2}{*}{Hospital service time (days)} 
        &  \multirow{2}{*}{\centering LN(\HEServiceMu, \HEServiceSigma)}  & \multirow{2}{4.45cm}{\cite{singh_national_2020, reinert_defining_2019}}& \multirow{2}{*}{{D}}\\  &&&&\\ \hline
        \multirow{2}{*}{(B)} & \multirow{2}{5cm}{\centering \ac{CJS} service time (days)} 
        &  \multirow{2}{*}{\centering LN(\ArrestServiceMu, \ArrestServiceSigma)} & \multirow{2}{4.45cm}{\cite{zhang_relationship_2022}}& \multirow{2}{*}{{D}}\\  &&&&\\ \hline
        \multirow{2}{*}{(C)} & \multirow{2}{5cm}{\centering OUD treatment service time (days)}
        &  \multirow{2}{*}{\centering LN(\TreatServiceMu, \TreatServiceSigma)}&  \multirow{2}{4.45cm}{\cite{division_of_care_and_teatment_services_2017_2018}}& \multirow{2}{*}{{D}}\\  &&&&\\ \hline
         \multirow{2}{*}{(D)} &  \multirow{2}{5cm}{\centering Inactive service time after release from {CJS} (days)} 
         &   \multirow{2}{*}{\centering LN(\InactiveArrestServiceMu, \InactiveArrestServiceSigma)} &  \multirow{2}{4.45cm}{\citet{bukten_high_2017, kinlock_study_2008}} & \multirow{2}{*}{{D}}\\  &&&&\\ \hline
         \multirow{2}{*}{(E)} &  \multirow{2}{5cm}{\centering Inactive service time after OUD treatment (days)}
         &   \multirow{2}{*}{\centering LN(\InactiveTreatServiceMu, \InactiveTreatServiceSigma)} &     \multirow{2}{4.45cm}{\citet{nunes_relapse_2018}} & \multirow{2}{*}{{D}}\\  &&&&\\ \hline
         \multirow{2}{*}{(F)} &  \multirow{2}{5cm}{\centering Inactive service time after hospital {or ED discharge} (days)} 
         &   \multirow{2}{*}{\centering LN(\InactiveHEServiceMu, \InactiveHEServiceSigma)}&  \multirow{2}{4.45cm}{\citet{chutuape_one-_2001}} & \multirow{2}{*}{{D}}\\  &&&&\\ \hline
         \multirow{2}{*}{(G)} &  \multirow{2}{5cm}{\centering Inactive service time after active state (days)} 
         &   \multirow{2}{*}{\centering LN(\InactiveActiveServiceMu, \InactiveActiveServiceSigma)}&    \multirow{2}{4.45cm}{{Calibrated}} & \multirow{2}{*}{{C}}\\  &&&&\\ \hline \hline
        \multirow{2}{*}{$p_D$} &   \multirow{2}{5cm}{\centering \% of Hosp. encounters leading to death}
        & \multirow{2}{*}{{Probability}(\ProbHEtoDeath)}  & \multirow{2}{4.45cm}{\cite{singh_national_2020}}& \multirow{2}{*}{{D}}\\  &&&&\\ \hline
        \multirow{2}{*}{$p_A$} & \multirow{2}{5cm}{\centering\% of Hosp. encounters leading to arrest} 
        & \multirow{2}{*}{{Probability}(\ProbHEtoArrest)} &  \multirow{2}{4.45cm}{Expert Opinion} & \multirow{2}{*}{{C}}\\  &&&&\\ \hline \hline
        \multirow{2}{*}{$p_{AD}$} & \multirow{2}{5cm}{\centering\% of Arrests leading to treatment} 
        & \multirow{2}{*}{{Probability}(0)}   & \multirow{2}{*}{Not Applicable} &\multirow{2}{*}{{NA}}  \\ &&&&\\ \hline
        \multirow{2}{*}{$p_{OD}$} & \multirow{2}{5cm}{\centering\% of Hosp. encounters leading to treatment} 
        & \multirow{2}{*}{{Probability}(\ProbHEtoTreat)}   & \multirow{2}{4.45cm}{\cite{singh_national_2020}} & \multirow{2}{*}{{D}}\\ &&&&\\ \hline
        \multirow{2}{*}{$p_{CM}$} & \multirow{2}{5cm}{\centering\% of CJS re-entries leading to treatment} 
        & \multirow{2}{*}{{Probability}(0)}  & \multirow{2}{*}{Not Applicable} &\multirow{2}{*}{{NA}}\\ &&&& \\\hline
        \multicolumn{5}{l}{\multirow{1}{20cm}{{Loc: Location Parameter, Trunc: Truncation Parameter}}} \\ \multicolumn{5}{l}{\multirow{2}{20cm}{Parameter Estimation Type has four possible types. D: Directly estimated. DM: Directly estimated with multiple estimation approaches considered. C: Calibrated using expert opinion as a starting point. NA: Not applicable.}}\\

    \end{tabular}
    }
    \label{tab:inputs}
\end{table}
}

\spacingset{1}
The publicly available sources used in this case study are {cited} in Table \ref{tab:inputs}. {These sources are used similarly in other opioid modeling studies such as} \citet{li_illuminating_2018}, \citet{homer_dynamic_2021}, and \citet{zarkin_benefits_2005}. {More comprehensive lists of all existing publicly available opioid-related \ac{US} data sources are listed in }\citet{national_academies_of_sciences_engineering_and_medicine_pain_2017} and \citet{jalali_data_2021}. The following describes our general strategy to estimate model inputs. First, we used Dane County data estimate inputs when possible. Second, if Dane County-specific data was unavailable, we used national or statewide data and interpolated Dane County estimates. This was done by estimating the proportion of opioid-related deaths in Dane County to opioid-related deaths nationally or statewide to interpolate Dane County\textquotesingle s ``share" of the national opioid epidemic. Third, if national or statewide data were unavailable, a search of relevant medical and criminology literature was done to provide initial estimates of parameters. Lastly, all parameters found in the literature and remaining estimates were verified and informed by community partners, such as medical doctors and former police chiefs, to obtain approximate estimates. Additional details on how inputs were calculated are in Appendix \ref{Appendix:ParamDesc}.

Table \ref{tab:yearly_event_costs} reports the 2017 USD average cost per event for {opioid-related-death, opioid-related-arrest, start OUD treatment, hospital encounter, active opioid use at year-end, and inactive opioid use at year-end}. We obtain the 2017 USD cost of a single opioid-related death of \$11,548,{462} from \citet{luo_state-level_2021}. 
{We estimate the cost of an individual starting OUD treatment in Wisconsin to be {$\$8,224$}. The per-event cost of starting OUD treatment was estimated by dividing the total estimated Wisconsin burden in 2017 due to OUD treatment (i.e., {$\$271.4$} million)  by the total number of 2017 OUD treatment episodes in the state of Wisconsin (i.e., $33,005$) obtained from \citet{wisconsin_department_of_health_services_opioids_2019}. {The total estimated Wisconsin burden in 2017 due to OUD treatment was 
 estimated by re-allocating the 2017 WI healthcare costs in \citet{luo_state-level_2021} to be 53.93\% healthcare cost and 46.07\% OUD treatment based on Table A.3 
 in \citetalias{substance_abuse_and_mental_health_services_administration_samhsa_2019_2019}. All hospitals, physicians, and other professionals were attributed to hospital encounter costs. Free-standing nursing homes, free-standing home health, and other residential, personal, and public health costs were attributed to OUD treatment episodes. Retail prescription drug costs and insurance administration were split evenly between hospital encounters and OUD treatment costs.
Therefore, we} estimated the cost of a hospital encounter to be $\${12,051}$. Similarly, the per-event cost of a hospital encounter was estimated by dividing the total estimated Wisconsin burden in 2017 due to health care (i.e., {$\$317.8$} million) by the total number of 2017 hospital encounters in the state of Wisconsin (i.e., $26,369$) from \citet{wisconsin_department_of_health_services_wish_2017}.}

{A single opioid-related arrest is estimated to be $\$55,726$. The opioid-related arrest cost was estimated from the 2017 Wisconsin CJS burden from \citet{luo_state-level_2021} (i.e., $\$383$ million) and dividing it by the opioid-related arrests in 2017 in Wisconsin (i.e.,  $6,872$) \citep{wisconsin_department_of_justice_uniform_2016}. The 2017 Wisconsin burden equals the direct CJS costs (i.e., $\$250.6$ million) plus reduced productivity cost due to incarceration (i.e., $\$132.4$ million). The loss in productivity due to incarceration makes up about 25\% of the total productivity loss due to opioids \citep{florence_economic_2021}.}
{The cost of an individual actively using opioids at year-end is $\$34,106$. This was estimated by first adding the 2017 Wisconsin burden of reduced quality of life due to opioid use (i.e., $\$6.59$ billion) plus the remaining 75\% of the 2017 Wisconsin reduced productivity cost due to opioids (i.e., $\$397$ million). The sum is then divided by the number of individuals that misused an opioid in 2017 in Wisconsin (i.e., $205,000$) (SAMHSA, \citeyear{substance_abuse_and_mental_health_services_administration_samhsa_national_2016}). We use the number of individuals in the active opioid use state at the end of the year as a conservative surrogate for opioid misuse in a year. 
Yearly societal costs for a given state are calculated by multiplying the number of events in a given year by the 2017 USD average cost per event in Table \ref{tab:yearly_event_costs}.} 

{\spacingset{1}
\begin{table}[htbp]
  \centering
  \caption{Average estimated yearly costs of a single opioid-related event per event type (USD)} 
    \begin{tabular}{|c||c|} 
      \hline
      \multirow{3}{*}{Event Type}  & \multicolumn{1}{c|}{\multirow{3}{3.5cm}{\centering 2017 USD Cost per Event}} \\ &\\&\\ \hline \hline 
      Opioid-related death       & \$11,548,{462}  \\ \hline 
      {Opioid-related} arrest  & {\$55,726}  \\ \hline 
      {Start} OUD treatment              & {\$8,224}  \\ \hline 
      Hospital encounter         & {\$12,051} \\ \hline 
      {Active opioid use at year-end}        & \$34,106 \\ \hline 
      {Inactive opioid use at year-end}         & \$0   \\ \hline
    \end{tabular}
  \label{tab:yearly_event_costs} 
\end{table}
}
\subsection{Initiating the Model} \label{sec:InitiatingModel}
We initiated the simulation with a starting population to shorten the warm-up period, where individuals are pre-generated into various states. 
First, the total starting population size is sampled using the triangular distribution reported in Table \ref{tab:inputs}. Then, {each individual's starting state} is determined by sampling a multinomial random variable corresponding to the possible starting states of the hospital, the \ac{CJS},  OUD treatment, inactive opioid use, and active opioid use. The starting state probabilities are determined as follows. 
These starting state probabilities are mutually exclusive and sum to one, and therefore, the remaining probability captures the multinomial probability that an individual starts in the inactive use state.
 
Once an individual is assigned a starting state through the above multinomial distribution, the associated service time distribution {assigns} a time in the given starting state. The time in the inactive use state depends on an individual's previous state. Therefore, for the inactive use starting state, a preceding state is also generated, in the same way as above, to decide which distribution, i.e., (D)-(G), to assign the time in service for the inactive state. 
{Derivation of the specific parameters used to generate the starting population is discussed in Appendix \ref{App:StartPop}.}

\subsection{Calibration and Validation of Model Outputs} \label{sec:validation}
For model calibration and validation, the simulation was run 600 times with different random seeds to construct 95\% joint \acp{PI} for the main simulation outputs of interest. We also estimate monetary costs for each of these states to approximate the cost savings{, not including implementation costs,} of a given policy. We use the common random numbers method to reduce variation in the simulation model when using each distribution in Table \ref{tab:inputs}. 

The model\textquotesingle s structure in Figure \ref{fig:DES}  was validated by consulting expert opinion via incorporating feedback from both the subject matter experts in policing and medicine. The policing subject matter expert provided high, low, and best guess estimates for the number of individuals in the Dane County jail system and $p_A$ in Table \ref{tab:inputs}. The medical subject matter expert provided high, low, and best guess estimates for the number of individuals currently in \ac{OUD} treatment and the hospital due to opioid-related hospital encounters. Appendix \ref{Appendix:Results} reports additional model outputs that were reviewed and validated by the subject matter experts in policing and addiction treatment.

The simulation has a starting population generated at the beginning of 2009 with a warm-up period of 5 years{. The simulation is run for an additional 20 years} and is evaluated between the years 2013 and 2032. We assume there are 365.25 days in a year. We note the model does not reach a steady state due to the exponential growth of opioid-related overdoses, deaths, and hospital encounters in the data. 
 
\begin{figure}[ht]
    \centering
    \includegraphics[width=1\textwidth]{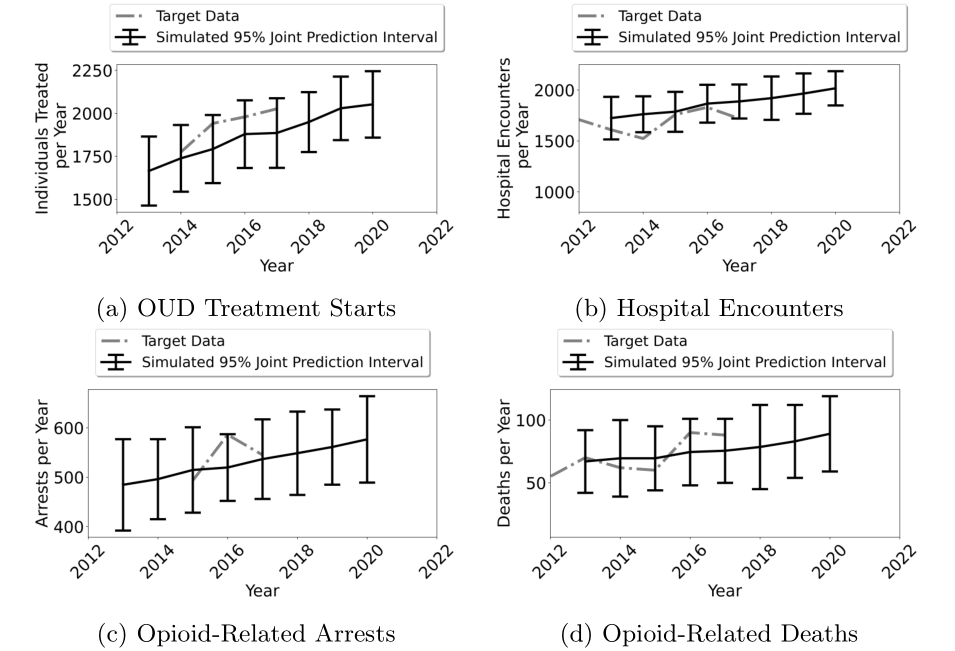}
    \caption{Model outputs compared with historical data under base model assumptions}
    \label{fig:CalibrationOutputs}
\end{figure}

Figure \ref{fig:CalibrationOutputs} shows the model validation, where the resulting calibrated model output \ac{PI}s are compared to the actual value of the corresponding target Dane County data. As shown in Figure \ref{fig:CalibrationOutputs}, the warm-up period was determined when the yearly outputs were similar to the 2013 to 2017 outcomes observed in the data sources. Only the pre-2018 target data points were used to estimate calibration error, since an arrest diversion and overdose diversion program were implemented in Dane County at the end of 2017 \citep{zgierska_pre-arrest_2021, wisconsin_voices_for_recovery_ed2recovery_2021}. Therefore, we use 17 calibration targets to calibrate the simulation. The targets are the number of OUD treatments from 2014 to 2017, the number of opioid-related deaths from 2013 to 2017, the number of opioid-related arrests from 2015 to 2017, and the number of opioid-related hospital encounters from 2013 to 2017, which are summarized as ``target data" in Figure \ref{fig:CalibrationOutputs}. We describe how the target data is collected in Appendix \ref{App:EventTime}.  The goodness-of-fit measures used are the number of target points within the simulation model 95\% PIs. The search algorithm we used was multi-approached. First, we adjusted the subject matter expert estimates by $\pm$ 5\% of the corresponding inputs that lay outside the 95\% PIs. {Parameters calibrated via this first method are denoted in Table \ref{tab:inputs} with ``C" as they were calibrated with an initial estimation that relied on expert opinion \citep{law_how_2005}. }Secondly, we used an alternative calculation of our parameter estimates. {Parameters calibrated via this second method are denoted in Table \ref{tab:inputs} with ``DM", since they have multiple direct estimations that are considered. }Additional details regarding our final parameter estimation are described in Appendix \ref{Appendix:ParamDesc}. As shown in Figure \ref{fig:CalibrationOutputs}, the model has an average calibration error of {11.76\%} 
among the 17 target Dane County data points, where only the number of Dane County hospital encounters in 2014 {and 2017} fell outside of the 95\% PI. 

\subsection{Validation of Cost Estimates} \label{Sec:CostCalc}
{We verified the accuracy of the simulation model's societal cost estimates by comparing the total simulated costs for 2017 with the estimated societal costs for Dane County. Table \ref{tab:valid_costs} summarizes the Dane County 2017 number of events and estimated total societal costs, as well as the simulated 2017 95\% PIs of total events and total simulated societal costs. We present a societal cost average and range for the active use state, representing the lowest and highest proportions observed in other states (e.g., OUD treatment, hospital encounter) multiplied by the estimated number of cases of opioid misuse in 2017 for Wisconsin (i.e., 205,000) (SAMHSA, \citeyear{substance_abuse_and_mental_health_services_administration_samhsa_national_2016}). This, therefore, calculates a mean and range for the total estimated 2017 number of active users at year-end and the associated societal costs for Dane County. To determine the total simulated societal costs for a specific scenario, we multiply the USD per event costs for 2017 by the yearly 95\% PI simulated event estimates, resulting in a 95\% PI total yearly cost estimate. As shown in Table \ref{tab:valid_costs}, the {total 2017 estimated costs are within the 95\% PIs of the simulated cost}. The simulated active use at year-end and the hospital encounter-related costs are near the lower end of the 95\% PIs, while opioid-related death and opioid-related arrest costs are at the high end. Overall, the total {simulated }societal costs of all states combined are on the low end of the estimated costs in 2017. This means a given policy scenario's actual cost savings are likely higher than the simulated {cost }values.}
{\spacingset{1}
\begin{table}[htbp]
  \caption{Validation of yearly simulated base model cost of the opioid costs in 2017 USD} 
  \label{tab:valid_costs}
  \centering
  \resizebox{\textwidth}{!}{
    \begin{tabular}{|c||c|c||c|c|}
      \hline
      \multicolumn{1}{|c||}{\multirow{4}{*}{Event Type}} & \multicolumn{1}{c|}{\multirow{4}{2.9cm}{ \centering Actual 2017 Dane County total events}} & \multicolumn{1}{c||}{\multirow{4}{2.5cm}{\centering 95\% PI Simulated 2017 Count of Event }} & \multicolumn{1}{c|}{\multirow{4}{2.75cm}{\centering Estimated Dane County 2017 total cost in millions USD }} & \multicolumn{1}{c|}{\multirow{4}{2.75cm}{\centering Total simulated cost in Millions USD (95\% PI)}} \\
       & & & &    \\
       & & & &    \\
       & & & &   \\ \hline
Active use at year-end & 17,630 (12,505, 28,085)      & {(14,277, 16,218)}	  & \$601.3 (\$426.5, \$957.9)	       & {(\$486.9, \$553.1)}            \\ \hline
Opioid-related death                  & 88             & {(48, 104)}	              & \$1016.3	      &{(\$554.3, \$1201)}                                                                       \\ \hline
Opioid-related arrest                 & 545           & {(452, 639)}	                           & \$30.4	                                            & {(\$25.2, \$35.6) }                             \\ \hline
Start OUD treatment                   & 2026          & {(1751, 2149)}                       & {\$16.7}	                                             & {(\$14.4, \$17.7)}                                \\ \hline
Hospital encounter                    & 1718         & {(1749, 2102)}	                      &{ \$20.7}	                                              & {(\$21.1, \$25.3) }                                 \\ \hline \hline
\multicolumn{3}{|r||}{Total cost:}                                & \$1686 (\$1511, {\$2042}) &	{(\$1,115, \$1,809)}   
                           \\ \hline
\end{tabular}
  }
\end{table}
}
\section{Results and Discussion}\label{s:results}
We evaluate treatment policies by modifying the simulation model to simulate policy implementation. Specifically, we evaluate the following three policies AD $(\dagger)$, OD $(\ddagger)$, and CM $(\dagger\dagger)$ shown in Figure \ref{fig:DES} at various scenarios of implementation. For clarity, we express the policy implementation values as a triplet in this section. For example, the base model implies the policy scenario (0, 22, 0), where there is no arrest diversion, 22\% of eligible individuals are successfully re-directed to overdose diversion, and there is no re-entry case management. All versions of the simulation were built using Python 3.9.0 and the SimPy package \citep{team_simpy_simpy_2002}. Each scenario uses the same set of 600 replications to create 95\% a \ac{CI} for each yearly output{, via the common random numbers method}. We compare each policy's simulated output \ac{CI}s to the base model's. Appendix \ref{Appendix:Replications} reports details regarding selecting the number of replications, checking for normality, and meeting an acceptable margin of error. 
{\subsection{10-Year Cumulative Policy Events: 2023 through 2032}}
\noindent{Table \ref{tab:AD_All_Results} reports several {10-year cumulative outputs from the start of 2023 through the end of} 2032. 
{\spacingset{1}
\begin{sidewaystable}
\caption{Simulated mean $\pm$ standard error (p-value) of {10-year} cumulative total events, cost, and savings for scenarios: 2023 through 2032}
\label{tab:AD_All_Results}
\resizebox{\textwidth}{!}{
\begin{tabular}{|c|c|c|c|c|c|c|c|c|}
\hline
\multicolumn{2}{|c|}{Scenario} & \multicolumn{1}{c|}{Opioid-Related Deaths} & \multicolumn{1}{c|}{Opioid-Related Non-Diverted Arrests} & \multicolumn{1}{c|}{Opioid-Related Hospital Encounters} & \multicolumn{1}{c|}{OUD Treatment Starts} & \multicolumn{1}{c|}{Active Use Starts} & \multirow{1}{*}{Mean Total Cost} & \multirow{1}{*}{Mean Cost  Difference}\\ \hline
\multicolumn{2}{|c|}{AD (\%), OD (\%), CM (\%)} & \multicolumn{1}{c|}{mean $\pm$ se (p-value)}  & \multicolumn{1}{c|}{mean $\pm$ se (p-value)}  & \multicolumn{1}{c|}{mean $\pm$ se (p-value)}  & \multicolumn{1}{c|}{mean $\pm$ se (p-value)}  & \multicolumn{1}{c|}{mean $\pm$ se (p-value)} & \$ in Millions (p-value) &  \$ in Millions  \\ \hline \hline
\multirow{2}{*}{\rotatebox[origin=c]{90}{\parbox[t]{8mm}{\spacingset{1} \centering Base \\ Model}}}& \multirow{2}{*}{0, 22, 0} & \multirow{2}{*}{1038.02 $\pm$ 1.42} & \multirow{2}{*}{6706.75 $\pm$ 4.99} & \multirow{2}{*}{23573.9 $\pm$ 14.4} & \multirow{2}{*}{24589.71 $\pm$ 15.64} & \multirow{2}{*}{192270.75 $\pm$ 112.0} & \multirow{2}{*}{21,478.0 $\pm$ 20.57} & \multirow{2}{*}{0.0}
\\ &&&&&&&&\\\hline \hline
\multirow{5}{*}{\rotatebox[origin=c]{90}{\parbox[t]{15mm}{\spacingset{1}\centering Arrest \\ Diversion}}} & \multirow{1}{*}{20, 22, 0} & \multirow{1}{*}{1036.17 $\pm$ 1.44(0.063)} & \multirow{1}{*}{\bf{5358.7$\pm$ 4.28$^{**}$ ($<$0.001)}} & \multirow{1}{*}{\bf{23553.95$\pm$ 14.37$^{**}$ ($<$0.001)}} & \multirow{1}{*}{\bf{25909.08$\pm$ 16.75$^{**}$ ($<$0.001)}} & \multirow{1}{*}{\bf{192105.33$\pm$ 112.68$^{**}$ ($<$0.001)}} & \multirow{1}{*}{\bf{21,376.0$\pm$ 20.73$^{**}$ ($<$0.001)}} & \multirow{1}{*}{\bf{-102.61$^{**}$}}
\\ & \multirow{1}{*}{40, 22, 0} & \multirow{1}{*}{1036.52 $\pm$ 1.43(0.135)} & \multirow{1}{*}{\bf{4014.55$\pm$ 3.37$^{**}$ ($<$0.001)}} & \multirow{1}{*}{\bf{23531.22$\pm$ 14.4$^{**}$ ($<$0.001)}} & \multirow{1}{*}{\bf{27216.7$\pm$ 17.3$^{**}$ ($<$0.001)}} & \multirow{1}{*}{\bf{191890.41$\pm$ 112.5$^{**}$ ($<$0.001)}} & \multirow{1}{*}{\bf{21,300.0$\pm$ 20.97$^{**}$ ($<$0.001)}} & \multirow{1}{*}{\bf{-178.4$^{**}$}}
\\ & \multirow{1}{*}{60, 22, 0} & \multirow{1}{*}{\bf{1033.43$\pm$ 1.44$^{**}$ ($<$0.001)}} & \multirow{1}{*}{\bf{2672.26$\pm$ 2.56$^{**}$ ($<$0.001)}} & \multirow{1}{*}{\bf{23504.16$\pm$ 14.19$^{**}$ ($<$0.001)}} & \multirow{1}{*}{\bf{28525.93$\pm$ 17.79$^{**}$ ($<$0.001)}} & \multirow{1}{*}{\bf{191696.84$\pm$ 111.35$^{**}$ ($<$0.001)}} & \multirow{1}{*}{\bf{21,186.0$\pm$ 20.96$^{**}$ ($<$0.001)}} & \multirow{1}{*}{\bf{-292.19$^{**}$}}
\\ & \multirow{1}{*}{80, 22, 0} & \multirow{1}{*}{\bf{1033.71$\pm$ 1.43$^{**}$ ($<$0.001)}} & \multirow{1}{*}{\bf{1335.86$\pm$ 1.61$^{**}$ ($<$0.001)}} & \multirow{1}{*}{\bf{23484.69$\pm$ 14.33$^{**}$ ($<$0.001)}} & \multirow{1}{*}{\bf{29828.8$\pm$ 18.72$^{**}$ ($<$0.001)}} & \multirow{1}{*}{\bf{191504.22$\pm$ 111.73$^{**}$ ($<$0.001)}} & \multirow{1}{*}{\bf{21,110.0$\pm$ 20.71$^{**}$ ($<$0.001)}} & \multirow{1}{*}{\bf{-368.34$^{**}$}}
\\ & \multirow{1}{*}{100, 22, 0} & \multirow{1}{*}{\bf{1032.11$\pm$ 1.47$^{**}$ ($<$0.001)}} & \multirow{1}{*}{\bf{0.0$\pm$ 0.0$^{**}$ ($<$0.001)}} & \multirow{1}{*}{\bf{23459.58$\pm$ 14.31$^{**}$ ($<$0.001)}} & \multirow{1}{*}{\bf{31135.28$\pm$ 19.47$^{**}$ ($<$0.001)}} & \multirow{1}{*}{\bf{191313.94$\pm$ 112.04$^{**}$ ($<$0.001)}} & \multirow{1}{*}{\bf{21,013.0$\pm$ 21.16$^{**}$ ($<$0.001)}} & \multirow{1}{*}{\bf{-465.0$^{**}$}}
\\\hline \hline
\multirow{5}{*}{\rotatebox[origin=c]{90}{\parbox[t]{15mm}{\spacingset{1}\centering Overdose \\ Diversion}}} & \multirow{1}{*}{0, 30, 0} & \multirow{1}{*}{1038.34 $\pm$ 1.4(0.849)} & \multirow{1}{*}{\bf{6693.43$\pm$ 4.97$^{**}$ ($<$0.001)}} & \multirow{1}{*}{\bf{23531.48$\pm$ 14.31$^{**}$ ($<$0.001)}} & \multirow{1}{*}{\bf{26359.38$\pm$ 16.66$^{**}$ ($<$0.001)}} & \multirow{1}{*}{\bf{191900.29$\pm$ 111.31$^{**}$ ($<$0.001)}} & \multirow{1}{*}{21,490.0 $\pm$ 20.45(0.61)} & \multirow{1}{*}{11.44}
\\ & \multirow{1}{*}{0, 45, 0} & \multirow{1}{*}{\bf{1027.26$\pm$ 1.38$^{**}$ ($<$0.001)}} & \multirow{1}{*}{\bf{6670.23$\pm$ 5.05$^{**}$ ($<$0.001)}} & \multirow{1}{*}{\bf{23453.72$\pm$ 14.29$^{**}$ ($<$0.001)}} & \multirow{1}{*}{\bf{29778.07$\pm$ 18.68$^{**}$ ($<$0.001)}} & \multirow{1}{*}{\bf{191221.15$\pm$ 112.24$^{**}$ ($<$0.001)}} & \multirow{1}{*}{\bf{21,344.0$\pm$ 20.15$^{**}$ ($<$0.001)}} & \multirow{1}{*}{\bf{-134.75$^{**}$}}
\\ & \multirow{1}{*}{0, 60, 0} & \multirow{1}{*}{\bf{1026.97$\pm$ 1.44$^{**}$ ($<$0.001)}} & \multirow{1}{*}{\bf{6645.81$\pm$ 4.98$^{**}$ ($<$0.001)}} & \multirow{1}{*}{\bf{23372.19$\pm$ 14.23$^{**}$ ($<$0.001)}} & \multirow{1}{*}{\bf{33166.83$\pm$ 20.71$^{**}$ ($<$0.001)}} & \multirow{1}{*}{\bf{190496.64$\pm$ 111.65$^{**}$ ($<$0.001)}} & \multirow{1}{*}{\bf{21,339.0$\pm$ 20.94$^{**}$ ($<$0.001)}} & \multirow{1}{*}{\bf{-139.51$^{**}$}}
\\ & \multirow{1}{*}{0, 75, 0} & \multirow{1}{*}{\bf{1023.01$\pm$ 1.35$^{**}$ ($<$0.001)}} & \multirow{1}{*}{\bf{6620.47$\pm$ 4.96$^{**}$ ($<$0.001)}} & \multirow{1}{*}{\bf{23289.89$\pm$ 14.14$^{**}$ ($<$0.001)}} & \multirow{1}{*}{\bf{36529.44$\pm$ 22.39$^{**}$ ($<$0.001)}} & \multirow{1}{*}{\bf{189767.75$\pm$ 111.32$^{**}$ ($<$0.001)}} & \multirow{1}{*}{\bf{21,281.0$\pm$ 19.85$^{**}$ ($<$0.001)}} & \multirow{1}{*}{\bf{-197.41$^{**}$}}
\\ & \multirow{1}{*}{0, 90, 0} & \multirow{1}{*}{\bf{1020.43$\pm$ 1.42$^{**}$ ($<$0.001)}} & \multirow{1}{*}{\bf{6598.3$\pm$ 4.97$^{**}$ ($<$0.001)}} & \multirow{1}{*}{\bf{23210.16$\pm$ 14.15$^{**}$ ($<$0.001)}} & \multirow{1}{*}{\bf{39876.39$\pm$ 24.42$^{**}$ ($<$0.001)}} & \multirow{1}{*}{\bf{189060.21$\pm$ 111.06$^{**}$ ($<$0.001)}} & \multirow{1}{*}{\bf{21,250.0$\pm$ 20.59$^{**}$ ($<$0.001)}} & \multirow{1}{*}{\bf{-228.2$^{**}$}}
\\\hline \hline
\multirow{5}{*}{\rotatebox[origin=c]{90}{\parbox[t]{20mm}{\spacingset{1}\centering Case \\ Management}}} & \multirow{1}{*}{0, 22, 20} & \multirow{1}{*}{\bf{1034.81$\pm$ 1.44$^{**}$ ($<$0.001)}} & \multirow{1}{*}{\bf{6686.57$\pm$ 5.0$^{**}$ ($<$0.001)}} & \multirow{1}{*}{\bf{23494.89$\pm$ 14.27$^{**}$ ($<$0.001)}} & \multirow{1}{*}{\bf{27532.79$\pm$ 16.96$^{**}$ ($<$0.001)}} & \multirow{1}{*}{\bf{191668.86$\pm$ 111.61$^{**}$ ($<$0.001)}} & \multirow{1}{*}{\bf{21,440.0$\pm$ 20.79$^*$ (0.003)}} & \multirow{1}{*}{\bf{-38.65$^{**}$}}
\\ & \multirow{1}{*}{0, 22, 40} & \multirow{1}{*}{\bf{1029.91$\pm$ 1.46$^{**}$ ($<$0.001)}} & \multirow{1}{*}{\bf{6662.63$\pm$ 5.01$^{**}$ ($<$0.001)}} & \multirow{1}{*}{\bf{23412.31$\pm$ 14.37$^{**}$ ($<$0.001)}} & \multirow{1}{*}{\bf{30449.56$\pm$ 18.28$^{**}$ ($<$0.001)}} & \multirow{1}{*}{\bf{191041.48$\pm$ 112.21$^{**}$ ($<$0.001)}} & \multirow{1}{*}{\bf{21,377.0$\pm$ 21.22$^{**}$ ($<$0.001)}} & \multirow{1}{*}{\bf{-101.72$^{**}$}}
\\ & \multirow{1}{*}{0, 22, 60} & \multirow{1}{*}{\bf{1027.84$\pm$ 1.44$^{**}$ ($<$0.001)}} & \multirow{1}{*}{\bf{6641.69$\pm$ 5.02$^{**}$ ($<$0.001)}} & \multirow{1}{*}{\bf{23330.99$\pm$ 14.21$^{**}$ ($<$0.001)}} & \multirow{1}{*}{\bf{33341.59$\pm$ 19.49$^{**}$ ($<$0.001)}} & \multirow{1}{*}{\bf{190453.61$\pm$ 111.12$^{**}$ ($<$0.001)}} & \multirow{1}{*}{\bf{21,353.0$\pm$ 20.55$^{**}$ ($<$0.001)}} & \multirow{1}{*}{\bf{-125.56$^{**}$}}
\\ & \multirow{1}{*}{0, 22, 80} & \multirow{1}{*}{\bf{1023.12$\pm$ 1.47$^{**}$ ($<$0.001)}} & \multirow{1}{*}{\bf{6620.54$\pm$ 4.94$^{**}$ ($<$0.001)}} & \multirow{1}{*}{\bf{23251.73$\pm$ 14.18$^{**}$ ($<$0.001)}} & \multirow{1}{*}{\bf{36208.12$\pm$ 20.8$^{**}$ ($<$0.001)}} & \multirow{1}{*}{\bf{189845.6$\pm$ 110.68$^{**}$ ($<$0.001)}} & \multirow{1}{*}{\bf{21,300.0$\pm$ 21.39$^{**}$ ($<$0.001)}} & \multirow{1}{*}{\bf{-178.81$^{**}$}}
\\ & \multirow{1}{*}{0, 22, 100} & \multirow{1}{*}{\bf{1020.25$\pm$ 1.4$^{**}$ ($<$0.001)}} & \multirow{1}{*}{\bf{6598.45$\pm$ 4.92$^{**}$ ($<$0.001)}} & \multirow{1}{*}{\bf{23173.76$\pm$ 14.14$^{**}$ ($<$0.001)}} & \multirow{1}{*}{\bf{39057.07$\pm$ 22.01$^{**}$ ($<$0.001)}} & \multirow{1}{*}{\bf{189257.63$\pm$ 111.38$^{**}$ ($<$0.001)}} & \multirow{1}{*}{\bf{21,259.0$\pm$ 20.53$^{**}$ ($<$0.001)}} & \multirow{1}{*}{\bf{-219.85$^{**}$}}
\\\hline \hline
\multirow{3}{*}{\rotatebox[origin=c]{90}{\parbox[t]{10mm}{\spacingset{1}\centering Policy \\ Mix}}} & \multirow{1}{*}{20, 40, 20} & \multirow{1}{*}{\bf{1026.72$\pm$ 1.37$^{**}$ ($<$0.001)}} & \multirow{1}{*}{\bf{5319.9$\pm$ 4.27$^{**}$ ($<$0.001)}} & \multirow{1}{*}{\bf{23381.8$\pm$ 14.34$^{**}$ ($<$0.001)}} & \multirow{1}{*}{\bf{32609.93$\pm$ 19.92$^{**}$ ($<$0.001)}} & \multirow{1}{*}{\bf{190681.75$\pm$ 111.51$^{**}$ ($<$0.001)}} & \multirow{1}{*}{\bf{21,267.0$\pm$ 20.56$^{**}$ ($<$0.001)}} & \multirow{1}{*}{\bf{-211.32$^{**}$}}
\\ & \multirow{1}{*}{40, 60, 40} & \multirow{1}{*}{\bf{1020.28$\pm$ 1.44$^{**}$ ($<$0.001)}} & \multirow{1}{*}{\bf{3956.64$\pm$ 3.39$^{**}$ ($<$0.001)}} & \multirow{1}{*}{\bf{23191.32$\pm$ 14.07$^{**}$ ($<$0.001)}} & \multirow{1}{*}{\bf{40510.23$\pm$ 23.65$^{**}$ ($<$0.001)}} & \multirow{1}{*}{\bf{189075.59$\pm$ 109.85$^{**}$ ($<$0.001)}} & \multirow{1}{*}{\bf{21,105.0$\pm$ 20.86$^{**}$ ($<$0.001)}} & \multirow{1}{*}{\bf{-373.84$^{**}$}}
\\ & \multirow{1}{*}{60, 80, 60} & \multirow{1}{*}{\bf{1009.8$\pm$ 1.44$^{**}$ ($<$0.001)}} & \multirow{1}{*}{\bf{2618.82$\pm$ 2.5$^{**}$ ($<$0.001)}} & \multirow{1}{*}{\bf{23023.47$\pm$ 14.08$^{**}$ ($<$0.001)}} & \multirow{1}{*}{\bf{47828.35$\pm$ 27.81$^{**}$ ($<$0.001)}} & \multirow{1}{*}{\bf{187662.98$\pm$ 110.06$^{**}$ ($<$0.001)}} & \multirow{1}{*}{\bf{20,894.0$\pm$ 21.18$^{**}$ ($<$0.001)}} & \multirow{1}{*}{\bf{-584.39$^{**}$}}
\\ \hline
\multicolumn{9}{l}{*{Statistically significant difference, with p-value $<0.05$, from the Base Model at a level of 0.05 using 2-sided paired t-test}}\\
\multicolumn{9}{l}{**{Statistically significant difference, with p-value $<0.001$, from the Base Model at a level of 0.05 using 2-sided paired t-test}}\\
\multicolumn{9}{l}{{Note: Overlapping confidence intervals that are marked statistically significant through a paired t-test can be attributed to Type one error \citep{knol_misuse_2011} }}
\end{tabular} 
}
\end{sidewaystable}
}The year 2032 is ten years from the potential implementation of the CM policy and 16 years from the potential implementation of the AD and OD policies.} Table  \ref{tab:AD_All_Results} reports 95\% \ac{CI} standard errors and mean difference p-values for the {10-year} cumulative number of opioid-related deaths, {non-diverted }opioid-related arrests, opioid hospital encounters, OUD treatment episodes, and individuals in the active use state at year end.  Additionally, Table  \ref{tab:AD_All_Results} reports each scenario's simulated total costs, cost difference from the base model, and cost mean difference p-values.
A two-sided paired t-test of the mean difference from the base model (i.e., (0, 22, 0)) is used to report statistically significant differences. P-values in boldface with a \textbf{($^*$)} are below a significance level of 0.05, and those in boldface with an \textbf{($^{**}$)} are below a level of 0.001. 

At the 0.001 level, Table \ref{tab:AD_All_Results} indicates that a policy that has at least 60\% AD, 45\% OD, or 20\% CM statistically significantly decreases the number of non-diverted opioid-related arrests, opioid-related deaths, active opioid use, and hospital encounters while statistically significantly increasing the amount of OUD treatment episodes. {All} policies show statistically significant cumulative reductions to {non-diverted opioid-related arrests, active opioid use, and hospital encounters, suggesting} that arrest diversion, overdose diversion, and case management can lead to cumulative reductions in opioid-related arrests, hospitalizations, and opioid use with any amount of policy implementation. The differences {between all policy outputs, including opioid-related deaths, from the base model} are greater when mixing multiple policies and at higher levels of policy implementation. For example, with the CM-focused policies of (0, 22, 20) with 20\% CM compared to (0, 22, 60) with 60\% CM, we see a decrease in the expected average number of opioid-related arrests from 6,687 to 6,642, respectively. If we combine multiple policies, we see an even larger reduction in {non-diverted} opioid-related arrests to {5,320} for the policy-mix (20, 40, 20). Therefore, when jurisdictions plan to implement a treatment policy, they should develop a plan to implement multiple policies over time and scale each policy within their communities, such as offering a treatment policy to more people.
{\subsection{Societal Costs}}
\noindent In terms of societal costs, Table \ref{tab:AD_All_Results} shows statistically significant 10-year societal cost savings through the Year 2032, compared to the base model, is between \$38.65 million at a (0, 22, {20}) CM-focused policy to \$584 million at the policy-mix of (60, 80, 60). At and above the (20, 22, 0) AD-focused policy, the (0, {45}, 0) OD-only policy, the (0, 22, 20) CM-focused policy, or a policy-mix of (20, 40, 20), we see that the cost savings from decreased opioid use, hospital encounters, and opioid-related arrests outweigh the added costs of additional OUD treatment. Decreased opioid use and opioid-related arrests increase the quality of life and productivity of the community \citep{florence_economic_2021}. Therefore, jurisdictions that invest in additional treatment capacity and divert individuals to treatment can expect to improve residents' lives, increase the active workforce, and reduce incarceration. 

\begin{figure}[b!]
\centering
\begin{subfigure}[b]{.49\linewidth}
    \includegraphics[width=\linewidth]{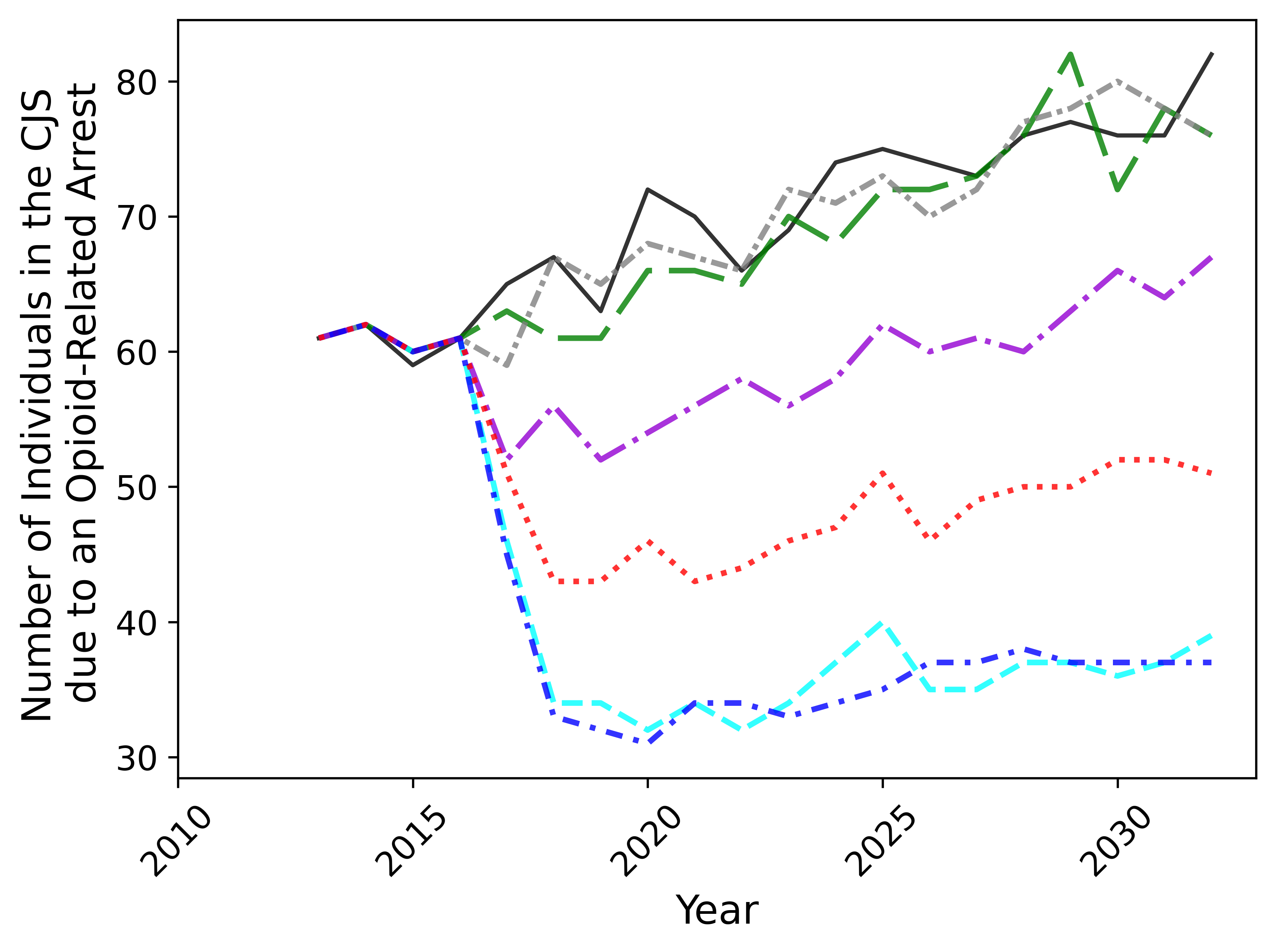}
    \caption{CJS usage - upper 95\% PI}
\end{subfigure}
\begin{subfigure}[b]{.49\linewidth}
    \includegraphics[width=\linewidth]{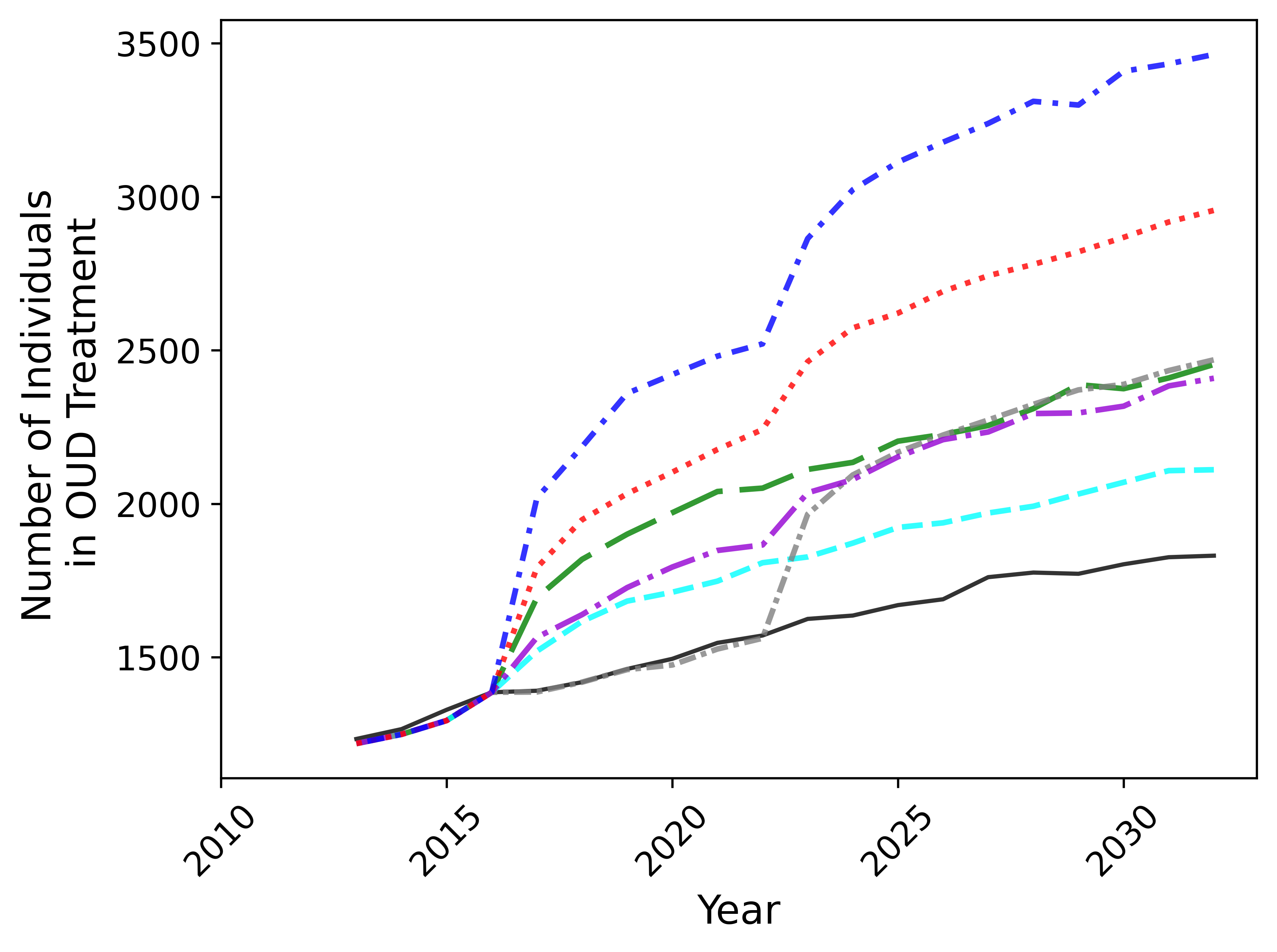}
    \caption{OUD treatment usage - upper 95\% PI}
\end{subfigure}
~
\begin{subfigure}[b]{.49\linewidth}
    \includegraphics[width=\linewidth]{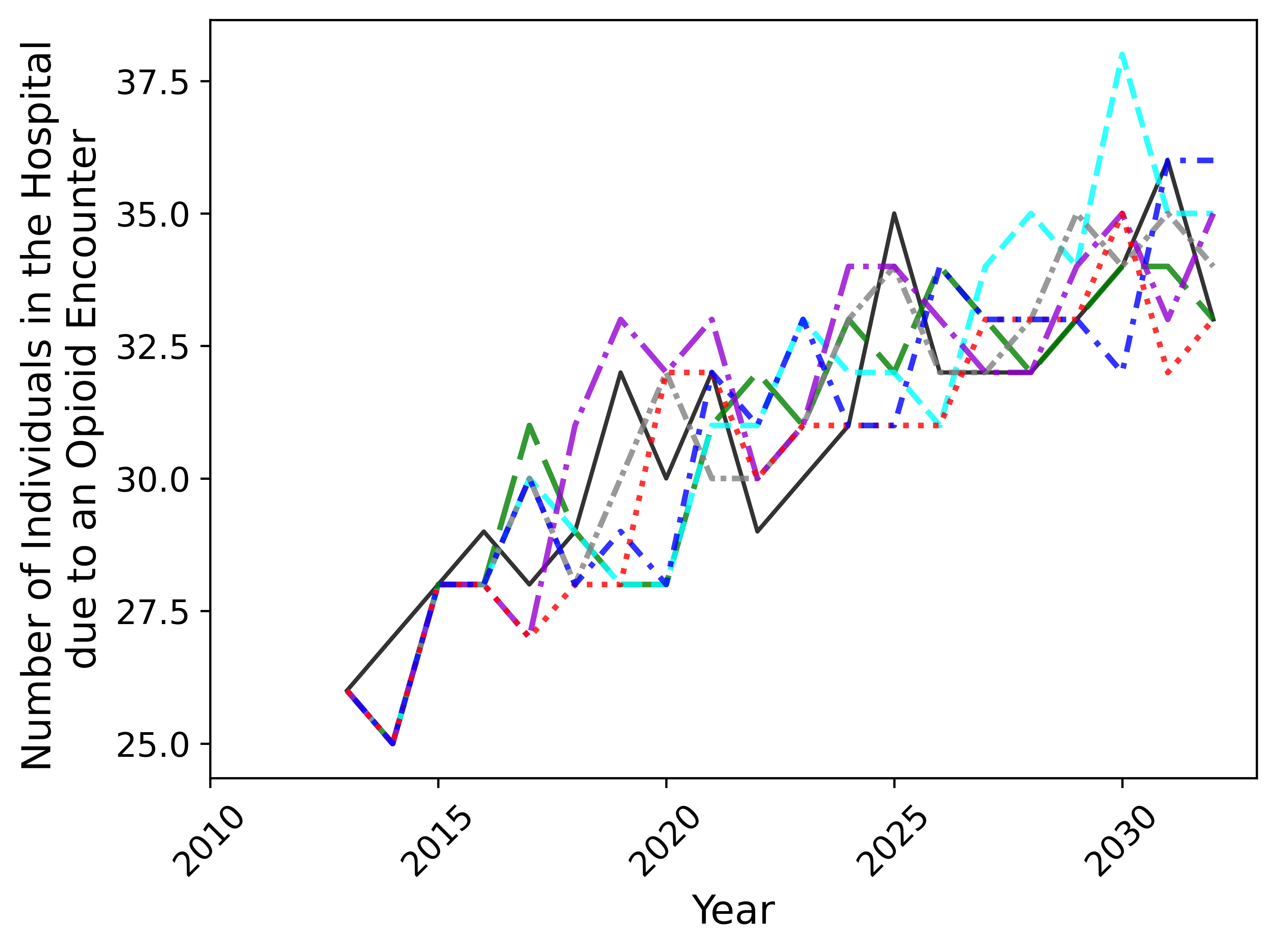}
    \caption{Hospital usage - upper 95\% PI}
\end{subfigure}
 \begin{subfigure}[b]{.48\linewidth}
\centering
    \includegraphics[width=.9\linewidth]{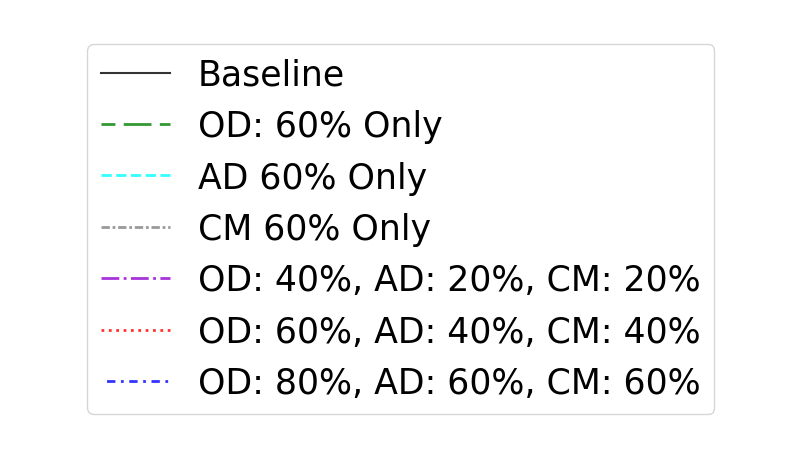}
\end{subfigure}
\caption{Upper 95\%  PI of resource usage over time}
\label{fig:Capacity}
\end{figure}

{Since the reported societal cost savings omits implementation costs, one could reduce the reported savings in this study by incorporating a fixed amount to initiate the program and yearly operating costs. Operating costs might include the costs of OUD treatment, program coordinators, additional OUD treatment staff, and overhead for OUD facilities. To provide some context, we estimate that implementing an intervention policy costs approximately 10 times less than the model's estimated cost reduction. For example, in 2016, the city of Madison, WI received \$700,000 in Federal funding to implement and study an arrest diversion program for two years \citep{city_of_madison_madison_2016}. MPD received $\$1.2$ million of funding in 2019 for two additional years to expand the arrest diversion program to cover the entirety of Dane County and incorporate additional pathways to treatment \citep{city_of_madison_madison_2019}. Extrapolating the program implementation cost at $1.2$ million for two years, we estimate the implementation would cost \$6 million over 10 years.
By comparison, our case study of Dane County, where Madison is located, estimates that successfully diverting 20\% of eligible individuals to arrest diversion would yield a mean cost savings of $\$102.61$ million over ten years or $10.26$ times the cost of Madison AD policy implementation. We assume that other interventions would be comparable in the ratio of implementation costs to societal costs.} 
{\subsection{CJS, OUD Treatment, and Hospital Usage}}
\noindent Figure \ref{fig:Capacity} shows the upper 95\% PI usage of the CJS due to opioid-related arrests, OUD treatment usage, and hospital usage due to an opioid encounter for each year. 
The upper 95\% PI can be interpreted as the predicted maximum resource usage at a given time. If the capacity of a resource is below the predicted maximum usage, there is a risk of resource shortages at some point in the year. Figure \ref{fig:Capacity}(a) shows that implementing an AD-focused policy reduces the number of individuals arrested for an opioid-related offense in the CJS. For example, implementing policy-mixes with AD at 60\%, i.e., (60, 22, 0) and (60, 80, 60), reduces the CJS maximum population by a magnitude of 35. Figure \ref{fig:Capacity}(b) shows that for any additional policy, usage of OUD treatment goes up and compounds with more/higher policies. For example, implementing the policy-mix of (60, 80, 60) requires 1.3 to 1.8 times the amount of OUD treatment capacity than the base model. 
Additionally, we observe a positive trend in the demand for OUD treatment for all policies, including the base model. Therefore, even if jurisdictions implement no treatment policies, due to the nature of the opioid crisis, demand for OUD treatment capacity is predicted to increase on average by 2\% to 5\% each year. Increasing OUD treatment capacity is important for jurisdictions to consider since most treatment centers are currently at or near capacity \citep{jones_national_2015}. Additionally, expansion of OUD treatment is needed when implementing a treatment policy that redirects individuals to facilities already at or near capacity. Lastly, Figure \ref{fig:Capacity}(c) shows that hospital usage does not significantly differ in any treatment policy.
{\subsection{Re-Hospitalization, Re-Arrest, and OUD Treatment Re-Start Rates}}
\noindent Table \ref{tab:perPerson_Results} shows the mean and  95\% standard error of the opioid-related re-arrest rate, opioid-related re-hospitalization rate, and OUD treatment re-start rate for the years 2023, 2027, and 2032. The opioid-related re-arrest rate includes both diverted and non-diverted arrests. These rates are calculated as follows for $ i\in \{\text{CJS, Hospital, OUD Treatment}\} $:
\begin{align*}
\text{rate}_{i} = \left(\frac{ \text{\# of }i \text{ episodes in a given year}}{\text{\# of individuals that used the } i  \text{ system in a given year}} -1 \right) * 100\%  
\end{align*}
Table \ref{tab:perPerson_Results} shows a statistically significant decrease in re-arrests for any AD and CM policy. For example, {the AD-focused p}olicy (60, 22, 0) reduces the re-arrest rate by {0.32} compared to the base model in 2032. We also see statistically significant decreases in re-hospitalization when implementing the OD policy at any level. All policies increase the number of individuals in and out of OUD treatment. Therefore, OUD treatment policies may help reduce ``frequent fliers" in the CJS and hospital systems while redirecting them to treatment resources.  
Additional results are in Appendix \ref{Appendix:Results}. 
{\spacingset{1}
\begin{sidewaystable}[htbp]
\centering
\caption{Simulated mean $\pm$ standard error (p-value) of Year{s' 2023, 2027, and }2032 opioid-related re-arrest, hospitalization, and treatment re-start rates for all scenarios}
\label{tab:perPerson_Results}
\resizebox{\textwidth}{!}{
\begin{tabular}{|c|c|c|c|c|c|c|c|c|c|c|c|c|c|c|c|c|c|c|}
\hline
\multicolumn{2}{|c|}{Scenario} & \multicolumn{3}{c|}{Re-Hospitalisation rate} & \multicolumn{3}{c|}{Re-Arrest rate} & \multicolumn{3}{c|}{OUD Treatment Re-Start rate} \\ \hline
\multicolumn{2}{|c|}{\multirow{2}{*}{AD (\%), OD (\%), CM (\%)}} & \multicolumn{3}{c|}{mean (\%) $\pm$ se (p-value)} & \multicolumn{3}{c|}{mean (\%) $\pm$ se (p-value)} & \multicolumn{3}{c|}{mean (\%) $\pm$ se (p-value)}  \\   \cline{3-11}
 \multicolumn{2}{|c|}{} & 2023 & 2027 & 2032 & 2023 & 2027 & 2032 & 2023 & 2027 & 2032\\ \hline \hline
\multirow{2}{*}{\rotatebox[origin=c]{90}{\parbox[t]{8mm}{\spacingset{1} \centering Base \\ Model}}} & \multirow{2}{*}{0, 22, 0} & \multirow{2}{*}{5.18 $\pm$ 0.02} & \multirow{2}{*}{5.19 $\pm$ 0.02 }& \multirow{2}{*}{5.22 $\pm$ 0.019 }& \multirow{2}{*}{0.76 $\pm$ 0.014} & \multirow{2}{*}{0.8 $\pm$ 0.014} & \multirow{2}{*}{0.76 $\pm$ 0.013 }& \multirow{2}{*}{0.46 $\pm$ 0.006} & \multirow{2}{*}{0.46 $\pm$ 0.006} & \multirow{2}{*}{0.46 $\pm$ 0.006}\\ &&&&&&&&&&\\
\hline \hline
\multirow{5}{*}{\rotatebox[origin=c]{90}{\parbox[t]{15mm}{\spacingset{1}\centering Arrest \\ Diversion}}} & \multirow{1}{*}{20, 22, 0} & 5.18 $\pm$ 0.02 & 5.17 $\pm$ 0.019 & 5.19 $\pm$ 0.019 &\bf{ 0.66 $\pm$ 0.013$^{**}$} &\bf{ 0.64 $\pm$ 0.014$^{**}$} &\bf{ 0.65 $\pm$ 0.013$^{**}$} &\bf{ 0.51 $\pm$ 0.006$^{**}$} &\bf{ 0.49 $\pm$ 0.006$^{**}$} &\bf{ 0.51 $\pm$ 0.006$^{**}$}\\
 & \multirow{1}{*}{40, 22, 0} & 5.19 $\pm$ 0.021 & 5.2 $\pm$ 0.021 & 5.18 $\pm$ 0.02 &\bf{ 0.55 $\pm$ 0.012$^{**}$} &\bf{ 0.54 $\pm$ 0.012$^{**}$} &\bf{ 0.56 $\pm$ 0.012$^{**}$} &\bf{ 0.55 $\pm$ 0.006$^{**}$} &\bf{ 0.56 $\pm$ 0.006$^{**}$} &\bf{ 0.56 $\pm$ 0.006$^{**}$}\\
 & \multirow{1}{*}{60, 22, 0} & 5.19 $\pm$ 0.021 & 5.21 $\pm$ 0.02 & 5.19 $\pm$ 0.019 &\bf{ 0.47 $\pm$ 0.011$^{**}$} &\bf{ 0.44 $\pm$ 0.011$^{**}$} &\bf{ 0.44 $\pm$ 0.01$^{**}$} &\bf{ 0.6 $\pm$ 0.006$^{**}$} &\bf{ 0.59 $\pm$ 0.006$^{**}$} &\bf{ 0.6 $\pm$ 0.005$^{**}$}\\
 & \multirow{1}{*}{80, 22, 0} & 5.22 $\pm$ 0.02 & 5.15 $\pm$ 0.02 & 5.18 $\pm$ 0.019 &\bf{ 0.34 $\pm$ 0.009$^{**}$} &\bf{ 0.33 $\pm$ 0.01$^{**}$} &\bf{ 0.35 $\pm$ 0.009$^{**}$} &\bf{ 0.65 $\pm$ 0.006$^{**}$} &\bf{ 0.65 $\pm$ 0.006$^{**}$} &\bf{ 0.65 $\pm$ 0.006$^{**}$}\\
 & \multirow{1}{*}{100, 22, 0} & 5.14 $\pm$ 0.02 & 5.15 $\pm$ 0.02 & 5.19 $\pm$ 0.019 &\bf{ 0.23 $\pm$ 0.008$^{**}$} &\bf{ 0.24 $\pm$ 0.007$^{**}$} &\bf{ 0.21 $\pm$ 0.007$^{**}$} &\bf{ 0.68 $\pm$ 0.006$^{**}$} &\bf{ 0.7 $\pm$ 0.006$^{**}$} &\bf{ 0.69 $\pm$ 0.006$^{**}$}\\
\hline \hline
\multirow{5}{*}{\rotatebox[origin=c]{90}{\parbox[t]{15mm}{\spacingset{1}\centering Overdose \\ Diversion}}} & \multirow{1}{*}{0, 30, 0} &\bf{ 4.81 $\pm$ 0.02$^{**}$} &\bf{ 4.79 $\pm$ 0.02$^{**}$} &\bf{ 4.84 $\pm$ 0.019$^{**}$} & 0.79 $\pm$ 0.015 & 0.77 $\pm$ 0.014 & 0.77 $\pm$ 0.013 &\bf{ 0.58 $\pm$ 0.006$^{**}$} &\bf{ 0.58 $\pm$ 0.006$^{**}$} &\bf{ 0.58 $\pm$ 0.006$^{**}$}\\
 & \multirow{1}{*}{0, 45, 0} &\bf{ 4.03 $\pm$ 0.019$^{**}$} &\bf{ 4.08 $\pm$ 0.018$^{**}$} &\bf{ 4.05 $\pm$ 0.017$^{**}$} & 0.78 $\pm$ 0.014 &\bf{ 0.73 $\pm$ 0.014$^{*}$} & 0.76 $\pm$ 0.014 &\bf{ 0.81 $\pm$ 0.007$^{**}$} &\bf{ 0.8 $\pm$ 0.007$^{**}$} &\bf{ 0.82 $\pm$ 0.007$^{**}$}\\
 & \multirow{1}{*}{0, 60, 0} &\bf{ 3.3 $\pm$ 0.017$^{**}$} &\bf{ 3.32 $\pm$ 0.016$^{**}$} &\bf{ 3.35 $\pm$ 0.015$^{**}$} & 0.76 $\pm$ 0.014 &\bf{ 0.76 $\pm$ 0.014$^{*}$} & 0.78 $\pm$ 0.014 &\bf{ 1.03 $\pm$ 0.007$^{**}$} &\bf{ 1.04 $\pm$ 0.007$^{**}$} &\bf{ 1.07 $\pm$ 0.007$^{**}$}\\
 & \multirow{1}{*}{0, 75, 0} &\bf{ 2.6 $\pm$ 0.014$^{**}$} &\bf{ 2.59 $\pm$ 0.014$^{**}$} &\bf{ 2.6 $\pm$ 0.013$^{**}$} & 0.76 $\pm$ 0.015 &\bf{ 0.74 $\pm$ 0.014$^{*}$} & 0.75 $\pm$ 0.013 &\bf{ 1.28 $\pm$ 0.008$^{**}$} &\bf{ 1.27 $\pm$ 0.008$^{**}$} &\bf{ 1.28 $\pm$ 0.007$^{**}$}\\
 & \multirow{1}{*}{0, 90, 0} &\bf{ 1.9 $\pm$ 0.012$^{**}$} &\bf{ 1.88 $\pm$ 0.012$^{**}$} &\bf{ 1.87 $\pm$ 0.011$^{**}$} & 0.75 $\pm$ 0.015 & 0.77 $\pm$ 0.014 & 0.76 $\pm$ 0.013 &\bf{ 1.51 $\pm$ 0.008$^{**}$} &\bf{ 1.5 $\pm$ 0.008$^{**}$} &\bf{ 1.51 $\pm$ 0.008$^{**}$}\\
\hline \hline
\multirow{5}{*}{\rotatebox[origin=c]{90}{\parbox[t]{20mm}{\spacingset{1}\centering Case \\ Management}}} & \multirow{1}{*}{0, 22, 20} & 5.19 $\pm$ 0.02 & 5.21 $\pm$ 0.02 & 5.2 $\pm$ 0.018 &\bf{ 0.66 $\pm$ 0.013$^{**}$} &\bf{ 0.63 $\pm$ 0.013$^{**}$} &\bf{ 0.65 $\pm$ 0.012$^{**}$} &\bf{ 0.53 $\pm$ 0.006$^{**}$} &\bf{ 0.52 $\pm$ 0.006$^{**}$} &\bf{ 0.53 $\pm$ 0.005$^{**}$}\\
 & \multirow{1}{*}{0, 22, 40} & 5.18 $\pm$ 0.021 & 5.17 $\pm$ 0.019 &\bf{ 5.17 $\pm$ 0.02$^{*}$} &\bf{ 0.54 $\pm$ 0.012$^{**}$} &\bf{ 0.53 $\pm$ 0.011$^{**}$} &\bf{ 0.54 $\pm$ 0.011$^{**}$} &\bf{ 0.67 $\pm$ 0.006$^{**}$} &\bf{ 0.66 $\pm$ 0.006$^{**}$} &\bf{ 0.66 $\pm$ 0.006$^{**}$}\\
 & \multirow{1}{*}{0, 22, 60} & 5.18 $\pm$ 0.022 & 5.18 $\pm$ 0.02 &\bf{ 5.17 $\pm$ 0.019$^{*}$} &\bf{ 0.44 $\pm$ 0.01$^{**}$} &\bf{ 0.42 $\pm$ 0.01$^{**}$} &\bf{ 0.43 $\pm$ 0.01$^{**}$} &\bf{ 0.84 $\pm$ 0.006$^{**}$} &\bf{ 0.83 $\pm$ 0.006$^{**}$} &\bf{ 0.83 $\pm$ 0.006$^{**}$}\\
 & \multirow{1}{*}{0, 22, 80} & 5.17 $\pm$ 0.021 &\bf{ 5.13 $\pm$ 0.019$^{*}$} &\bf{ 5.16 $\pm$ 0.019$^{*}$} &\bf{ 0.3 $\pm$ 0.009$^{**}$} &\bf{ 0.3 $\pm$ 0.008$^{**}$} &\bf{ 0.3 $\pm$ 0.008$^{**}$} &\bf{ 1.03 $\pm$ 0.007$^{**}$} &\bf{ 1.01 $\pm$ 0.007$^{**}$} &\bf{ 1.02 $\pm$ 0.007$^{**}$}\\
 & \multirow{1}{*}{0, 22, 100} &\bf{ 5.13 $\pm$ 0.02$^{*}$} & 5.17 $\pm$ 0.021 &\bf{ 5.17 $\pm$ 0.02$^{*}$} &\bf{ 0.2 $\pm$ 0.007$^{**}$} &\bf{ 0.19 $\pm$ 0.007$^{**}$} &\bf{ 0.19 $\pm$ 0.007$^{**}$} &\bf{ 1.26 $\pm$ 0.007$^{**}$} &\bf{ 1.23 $\pm$ 0.007$^{**}$} &\bf{ 1.22 $\pm$ 0.007$^{**}$}\\
\hline \hline
\multirow{3}{*}{\rotatebox[origin=c]{90}{\parbox[t]{10mm}{\spacingset{1}\centering Policy \\ Mix}}} & \multirow{1}{*}{20, 40, 20} &\bf{ 4.27 $\pm$ 0.02$^{**}$} &\bf{ 4.33 $\pm$ 0.018$^{**}$} &\bf{ 4.3 $\pm$ 0.018$^{**}$} &\bf{ 0.58 $\pm$ 0.013$^{**}$} &\bf{ 0.58 $\pm$ 0.012$^{**}$} &\bf{ 0.53 $\pm$ 0.011$^{**}$} &\bf{ 0.87 $\pm$ 0.007$^{**}$} &\bf{ 0.85 $\pm$ 0.006$^{**}$} &\bf{ 0.86 $\pm$ 0.007$^{**}$}\\
 & \multirow{1}{*}{40, 60, 40} &\bf{ 3.29 $\pm$ 0.016$^{**}$} &\bf{ 3.28 $\pm$ 0.017$^{**}$} &\bf{ 3.29 $\pm$ 0.015$^{**}$} &\bf{ 0.41 $\pm$ 0.011$^{**}$} &\bf{ 0.42 $\pm$ 0.01$^{**}$} &\bf{ 0.4 $\pm$ 0.01$^{**}$} &\bf{ 1.31 $\pm$ 0.008$^{**}$} &\bf{ 1.3 $\pm$ 0.007$^{**}$} &\bf{ 1.3 $\pm$ 0.007$^{**}$}\\
 & \multirow{1}{*}{60, 80, 60} &\bf{ 2.33 $\pm$ 0.014$^{**}$} &\bf{ 2.3 $\pm$ 0.013$^{**}$} &\bf{ 2.32 $\pm$ 0.013$^{**}$} &\bf{ 0.29 $\pm$ 0.009$^{**}$} &\bf{ 0.3 $\pm$ 0.009$^{**}$} &\bf{ 0.3 $\pm$ 0.009$^{**}$} &\bf{ 1.76 $\pm$ 0.008$^{**}$} &\bf{ 1.75 $\pm$ 0.008$^{**}$} &\bf{ 1.75 $\pm$ 0.007$^{**}$}\\

 \hline 
\multicolumn{11}{l}{*{Statistically significant difference, with p-value $<0.05$, from the Base Model at a level of 0.05 using 2-sided paired t-test}}\\
\multicolumn{11}{l}{**{Statistically significant difference, with p-value $<0.001$, from the Base Model at a level of 0.05 using 2-sided paired t-test}}
\end{tabular} 
 }  
 \end{sidewaystable} 
 }
{\subsection{Discussion}}
 \noindent We note that for all policies, OUD treatments, opioid-related arrests, hospitalizations, deaths, and individuals actively using opioids increase over time. (See Figure \ref{fig:Full_Results} in Appendix \ref{Appendix:Results} for more details). This suggests that treatment policies alone are insufficient to reverse the positive trend of yearly active users, deaths, and hospital encounters. Due to a large number of individuals in the active use state that never interact with the \ac{CJS}, hospital, or treatment systems, prevention policies that aim to reduce the initiation of opioid use appears to be a valuable strategy \citep{united_nations_office_on_drugs_and_crime_international_2020}. In fact, \citet{ansari2024curbing} used their decision support tool that incorporated both a Susceptible, Infection, and Recover model and a Markov decision process to find the optimal budget mix between prevention and mitigating strategies. They find that prevention strategies should dominate the budget mix until the number of opioid-related fatality rates is greater than opioid access rates.  Other mitigating policies that could help reduce the worst effects of the opioid crisis (e.g., deaths) could be finding additional ways to connect individuals actively using opioids to treatment and harm reduction strategies \citep{greer_harm_2019}.
\subsection{Model Sensitivity}
 To assess model sensitivity, we used the (60, 80, 60) scenario and sampled 1024 Sobol sets across 41 parameters, ranging each parameter $\pm$ 5\% of their base model value. We ran each set with three replications for a total of 3,072 simulation runs. The sensitivity analysis in Appendix \ref{Appendix:Sensitivity} suggests that this paper's conclusions are not sensitive to model inputs, since the sensitivity analysis shows that the longer the model is run, the less sensitive the model is to the starting population. Additionally, the relationship between the model inputs and outputs is expected due to the model structure. 
 Care should be taken when estimating distributions for arcs (2)-(6) and (G) since these have the largest effect on base model outputs, particularly $\mu$ for these distributions. Overall, the sensitivity analysis suggests that changes to the model parameters would not affect the conclusions of this paper between the base model and treatment policy scenarios regarding opioid-related arrests, hospital encounters, treatments, and active use.
 \subsection{Policy Effect Comparison}
To compare policies and assess marginal returns, we evaluate the comparative effects between our three policies AD, OD, and CM, and each of the simulation main outputs: overdose-related deaths, number of hospital encounters, number of non-diverted arrests, number of OUD treatment, and number of individuals in the active state at the end of a given year. We use Ordinary Least Squares (OLS) regression, where the policies are variables against each output in years 2023, 2027, and 2032.  
Table \ref{tab:policycorr} shows the regression coefficients against each policy and the model outputs across 600 replications for each of the scenarios listed in Tables \ref{tab:AD_All_Results} and \ref{tab:perPerson_Results}. {The coefficients are interpreted as the improvement in the 1-year, 5-year and 10-year cumulative outcome per a 1\% increase of the policy implementation level.} 

{As shown in Table \ref{tab:policycorr}, all of the policies are consistent in their direction of influence over model outputs. This means the policies are all negative or all positive for a given year and model output. This means that the policies are complementary in nature and that increasing any policy will likely improve model outputs. We also see that the OD policy has the highest magnitude of coefficients, followed by the CM policy. Therefore, {if one could only improve a single policy{, assuming the same level of implementation increase in all policies}, an improvement in OD would likely have the largest impact, followed by the CM policy and then the AD policy.} This makes sense, given the model input parameters, since a larger number of individuals experience a hospital encounter than with the CJS It also makes sense that the CM policy has a larger impact than the AD policy since it can engage any individual in the CJS with past opioid misuse rather than just individuals who are in the CJS for an opioid-related crime. } 

{\spacingset{1}
\begin{table}[htbp]
\centering
\caption{OLS regression of policies vs. model outputs post policy implementation}
\label{tab:policycorr}
\resizebox{\textwidth}{!}{
\color{black}\begin{tabular}{|c|c|c|c|c|c|}
\hline
\multicolumn{1}{|c|}{\multirow{2}{*}{Year}} & \multicolumn{1}{c|}{\multirow{2}{*}{Model Output}} & \multicolumn{4}{c|}{Regression Coefficients (p-value)}\\  \cline{3-6}  
 & & \multicolumn{1}{c|}{Intercept (p-value)} & \multicolumn{1}{c|}{OD (p-value)} & \multicolumn{1}{c|}{AD (p-value)}  & \multicolumn{1}{c|}{CM (p-value)}  \\  \hline \hline
 2023 & Opioid-related death & 96.42 ($<$ 0.001)&-0.03 ($<$ 0.001)&-0.0(0.29)&-0.02 ($<$ 0.001) \\ \hline
 2023 & Opioid-related arrest & 625.2 ($<$ 0.001)&-0.16 ($<$ 0.001)&-0.04 ($<$ 0.001)&-0.08 ($<$ 0.001) \\ \hline
 2023 & Hospital encounters & 2198.77 ($<$ 0.001)&-0.55 ($<$ 0.001)&-0.09 ($<$ 0.001)&-0.48 ($<$ 0.001) \\ \hline
 2023 & Treatment & 1888.38 ($<$ 0.001)&19.2 ($<$ 0.001)&5.18 ($<$ 0.001)&12.98 ($<$ 0.001) \\ \hline
 2023 & Active use & 17859.79 ($<$ 0.001)&-4.82 ($<$ 0.001)&-0.85 ($<$ 0.001)&-3.29 ($<$ 0.001) \\ \hline\hline
 2027 & Opioid-related death & 499.13 ($<$ 0.001)&-0.12 ($<$ 0.001)&-0.02 ($<$ 0.001)&-0.09 ($<$ 0.001) \\ \hline
 2027 & Opioid-related arrest & 3238.95 ($<$ 0.001)&-0.77 ($<$ 0.001)&-0.14 ($<$ 0.001)&-0.58 ($<$ 0.001) \\ \hline
 2027 & Hospital encounters & 11388.14 ($<$ 0.001)&-2.56 ($<$ 0.001)&-0.45 ($<$ 0.001)&-2.3 ($<$ 0.001) \\ \hline
 2027 & Treatment & 9812.7 ($<$ 0.001)&100.18 ($<$ 0.001)&26.93 ($<$ 0.001)&65.73 ($<$ 0.001) \\ \hline
 2027 & Active use & 92765.62 ($<$ 0.001)&-22.67 ($<$ 0.001)&-3.96 ($<$ 0.001)&-16.05 ($<$ 0.001) \\ \hline\hline
 2032 & Opioid-related death & 1042.9 ($<$ 0.001)&-0.26 ($<$ 0.001)&-0.05 ($<$ 0.001)&-0.17 ($<$ 0.001) \\ \hline
 2032 & Opioid-related arrest & 6736.45 ($<$ 0.001)&-1.5 ($<$ 0.001)&-0.27 ($<$ 0.001)&-1.02 ($<$ 0.001) \\ \hline
 2032 & Hospital encounters & 23670.73 ($<$ 0.001)&-4.93 ($<$ 0.001)&-0.89 ($<$ 0.001)&-3.78 ($<$ 0.001) \\ \hline
 2032 & Treatment & 20451.01 ($<$ 0.001)&209.32 ($<$ 0.001)&56.29 ($<$ 0.001)&135.98 ($<$ 0.001) \\ \hline
 2032 & Active use & 193153.61 ($<$ 0.001)&-44.0 ($<$ 0.001)&-7.79 ($<$ 0.001)&-28.5 ($<$ 0.001) \\ \hline
\multicolumn{6}{l}{{}}
\end{tabular} 
 }  
 \end{table} 
 }

{\subsection{Study Limitations}}
The model makes various assumptions based on data availability and to manage model complexity. The model does not generate individual characteristics except for age, which is used to estimate the time until a non-opioid-related death. The model incorporates some individual historical dependencies. The time in the inactive state depends on whether the individual came from the OUD treatment, \ac{CJS}, {hospital or ED}, or active use state. Therefore, four separate distributions estimate the time in the inactive opioid use state.  However, when individuals start using opioids  (i.e., move to the active use state), they use the same distributions as stated in Table \ref{tab:inputs} to generate the next time until they are arrested, start OUD treatment, experience a hospital encounter, stop using opioids on their own, and experience an opioid-related death. {While the model accounts for the individual's previous state,  it assumes ``memoryless'' cumulative state durations 
that do not account for the cumulative effects of treatment, hospitalizations, and arrests. {In reality, some state transitions may be history-dependent. For example,} \citet{nosyk_characterizing_2014} conclude that individuals who use heroin show cumulative time incarcerated delays cessation from using heroin, and \citet{nosyk_proportional_2009} conclude that subsequent treatment episodes of \ac{MAT} with methadone tend to be longer in duration. Therefore, our model may overestimate active use to treatment enrollment for individuals with larger criminal histories while underestimating the duration of treatment for individuals in recurrent \ac{OUD} treatment. We acknowledge there may be other cumulative impacts that are still being studied for individuals with opioid use that are not incorporated in this model.} Due to the sparsity of data, no other distributions take individual characteristics into account. As more data becomes available, future model versions could incorporate more individualistic data on risk and protective factors. To help obtain that level of data, this study could be used to justify additional data collection and information sharing.

{For modeling opioid-related deaths, we estimate county-level opioid use prevalence by redistributing the NSDUH state-level opioid use prevalence based on county-level overdose deaths. We acknowledge that the proportion of opioid use prevalence and opioid-related deaths likely do not exactly occur at consistent rates in the state of Wisconsin, which could impact the number of opioid-related deaths in Dane County estimated by the model. Additionally, NSDUH has been shown to underestimate the prevalence of drug use \citep{reuter_heroin_2021}. Therefore, the NSDUH data is another source of underestimation in the model when estimating both the amount of opioid use prevalence and opioid-related deaths since it relies on prevalence. This limitation underscores the importance of jurisdictions collecting accurate data regarding opioid use prevalence.}

Due to a lack of criminal justice and treatment data differentiating between misused prescription opioids and illicitly manufactured opioids at the county level, we model opioid use generally. Therefore, opioid use and OUD are defined to encompass the effects of all opioids (e.g., prescription opioids, heroin, fentanyl) for all estimates. Similarly, we do not model seasonality; instead, we utilize and report yearly and cumulative estimates. Additionally, this study does not examine different types of OUD treatment and models OUD treatment in general. As discussed, various treatment options have various success rates and widespread use. As better treatments become more widely available, we expect even better outcomes as individuals stay inactive for longer. 

Individuals are assumed to complete their proven intervention successfully. However, in practice, individuals can be referred to a program but not complete it. In the context of an arrest, this is problematic since this would delay their time to the CJS state, staying in the active opioid use state for longer and leaving them exposed to additional risk of an opioid-related overdose, arrest, or death that is not accounted for in our current model. Additionally, individuals may continue to use opioids while in \ac{OUD} treatment. While those in treatment, especially those using \ac{MAT}, have better outcomes that lower the risk of overdose and death than those not in treatment \citep{ma_effects_2019}, a meta-analysis showed overdose mortality was 2.6 and 1.4 per 1000 persons in treatment using methadone and buprenorphine, respectively \citep{sordo_mortality_2017}. Therefore, our model may underestimate opioid-related deaths and active opioid use among individuals in \ac{OUD} treatment.

Lastly, we do not include the cost of implementing various diversions; we only include their cumulative savings of societal costs after policy implementation. {Additional fixed and variable cumulative policy implementation costs could be subtracted from the expected societal cost savings to get a total savings estimate that accounts for implementation.} 

\section{Conclusion}\label{s:conclusion}
The opioid crisis has led to hundreds of thousands of opioid-related deaths, overdoses, arrests, and instances of substance use disorder while straining many organizations, systems, and personnel in hospitals, treatment facilities, and police departments. As a result, many individuals who use opioids never receive or finish the treatment they need and instead may have many interactions with hospitals or the \ac{CJS}. Each of these interactions provides an opportunity to divert individuals who use opioids to \ac{OUD} treatment.

This paper introduces a new DES model to evaluate the combinations of three OUD treatment policies that divert individuals who use opioids to OUD treatment{:} \textit{arrest diversion}, \textit{overdose diversion}, and \textit{re-entry case management}. Through analyzing a variety of policy-mix implementations, the study offers a versatile framework for evaluating policy efficacy at different implementation levels. This study projects opioid-related outcomes within a community through metrics such as opioid use, fatalities, hospital encounters, OUD treatment starts, and societal costs over time. We use public and community-sourced data to populate the case study based on Dane County, Wisconsin. The outlined data collection and estimation methods specifically designed for community-level simulation projects have enhanced the model's applicability. 

The results demonstrate that policies that create new pathways and programming by utilizing treatment services can lead to more opioid-resilient communities (e.g., communities with fewer adverse opioid-related outcomes). All three treatment policies can successfully create observable reductions in opioid-related arrests, hospital encounters, and opioid use when successfully diverting at least 20\% of eligible individuals. 
These findings offer critical insights into the economic implications of different intervention strategies and can guide resource allocation and implementation decisions for jurisdictions. The three policies are shown to be complementary, with the largest impacts from overdose diversion, followed by case management, and then arrest diversion{, assuming the same level of implementation increase in all policies}. When all three are implemented, it could lead to potential 10-year cumulative societal savings of over {\$500 million}, less implementation costs{, which we estimated to be approximately one tenth of the savings of the case study's arrest diversion program.} 

It is important to acknowledge the limitations of the model and analyses.
 The simplifications inherent in the model may not capture the full complexity of the real-world dynamics of the opioid crisis. This simulation could be improved by linking the data of individuals who enter and exit throughout these various systems. As data improves, a future version of the model could incorporate separate distributions based on individual state histories, such as increasing the likelihood of an arrest for an individual with a previous arrest. In addition, the sensitivity analysis in this paper can guide future research and highlight areas where further data collection and analysis are needed to improve the accuracy and reliability of the model. 

Extending simulations to encompass various opioid-related strategies like harm reduction, supply reduction, and prevention may hold the key to mitigating the local opioid crisis. Hence, decision-makers could allocate resources to a mix of policies for optimal impact in reducing opioid-linked fatalities, overdoses, and arrests. Subsequent research could concentrate on long-term policy effects, cross-policy and regional comparisons, and incorporating social determinants of health. These inquiries will advance our knowledge of intervention sustainability and evidence-based choices to address the broader opioid crisis. 

\section*{Acknowledgments}
 The authors would like to thank Captain Joseph Balles (retired) from the Madison Police Department and Dr. Aleksandra Zgierska, Professor of Family and Community Medicine, Public Health Sciences, and Anesthesiology and Perioperative Medicine at Penn State College of Medicine, for their subject matter expertise in providing information and feedback related to the data, and its analysis and modeling. The authors would like to thank the anonymous reviewers and the associate editor for their valuable comments and suggestions to improve the quality of the paper.
\section*{Disclosure Statement}
The authors report there are no competing interests to declare.
\section*{Data availability statement} 
The data and code that support the findings of this study are available at 
\url{https://doi.org/10.5281/zenodo.15498919}. These data were derived from resources available in the public domain listed and cited in Table \ref{tab:inputs}.
\section*{Funding Details}
This work was funded by the National Science Foundation Award 1935550. The views and conclusions contained in this document are those of the authors and should not be interpreted as necessarily representing the official policies, either expressed or implied, of the National Science Foundation.

\bibliographystyle{apalike} 
\spacingset{1}

\appendix
  \acrodef{AD}{arrest diversion}
 \acrodef{aOR}{adjusted Odds Ratio}
 \acrodef{CACE}{compiler average causal effect}
  \acrodef{CAD}{computer-aided-dispatch}
   \acrodef{CARES}{Community Alternative Response Emergency
Services}
    \acrodef{C-CRM}{conditional crisis response model}
  \acrodef{CDF}{cumulative density function}
  \acrodef{CI}{confidence interval}
 \acrodef{CIT}{crisis intervention team}
 \acrodef{CJS}{criminal justice system}
 \acrodef{CM}{case management}
 \acrodef{CRM}{crisis response model}
 \acrodef{CCSQ}{conditional crisis service quality}
 \acrodef{CSQ}{crisis service quality}
 \acrodef{DES}{discrete event simulation}
 \acrodef{DSOR}{decision science and operations research}
 \acrodef{EBP}{evidence-based policing}
  \acrodef{ED}{emergency department}
 \acrodef{EMS}{emergency medical services}
  \acrodef{FIFO}{first in, first out}
  \acrodef{FRM}{follow-up response model}
 \acrodef{HC}{historical comparison}
  \acrodef{HE}{hospital encounter}
 \acrodef{ITT}{intention-to-treat}
 \acrodef{JRM}{joint response model}
  \acrodef{LN}{log-normal}
 \acrodef{MARI}{Madison Addiction Recovery Initiative}
 \acrodef{MAT}{medication-assisted treatment}
 \acrodef{MCIT}{mobile crisis intervention team}
 \acrodef{MPD}{Madison Police Department}
 \acrodef{OD}{overdose diversion}
 \acrodef{OUD}{opioid use disorder}
 \acrodef{PD}{Police Department}
  \acrodef{PI}{prediction interval}
  \acrodef{P-P}{probability-probability}
  \acrodef{PRM}{police response model}
 \acrodef{PWMI}{persons with mental illness}
 \acrodef{PRM}{police response model}
 \acrodef{SPD}{Seattle Police Department}
 \acrodef{SUD}{substance use disorder}
 \acrodef{Q-Q}{quantile-quantile}
  \acrodef{UK}{United Kingdom}
 \acrodef{US}{United States}
 \acrodef{USD}{United States Dollar}
  \acrodef{WI}{Wisconsin}
   \acrodef{WA}{Washington}
\section{Parameter Details}
\spacingset{1}
\label{Appendix:ParamDesc}
{This appendix describes in detail how each of the parameters and distributions used in the simulation were estimated.}
\subsection{Estimating Dane County Opioid Use Prevalence} \label{App:DaneCounty}
This section describes how we estimated Dane County opioid use prevalence, i.e., ${p}^{\textit{DC}}_i$ for all years $i = \text{2016, ..., {2020}}$, which aided in the estimation of various input parameters for the case study. {Opioid use prevalence could be described as the number of individuals that are actively using opioids in a given time frame.} The SAMSHA National Survey of Drug Use and Health Survey (NSDUH) estimates ``Misuse in the Past Year, 12+, Wisconsin, Estimated numbers (in Thousands)" where heroin use prevalence, ${p}^{\textit{WI,H}}_i$, and pain reliever misuse prevalence, ${p}^{\textit{WI,P}}_i$ are estimated (SAMHSA, \citeyear{substance_abuse_and_mental_health_services_administration_samhsa_national_2016}). This data was collected from 2016 to {2020},  since in the 2016 survey the NSDUH survey revised its pain reliever sub-types. Note these are state estimates, and therefore, we obtain Dane county\textquotesingle s specific estimates by multiplying the statewide prevalence estimates by Dane County\textquotesingle s ``share" of Wisconsin opioid-related deaths, $s^{WI,H}_i$ and $s^{WI,P}_i$ for $i = 2016 $, ..., ${2020}$ (i.e., the proportion of opioid-related heroin and prescription-related deaths, respectively),  as shown in equation (\ref{eq:A2}).
\begin{align}
    {p}^{\textit{DC,H}}_i & = {p}^{\textit{WI,H}}_is^{\textit{WI,H}}_i, & \qquad 
    {p}^{\textit{DC,P}}_i &= {p}^{\textit{WI,P}}_is^{\textit{WI,P}}_i, & \text{for } i = \text{2016, ..., {2020.}} \label{eq:A2}
\end{align}

To calculate the Dane County ``share", the number of opioid-related deaths in Dane County and WI are taken from the CDC Wonder online database of multiple cause of death 1999-{2020} \citep{national_center_for_health_statistics_cdc_2022}. For a given year $i = \text{2016, ..., {2020}}$, the number of opioid-related deaths involving a prescription opioid, $d_i^{P}$, is defined as death at age 12+ and had an ICD-10 death code indicating drug-induced causes as an underlying cause of death (i.e., X40-X44, X60-X64, X85, or Y10-Y14), and a contributing cause of death of T402 (natural and semi-synthetic opioid), T403 (methadone), T404 (synthetic opioid other than methadone), or T406 (unknown). Similarly, the number of opioid-related deaths involving heroin, $d_i^{H}$, is defined as that of a prescription opioid, but with a contributing cause of death of T401 (heroin). As shown in equation (\ref{eq:A1}), $s_i^{WI,H}$ and $s_i^{WI,P}$, for a given year $i = \text{2016, ..., {2020}}$, can be solved from the number of deaths in Dane County and WI.
\begin{align}
    s^{\textit{WI,H}}_i & = \frac{d^{\textit{DC,H}}_i}{d^{\textit{WI,H}}_i}, & \qquad s^{\textit{WI,P}}_i &= \frac{d^{\textit{DC,P}}_i}{d^{\textit{WI,P}}_i}, &\text{for } i = \text{2016, ... , {2020.}}  \label{eq:A1}
\end{align}

Finally, equation (\ref{eq:A3}) shows that for a given year, total opioid use prevalence can be estimated by the sum of Dane County heroin ${p}^{\textit{DC,H}}$, and prescription opioid use, ${p}^{\textit{DC,P}}$, prevalence. 
\begin{align}
    &&{p}^{\textit{DC}}_i &= {p}^{\textit{DC,H}}_i + {p}^{\textit{DC,P}}_i, & \text{for } i = \text{2016, ... , {2020.}}  \label{eq:A3}
\end{align}

{\subsection{Starting Population Size and State} \label{App:StartPop}}
{\noindent The number of people in the starting population is modeled using a triangular distribution. Parameters for the triangular distribution are estimated using the Dane County prevalence, i.e., $p_i^{DC}$ for $i=2016,...,2019$, which is described in Appendix \ref{App:DaneCounty}. The starting population had a minimum of $\min(p_i^{DC} )=27,299$ people, a mode estimated as $\text{avg}(p_i^{DC} )=34,224$ people, and a maximum of  $\max(p_i^{DC} )=44,087$ people, for $i=2016,...,2019$. We estimate the mode of the triangular distribution as the average of the four years of data available.}

{Expert opinion and the total number of individuals in the starting population for a given simulation run are used to estimate the total number of individuals starting in a specific state.} Expert opinion is used to estimate the total number of individuals in the hospital {or ED}, the \ac{CJS}, and OUD treatment states. Expert opinion is used to estimate triangular distributions with parameters (min, mode, max) for the number of individuals starting in these states, with parameters (5, 11, 15) for individuals in the hospital {or ED} due to opioids (15, 25, 50) for individuals in the CJS state due to an opioid-related arrest, and (300, 450, 500) for individuals in OUD treatment. This estimate is divided by the starting population to set the multinomial probability that an individual in the starting population starts in the hospital {or ED}, the \ac{CJS}, and OUD treatment states.
Neither subject matter expert in policing nor medicine were able to give a reasonable estimate for the number of individuals actively using opioids nor the number of individuals that had used opioids in the past. There is also a gap in public data regarding the number of people who have used opioids and are actively using opioids. Therefore, we sampled a triangular distribution with parameters (20\%, 40\%, 80\%) to set the multinomial probability that an individual in the starting population starts in the active opioid use state.

{\subsection{Estimating the Log-normal Distributions with Limited Data}\label{App:EstLN}}
\noindent Due to a lack of data sources for {many model parameters}, we use the method described in Section 6.11 in \citet{law_simulation_2007} to estimate $\mu$ and $\sigma$ for LN distributions (2)-(6) and (8). The method is shown in equations (\ref{eq:muEst})-(\ref{eq:c}), where $m$ is the most likely value, $\gamma$ is the location parameter, $z_q$ is the $q$-quantile of the normal distribution. The inputs are estimated from the sources in Table \ref{tab:inputs} and are described in detail in the following sections. This method is recommended for estimating log-normal distributions with parameters $\Tilde{\mu}$ and $\Tilde{\sigma}$ when limited data is available. When using this method, it is best to use a larger quantile (e.g., $z_{0.9}$ ) to prevent $\Tilde{\sigma}$ from being underestimated \citep{law_simulation_2007}. 
\begin{align}
    \Tilde{\mu} =  &  \ln(m - \gamma) + \Tilde{\sigma}^2,   \label{eq:muEst} \\
    \Tilde{\sigma} =  & \frac{-z_q + \sqrt{z_q^2 - 4c}}{2}, \label{eq:sigma}\\
    c =       & \ln{\frac{m-\gamma}{x_q-\gamma}}. \label{eq:c}
\end{align}

{\subsection{Estimating Initiation and Prevalence Age} \label{App:Age}}
{\noindent We estimated two separate distributions to sample individuals' ages. Individuals generated in the starting population, described in Appendix A.2, use the prevalence age distribution. In contrast, individuals not in the starting population (i.e., during the warm-up and simulation periods) use the first initiation age distribution. Both age distributions use a log-normal distribution with a location parameter equal to 12 since we assume individuals are at least 12 years of age. {We also assume a maximum enter age of 105 years; this is done by truncating the age log-normal distributions at 105 years by resampling until a value below 105 years is generated. We estimate $\mu$ and $\sigma$ using \citetalias{substance_abuse_and_mental_health_services_administration_samhsa_2019_2019} and equations (\ref{eq:muEst})-(\ref{eq:c}).  From \citetalias{substance_abuse_and_mental_health_services_administration_samhsa_2019_2019}, we use Table 1.19A to estimate the prevalence age distribution. We can estimate mode $m=32$ and quantile ${x}_{0.933}=65$, resulting in  $\mu=3.74$ and $\sigma=0.49$. The mode was selected by taking the midpoint of the age range with the largest population percentage (e.g., $\frac{30+34}{2}=32$) in 2018. The quartile was selected by summarizing the percentage of all age groups below 65 years old, of which ``65 and older" is the maximum age reported in Table 1.19A. The year 2018 was selected as it had the largest percentile of underlying populations between reported years. Using similar logic, we use Tables 4.5, 4.7, and 4.8 to estimate the initial age of individuals when they enter the simulation through Distribution (1), i.e., the age at the time of initiation of opioid use. We can estimate for the first initiation age distribution using $m=21.5$ and ${x}_{0.403}=26$ in 2019, resulting in $\mu=2.08$ and $\sigma=0.76$.}
}

\subsection{Event Time Estimations} \label{App:EventTime}
We now describe the parameter and distribution decisions for distributions (1)-(8) in Figure \ref{fig:DES} and Table \ref{tab:inputs}. {For distributions (2)-(6) and (8), the associated parameters are estimated through two different methods. The first method estimated the log-normal $\mu$ and $\sigma$ using equations (\ref{eq:muEst}) - (\ref{eq:c}). The second method estimated the log-normal $\mu$ from the dataset and the log-normal $\sigma$ using equation (\ref{eq:muEst}). During model calibration, the better fit between the two estimates was selected through visual inspection of the 95\% PIs of the simulation model compared to the observed Dane County estimates. This was done by changing associated parameters one by one from the first method estimate to the second method estimate. For brevity, we only detail the estimated parameters for the selected method. Distributions (2), (5), and (6) used the first method, and distributions (3), (4) and (8) used the second method. We also gain confidence in this estimate from the model validation in Figure \ref{fig:CalibrationOutputs}, which shows that 16 of the 17 data points lie within the simulation 95\% PIs. }
{\subsubsection{Distribution (1) - Time of next arrival}}
For distribution (1), inter-arrival times are modeled using an exponential distribution with parameter $\lambda_1 =$ {10.87} days per opioid initiation. The $\lambda_1$ parameter is estimated using two data sources. We used the 2015-{2020} SAMSHA National Survey of Drug Use and Health Survey (NSDUH) estimates of ``Past Year Initiation of Drug Use among Persons Ages 12 or Older: Numbers in Thousands" for both heroin and pain reliever use. Substance initiation is defined as ``the use of a substance for the first time (new use)" \citepalias{substance_abuse_and_mental_health_services_administration_samhsa_2019_2019}. Note these are national estimates, therefore we multiplied these national estimates by Dane County\textquotesingle s ``share" of U.S. opioid-related deaths, which is estimated as in Equation (\ref{eq:A1}), but with U.S. opioid-related deaths instead of Wisconsin. The estimates for pain reliever use initiation and heroin use initiation are added together and averaged over the five years of data available. This provides a yearly estimate of the number of initiations of opioid use. Finally, we estimate the resulting arrival rate, $\lambda_1$, by taking the reciprocal of the yearly estimate and multiplying by 365.25, i.e., the average number of days in a year.
{\subsubsection{Distribution (2) - Time in the active state given next event is an opioid-related death}}
{Distribution (2) parameters are calculated separately for pre- and post-2019 due to the rapid increase in opioid-related deaths due to fentanyl in 2019 \citep{pardo_future_2019}}. Distribution (2) is estimated via log-normal distributions with parameters $\mu_{2}^{{{\text{pre19}}}}$ and  $\sigma_{2}^{{{\text{pre19}}}}$ for years before 2019 and {  $\mu_{2}^{{{\text{post19}}}}$ and $\sigma_{2}^{{{\text{post19}}}}$} for years 2019 and later. To estimate $\mu_{2}^{{{\text{pre19}}}}$ {and $\mu_{2}^{\text{post19}}$}, we { use equations (\ref{eq:muEst}) - (\ref{eq:c}), where $\gamma=0$ and $m=1$ since individuals can experience a fatal overdose within the first day of use. We again assume ${x}_q =365.25$ days in a year. To estimate the respective quantile{s}, $q^{{{\text{pre19}}}}$ {and $q^{\text{post19}}$} as the percentile of individuals that suffer an opioid-related death within one year, we first} use the CDC wonder database and query for Dane County \citep{national_center_for_health_statistics_cdc_2022} to provide {the number of deaths per year, i.e., $d_{i}^{\textit{DC}} $, and divide it by the prevalence of active opioid use in Dane County, i.e., $p^{DC}_{i}$ for all years $i$ that are available. Both the prevalence and death estimates are further described in Appendix A.1. Therefore, the quantiles represent the percentage of individuals with active opioid use who suffered an opioid-related death {within a year} and can be calculated as:  
\begin{align}
        q^{\text{pre19}} = \frac{\sum_{i=2016}^{2018} \nicefrac{d^{DC}_i}{p}^{DC}_{i}}{3} & \quad \text{ and } \quad
        q^{\text{post19}} = \frac{\sum_{i=2019}^{2020} \nicefrac{d^{DC}_i}{p}^{DC}_{i}}{2}.\label{eq:quant}
\end{align}
This results in $q^{\text{pre19}}=0.0027$ and $q^{\text{post19}} = 0.0033$. We then estimated the log-normal distribution parameters using the equations (\ref{eq:muEst}) - (\ref{eq:c}), resulting in $\mu_{2}^{\text{pre19}}=17.60$, $\sigma_{2}^{\text{pre19}}=4.19$,  $\mu_{2}^{\text{post19}}=17.13$, and $\sigma_{2}^{\text{post19}}=4.14$.} We note that a very small {value of} $q$ {is estimated in} (\ref{eq:quant}); {which may} limit our study by underestimating the variability in individuals' time to opioid-related death. {As a result, the simulated times may be more clustered or have a narrower range than what we would observe in the real world.} However, since {our evaluation focuses on} population{-level outcomes rather than individual trajectories}, we believe the underestimation would have a limited impact on our results and conclusions.

 {To aid the interpretation of the quantile, consider the annual overdose death rate among the active use population, $\nicefrac{d^{DC}_{2019}}{p^{DC}_{2019}}$. The reciprocal of $\nicefrac{d^{DC}_{2019}}{p^{DC}_{2019}}$ can be interpreted as the average time until overdose death, since the start of 2019. When we average this over multiple years, the reference point of ``since the start of the year" becomes an arbitrary year. We instead interpret ``starting active opioid use" as the beginning of the time period. This yields an estimate of time-to-overdose death from the onset of use. However, this estimate represents a lower bound, as the prevalence data likely underestimates the true number of active users.} 

 {The parameters $\mu_{2}^{\text{pre19}}$ and $\mu_{2}^{\text{post19}}$ can be interpreted as the average number of log-normal days until the next opioid-related death per individual with active opioid use. Further, if an individual samples a time that is longer than the remaining length of the simulation, this can be interpreted that the individual will not suffer an opioid-related death. We note that ideally, we would directly estimate how long it takes for an individual to have an opioid-related death since most recently starting to use an opioid. However, this would be difficult to estimate as the vast majority of individuals who use opioids do not experience an opioid-related death (SAMHSA, \citeyear{substance_abuse_and_mental_health_services_administration_samhsa_nsduh_2014}). Therefore, estimating the time until opioid death only among those who experienced an opioid-related death would heavily skew the actual estimate among all individuals to the left. An alternative could be possible, such as only estimating an opioid-related death time for some percent of individuals, however, this would add additional parameters and variance to the model. In contrast, our estimation approach uses only four parameters, i.e., $\mu_{2}^{{{\text{pre19}}}}$, $\sigma_{2}^{{{\text{pre19}}}}${,  $\mu_{2}^{{{\text{post19}}}}$, and $\sigma_{2}^{{{\text{post19}}}}$} for a given set of years, which can estimate a time for all individuals. } 
{\subsubsection{Distribution (3) - Time in the active state given next event is a hospital encounter}}
Distribution (3) is modeled as a log-normal distribution with parameters $\mu_3$ and $\sigma_3$ as specified in Table \ref{tab:inputs}. To estimate $\mu_3$, we use the \citet{wisconsin_department_of_health_services_wish_2017} Hospital Encounter Data via the WISH Query. We query the rate for Dane County and ages 14+ since ages 12+ are not available. This data provides the number of hospital encounters $x_{i,3}$ for years $i =$ 2018 and 2019. We note that data from 2016 and 2017 were excluded during the calibration process to ensure the hospital encounter calibration targets were met. We then convert the data to reflect the mean number of days until the next hospital encounter {among the active opioid use population} by multiplying the reciprocal by the number of days in the given year{, $D_i$}. {To obtain a $\mu_3$  that estimates the} days until the next hospital encounter for {the average} individual actively using opioids, {rather than among the active opioid use population, we multiply} by the prevalence of active opioid use in Dane County, or $p^{DC}_{i}$ for all years $i$ that are available. Lastly, we average each yearly estimate and take the natural logarithm to obtain the final estimate for $\mu_{i,3}$. The calculation to estimate $\mu_{i,3}$ is: 
\begin{align}
    \mu_{3} = \frac{\sum_{i=2018}^{2019}\ln\left(\nicefrac{D_ip^{DC}_{i}}{x_{i,3}}\right)}{2}, \label{eq:A4}
\end{align}
where $D_i$ is the number of days in a given year $i$. As with an opioid-related death, a sampled value that is beyond the simulation length would be interpreted as an individual not having a hospital encounter. 
To obtain an estimate for $\sigma_3$ we solved equation (\ref{eq:muEst}) for $\sigma_3$, with inputs $m = 1, \gamma=0,$ and $\mu_3=9.07$. The location parameter and most likely value are chosen as zero and one, respectively, since an individual might present to a healthcare provider on the first day of taking an opioid. This resulted in $\sigma_3= 3.01$.

{For example, if the average number of days until the next hospital encounter is 40 days (i.e., $D_i/x_{i,3} = 40$), then the mean number of days for an individual's next hospital encounter is 400 days (i.e., $D_ip_i/x_{i,3} =4000$), or more than 10 years. As discussed in Section A.5.2, most individuals who use opioids will not experience an opioid-related death within the simulation run time. The evidence is similar for hospital encounters, opioid-related arrests, and OUD treatment starts. Therefore, if an individual is scheduled for a hospital encounter time beyond the end of the simulation, this would be interpreted as they did not experience a hospital encounter. Justification for this modeling choice can be seen in Table 3 by comparing the estimated prevalence of 2017 opioid use against 2017 opioid-related deaths, hospital encounters, opioid-related arrests, and OUD treatment starts.}
{\subsubsection{Distribution (4) - Time in the active state given next event is an opioid-related arrest}}
 When modeling distribution (4), we used SRS Wisconsin Drug Offense Data \citep{wisconsin_department_of_justice_uniform_2016}. We used the data containing Adults and Juveniles in Dane County for Drug Arrests 2015-2019 data summaries, filtered data for Dane County, and the following drug types: ``Opium" or ``cocaine and their derivatives (morphine, heroin, codeine)," ``Synthetic narcotics," ``manufactured narcotics which cause true drug addiction (demerol, methadones)," and ``Unknown." We then averaged the yearly arrests to estimate the parameter $\mu_4$ for the following arrest types: ``Drug - Unknown", ``Drug Possession - Opium/Cocaine," ``Drug Possession - Synthetic," ``Drug Sale - Opium/Cocaine," ``Drug Sale - Synthetic." 
 {The log-normal distribution (4) parameters $\mu_4$ and $\sigma_4$ are estimated in the same way as the distribution (3) parameters, using equation (\ref{eq:A4}). Where the SRS data provided the yearly number of opioid-related arrests, $x_{i,4}$ for $i= 2016,...,2019$. We also use $\gamma = 0$, and $m=90$. The most likely value $m$ is selected through model calibration by adding approximately $\pm$5\% from a starting value of 30. This resulted in $\mu_4=10.04$ and $\sigma_4 = 2.35$.}
{\subsubsection{Distribution (5) - Time in the active state given next event is start OUD treatment}}
 For distribution (5), we used data from the Wisconsin Department of Health Services ``Opioids: Treatment Data by County Dashboard" \citep{wisconsin_department_of_health_services_opioids_2019}. We filtered for Dane County Only and summed the Number of People Treated with Opioid Use Disorder who paid for treatment via County-Authorized Treatment and Medicaid Treatment. We did not include private insurance claims{,} since this data are in terms of the number of episodes, could not be filtered for Dane County, and only made up a small fraction for yearly claims.
 
  {The distribution (5) parameters $\mu_5$ and $\sigma_5$ are estimated in the same way as the distribution (2) parameters. The Wisconsin DHS data provided the yearly number of OUD treatments, $x_{i,5}$ for $i= 2016$ and $2017$. This resulted in  ${x}_{0.063} =365.25$. We also use $\gamma = 0$, and $m=610$ days. The most likely value $m$ is selected through model calibration by adding approximately $\pm$5\% from a starting value of 365. This resulted in $\mu_5=7.48$ and $\sigma_5 = 1.03$.  }
{\subsubsection{Distribution (6) - Time in the active state given next event is stop opioid use}}
{
Distribution (6) parameters were estimated using equations (\ref{eq:muEst})-(\ref{eq:c}) and two analytical studies using NSDUH data (SAMHSA, \citeyear{substance_abuse_and_mental_health_services_administration_samhsa_nsduh_2014}). The location and likelihood parameters were set to 0 and 1 day, respectively, assuming that an individual is most likely to use an opioid once recreationally with no plans or desire to use an opioid in the foreseeable future (i.e., their opioid use would become inactive). }

{The quantile $q$ was estimated from \citet{rivera_risk_2018}, where they estimated that the mean confidence interval for the proportion of individuals developing heroin dependence was between .167 and .506 in 2016. This year was chosen because it aligned with the time frame from most of the dataset. We assume that if someone fails to develop dependence, this could alternatively be interpreted as the individual stopping opioid use. Therefore, we can use the range $(1-0.506, 1 - 0.167)$ to estimate the proportion of individuals that move to the inactive use state. 
To estimate the ${x}_q$, or the timeframe individuals move to inactive use, we used \citet{bauer_contributions_2019}, who concluded that ``heroin problems and experiences now begin to coalesce into heroin use disorder syndrome within 90-120 days after first heroin use". During model calibration, we tested a baseline, low, and high mix of these ranges. We ended up with our high estimate where we estimated about 49.4\% of individuals would be in the inactive use state within 120 days after using an opioid since moving to the active use state. This resulted in a $\gamma=0$, $m=1$, and ${x}_{0.494}=120$ days to obtain a $\mu_6=4.82$ and $\sigma_6=2.20$.
}
{\subsubsection{Distribution (7) - Time in the active state given next event is non-opioid related death}} \label{App:Dist7}
Distribution (7) is an empirical distribution of the length of remaining life based on the individual's current age. Age estimation is described in Section \ref{App:Age}. We {use} ``Table 1. Abridged life tables for all causes of death combined and eliminating specified causes, for the total population: United States, 1999-2001" from \citet{arias_united_2013} {to estimate the number of deaths that occur in each age interval when eliminating deaths due to Drug-induced causes (i.e., death codes F11.0–F11.5, F11.7–F11.9, F12.0–F12.5, F12.7–F12.9, F13.0–F13.5, F13.7–F13.9, F14.0–F14.5, F14.7–F14.9, F15.0–F15.5, F15.7–F15.9, F16.0–F16.5, F16.7–F16.9, F17.0, F17.3–F17.5, F17.7–F17.9, F18.0–F18.5, F18.7–F18.9, F19.0–F19.5, F19.7–F19.9, X40–X44, X60–X64, X85, and Y10–Y14). From there, we create a survival function using the Kaplan-Meier Estimator \citep{kaplan_nonparametric_1958}, implemented through the Python KaplanMeierFitter function as part of the lifelines package \citep{davidson-pilon_lifelines_2024}. Therefore, we estimate the expected age at which an individual will experience a non-opioid-related death and the number of days until that death from when the individual enters the simulation, given the current simulation time and the estimated current age of the individual.}

{\subsubsection{Distribution (8) - Time in the active state given next event is non-opioid related arrest}} \label{App:Dist8}

When modeling distribution (8), we used the same data as for distribution (4), i.e., SRS Wisconsin Drug Offense Data \citep{wisconsin_department_of_justice_uniform_2016}. We filtered the Dane County Adult and juvenile data by removing the following arrest types: ``Drug - Unknown," ``Drug Possession - Opium/Cocaine," ``Drug Possession - Synthetic," ``Drug Sale - Opium/Cocaine," ``Drug Sale - Synthetic." The distribution (8) LN parameters $\mu_8$ and $\sigma_8$ are estimated similarly as the distribution (3) parameters. The main difference is that instead of multiplying by $p_i^{DC}$, we multiply by the Dane County population estimate. Therefore, we implicitly assume that all individuals in Dane County are equally likely to commit a non-opioid-related crime. The SRS data provides the yearly number of non-opioid-related arrests, $x_{i,8}$ for $i= 2016,...,2019$. We also use the location parameter, $\gamma = 0$, and most likely value, $m= 9$ days, to estimate the LN parameters. The most likely value $m$ was estimated from an arrest diversion case study raw dataset \citep{zhang_relationship_2022}. Data collection was started for individuals 12 months before their diversion-eligible crime. From that beginning, nine days was the mode of the number of days until the individual first arrest of any type.  This resulted in $\mu_8=7.88$ and $\sigma_8 =2.38$.  
\subsection{Time in State Estimations} \label{App:TimeState}
We now describe the parameter and distribution decisions for distributions (A)-(G) are estimated using equations (\ref{eq:muEst})-(\ref{eq:c}). {Due to a lack of data, we relied on various studies to estimate distribution parameters.}
{\subsubsection{Distribution (A) - Time in the hosptital or ED}} \label{App:DistA}
{Distribution (A) parameters were estimated using the study done by \citet{singh_national_2020}, which analyzed U.S. National Inpatient Sample (NIS) data from 1998–2016 to estimate OUD hospitalization rates. OUD-hospitalization characteristics by the U.S. hospital region report a median length of stay in the Midwest (total Midwest hospital stays was 182,787), where Dane County is located, of 1.8 days. Additionally, they report 72.3\% of patients were in the hospital for less than or equal to $3$ days. Therefore, we estimated our model parameters using $\gamma=0$, $m=1.8$, and ${x}_{0.723}=3$ to obtain  $\mu_A=0.82$ and $\sigma_A=0.48$.}
{\subsubsection{Distribution (B) - Time in the CJS}} \label{App:DistB}
{Distribution (B) parameters were estimated directly from the dataset collected apart of the study done by \citet{zhang_relationship_2022}, which compares police contacts, arrests, and incarceration between a historical control group  ($N=52$) and an arrest diversion group ($N=263$). The data showed that for the historical comparison group, 47 of the individuals spent less than 57 days in jail after being arrested for an offense that would have qualified them for an illicit drug arrest diversion program. The mode was one day. Therefore, we estimated our model parameters using $\gamma=0$, $m=1$, ${x}_{0.9}=57$ days to obtain $\mu_B=2.16$ and $\sigma_B=1.47$.} {
The same estimates were used time in the CJS due to non-drug-related crime for two reasons. First, among the 79 non-opioid related crimes 0 to 12 months before the arrest diversion eligible crime, 75 of the arrests resulted in spending less than 91 days in jail, with a mode of one day, which is well within the range considered for distribution (B) in our sensitivity analysis. Secondly, limiting the number of estimated distributions reduces model complexity.}
{\subsubsection{Distribution (C) - Time in OUD treatment}} \label{App:DistC}
{Distribution (C) parameters were estimated from Table 51 in \citet{division_of_care_and_teatment_services_2017_2018}, which indicates the percentage of patients receiving at least 90 days of any SUD treatment in Dane County was 59.5\%. The most likely value, $m$, is selected through model calibration by adding approximately
$\pm5$\% from a starting value of 30. Therefore, we estimated our model parameters using $\gamma=0$, $m=30$, and ${x}_{1 -.595}=90$ days to obtain $\mu_C=4.78$ and $\sigma_C=1.18$.} 
{\subsubsection{Distribution (D) - Time in the inactive state given the previous state was the CJS}} \label{App:DistD}
{Distribution (D) parameters were estimated from two separate studies. In Table 2 of \citet{kinlock_study_2008}, $\nicefrac{48}{62}=77.4$\% of males who were released from prison or jail in Baltimore and received counseling only for their opioid use during prison or jail had used heroin within 90 days from release. In Figure 1 from \citet{bukten_high_2017}, individuals ($N = 91,090$) from the 15-year cohort study who were released from prison in Norway were most likely to relapse (and die of an overdose death) on the second day. Therefore, we estimated our model parameters using $\gamma=0$, $m=2$, and ${x}_{0.774}=90$ days to obtain $\mu_D=3.29$ and $\sigma_D=1.61$. These estimates were used for opioid and non-opioid-related arrests that led to time in the CJS.}
{\subsubsection{Distribution (E) - Time in the inactive state given the previous state was OUD treatment}} \label{App:DistE}
Distribution (E) parameters were estimated from \citet{nunes_relapse_2018}, where the study had two types of treatment, naltrexone, and treatment of OUD as usual. It was further split into long and short inpatient and outpatient care. The most likely value, $m$, was estimated from Figure 1, where the fourth week (i.e., 28 days) had the largest drop in treatment retention {when considering all treatment groups. This is driven by the outpatient and short-term treatment groups, noting that the outpatient group is approximately four times the size of the short-term and long-term treatment groups.} We also estimated from Table 3 the cumulative opioid use at the end of 26 weeks (i.e., 182 days) that $\nicefrac{226}{308}= 0.734$ of patients in all treatment groups had used an opioid within 182 days. Therefore, we estimated our model parameters using $\gamma=0$, $m=28$, $q=0.734$, and ${x}_{0.734}=182$ days to obtain $\mu_E=4.52$ and $\sigma_E=1.09$.
{\subsubsection{Distribution (F) - Time in the inactive state given the previous state was the hospital or ED}} \label{App:DistF}
{Distribution (F) parameters were estimated from \citet{chutuape_one-_2001}, where 21.3\% of individuals ($N=116$) self-reported not using after 30 days following release of a three-day inpatient opioid detoxification. However, the researchers reported that urine samples indicate this is closer to 12-15\%, as stated on page 31. The most likely value, $m$, is informed from Figure 2, where 26\% of 66 individuals reported using heroin on the day of discharge. Therefore, we estimated our model parameters using $\gamma=0$, $m=1$, and ${x}_{0.85}=30$ days to obtain $\mu_F=1.95$ and $\sigma_F=1.40$.}
{\subsubsection{Distribution (G) - Time in the inactive state given the previous state was active opioid use}} \label{App:DistG}
{Distribution (G) parameters were calibrated. The quantile ${x}_q$ was selected through model calibration by adding approximately $\pm$5\% from a starting value assessment that 90\% of individuals used opioids within 20 years. The most likely value $m$ is selected through model calibration by adding approximately $\pm$5\% from a starting value 30. The final model parameters were estimated using $\gamma=0$, $m=1$, and ${x}_{0.7}=5.5*365.25$ days per year to obtain $\mu_G=6.29$ and $\sigma_G=2.51 $.}

\section{Determining Number of Replications}
\label{Appendix:Replications}
We calculated the number of replications, $n$,  required to achieve a 95\% confidence interval using equation (\ref{eq:n}). The parameter
$\hat{s}$ is the estimated standard deviation. The parameter
$t_{1-(\nicefrac{\alpha}{2}), n-1}$ is the t-critical value for a two-tailed $1-\alpha$ confidence interval. Lastly, the parameter
$h$ is the confidence interval's half-width (i.e., margin of error). Then,
\begin{align}
n \geq \bigg( \frac{\hat{s}\;t_{1-(\nicefrac{\alpha}{2}), n-1}}{h}   \bigg)^2 .\label{eq:n} 
\end{align}
To determine the total number of replications, we first estimated $\hat{n}$ for each of the outputs in the main analysis of the paper. The outputs are for the Year 2032, which is the final year of the simulation. Table \ref{tab:n} reports $\hat{n}$ for each output with the desired half-lengths, $h$, and standard deviation $\hat{s}$. According to the table, at least $n=563$ replications are needed. We rounded up to 600 replications for the main analysis. 

{\spacingset{1} 
\centering
\begin{table}[tp]
\caption{Half width, $h$, number of replications, $n$, for simulation outputs in Year 2032}
\label{tab:n}
    \resizebox{\textwidth}{!}{
\begin{tabular}{|c||c|c|c|c|c|}
\hline
\multirow{5}{*}{\centering  Steps in estimating $\hat{n}$} & 
\multirow{5}{2cm}{\centering  \#  in the  Active State at the end of Year 2032 } 
& \multirow{5}{2cm}{\centering \#  of Deaths in  Year 2032  }
& \multirow{5}{2cm}{\centering\#   of Opioid-Related Arrests in Year 2032 }
&\multirow{5}{2cm}{\centering \#   of OUD Treatments in Year 2032 }
& \multirow{5}{2.5cm}{\centering \#   of Opioid-Related Hospitalizations in Year 2032 } \\
&&&&& \\
&&&&& \\
&&&&& \\  
&&&&& \\ \hline \hline
desired $h$& 100 & 1 & 5 & 5 & 5 \\ \hline
$\hat{s}$ at $n=600$ & 2{57.10} & 10.{39} &  28.{71}& 5{9.03} & 60{.38}\\ \hline
$h$ at $n=600$ & 2{0.57} & 0.8{3} & 2.{30} & 4.{72} & 4.{83} \\ \hline \hline
$\hat{n}$ & 2{6} & 432 & 12{8} &5{38} &  \textbf{563} \\ \hline 
\end{tabular}
}
\end{table}
}

\begin{figure}[!ht]
    \centering
    \includegraphics[width=\linewidth]{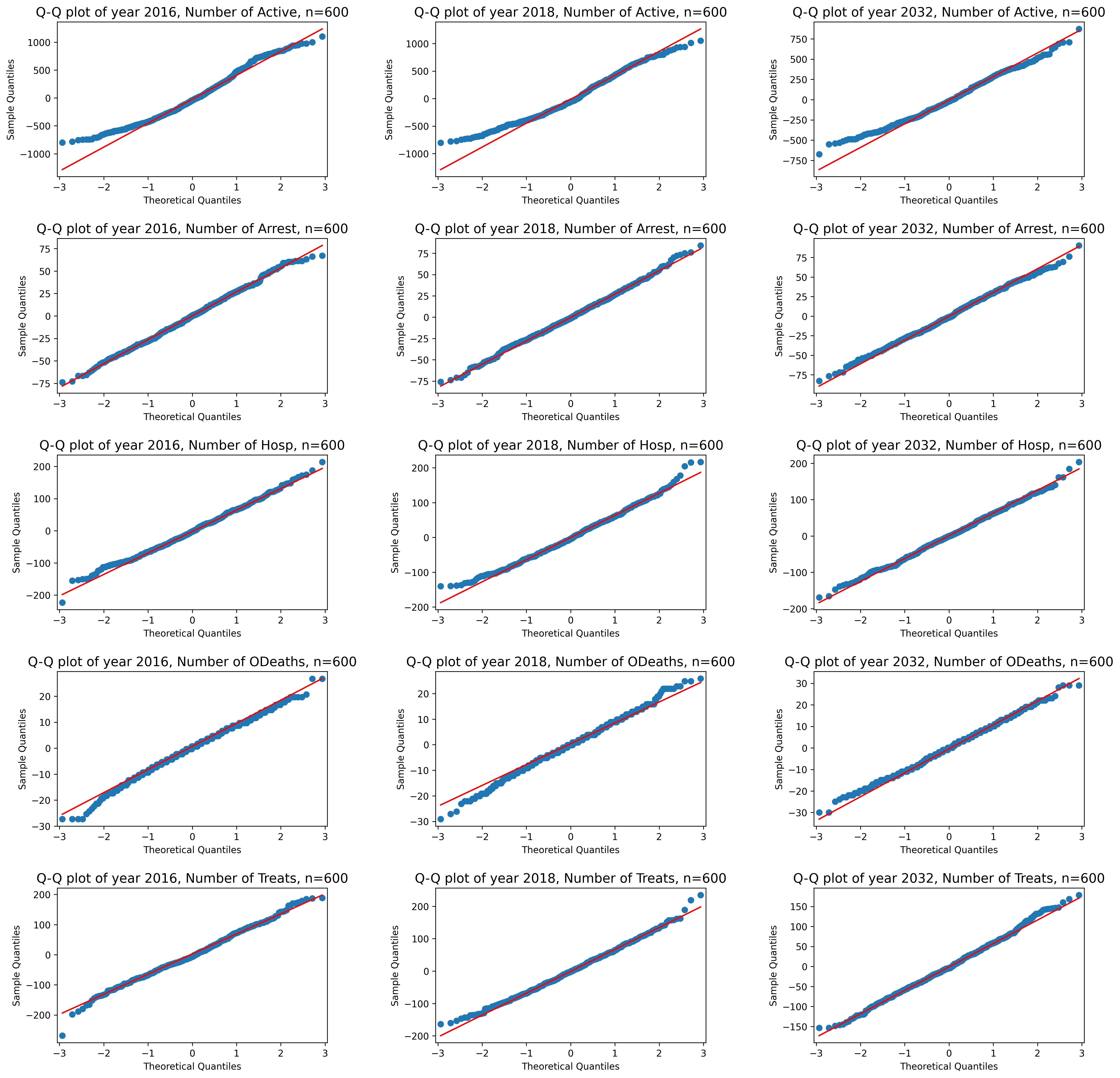}
    \caption{{Residual Q-Q plots of the number of individuals in the active state,opioid-related deaths, arrests, treatments, and hospital encounters for years 2016, 2018, and 2032}}
    \label{fig:QQ}
\end{figure}

Additionally, we checked the residuals of the main outputs for normality by plotting and observing Q-Q plots and performing K-S tests of the main outputs. Table \ref{tab:NormTests} reports the K-S test with Lilliefors correction and skewness. The K-S test fails to reject the null hypothesis that the number of arrests, number of hospital encounters, and number of treatments in 2032 follow a normal distribution. The K-S test rejects the null hypothesis for the number of individuals in the active use state and the number of opioid-related deaths at the end of 2032. Looking at the skewness in Table \ref{tab:NormTests} we see that the active use state {and opioid-related deaths} skews slightly positively, meaning the mean estimates used in the main analysis are above the median of all $n$ runs. {Figure \ref{fig:QQ} shows the Q-Q plots for the five main outputs of the model for years 2016, 2018, and 2032}. Based on the relatively small skewness ($<0.75$) and Q-Q plots we conclude that all the main outputs are sufficiently normal to use confidence intervals as the main results, yet keep skewness in mind when interpreting the results. { We also note that the normality assumption is valid for our cumulative results since the sum of two jointly normal random variables also follows a normal distribution.In fact, following the same process as above with cumulative outputs, showed the cumulative outputs required fewer than 400 replications to achieve a 95\% CI. We have omitted the cumulative results due to their similarity to the yearly outputs.}
{\spacingset{1} 
\begin{table}[h!]
    \caption{Normality tests for simulated residual outputs for 2032 $n=600$}
    \resizebox{\textwidth}{!}{
        \begin{tabular}{|c||c|c|c|c|c|}
            \hline
            \multirow{3}{3cm}{\centering  Normality Test}
            & \multirow{3}{2cm}{\centering  Active} 
            & \multirow{3}{2.5cm}{\centering  {Opioid-related} Deaths} 
            & \multirow{3}{2.5cm}{\centering  {Opioid-related} Arrests} 
            & \multirow{3}{2.5cm}{\centering  {OUD} Treatments} 
            & \multirow{3}{2.5cm}{\centering  {Opioid-related} Hospital Encounters} \\
            &&&&& \\
            &&&&& \\ 
            \hline \hline
            \multirow{1}{3.5cm}{K-S Test$^\alpha$ (p-value)} & 
            \multirow{1}{*}{{0.040}} & 
            \multirow{1}{*}{{0.030}} & 
            \multirow{1}{*}{{0.190}} & 
            \multirow{1}{*}{{0.260}} & 
            \multirow{1}{*}{{0.580}} \\ \hline
            Skewness & {0.240} & {0.060} & {0.020} & {0.140} & {0.70} \\ \hline
            \multicolumn{6}{l}{$^\alpha$\footnotesize{with Lilliefors correction }} 
        \end{tabular}
    }
    \label{tab:NormTests}
\end{table}
}

\section{Additional Simulated Results}
\label{Appendix:Results}
This section reports additional results from the DES model. Table \ref{tab:output} shows the simulated 95\% CI service times for each state in the simulation model with 600 replications. On average, individuals use opioids for 233 days following their first opioid use. On average{,} individuals are in OUD treatment for 238 days, the \ac{CJS} for 25 days, and the hospital for 2.5 days. Individuals are {inactive} for 12,945 days following the active use state, 169 days after the OUD treatment state, 98 days following the \ac{CJS} state, and 19 days following the hospital state.
\newcommand*{\FirstAServiceLow}{230.41}
\newcommand*{\TServiceLow}{234.26}
\newcommand*{\CJServiceLow}{24.69}
\newcommand*{\HEServiceLow}{2.53}
\newcommand*{\IAOServiceLow}{12,260.89}
\newcommand*{\IATServiceLow}{166.30}
\newcommand*{\IACServiceLow}{97.62}
\newcommand*{\IAHServiceLow}{18.68}
\newcommand*{\AgeServiceLow}{32.69}
\newcommand*{\FirstActiveServiceHigh}{235.71}
\newcommand*{\TServiceHigh}{241.40}
\newcommand*{\CJServiceHigh}{25.95}
\newcommand*{\HEServiceHigh}{2.55}
\newcommand*{\IAOServiceHigh}{13,629.31}
\newcommand*{\IATServiceHigh}{171.26}
\newcommand*{\IACServiceHigh}{98.34}
\newcommand*{\IAHServiceHigh}{19.55}
\newcommand*{\AgeServiceHigh}{32.84}
\setlength{\tabcolsep}{4pt} 

\begin{table}[ht]
    \centering
    \caption{Summary of time (days) in each state for base model}
        \begin{tabular}{|c||c|}
            \hline
            \multirow{1}{*}{\centering State} & \multirow{1}{*}{\centering Simulated mean $95\%$ CI (days)} \\
            \hline \hline
            \multirow{1}{7cm}{\centering Active following opioid initiation} & \multirow{1}{4cm}{\centering(\FirstAServiceLow, \FirstActiveServiceHigh)} \\
            \hline
            \multirow{1}{*}{Treatment} & \multirow{1}{3cm}{\centering(\TServiceLow, \TServiceHigh)} \\
             \hline
            \multirow{1}{*}{\ac{CJS}} & \multirow{1}{3cm}{\centering(\CJServiceLow, \CJServiceHigh)} \\
            \hline
            \multirow{1}{*}{Hospital} & \multirow{1}{3cm}{\centering(\HEServiceLow, \HEServiceHigh)} \\
            \hline
            \multirow{1}{*}{Inactive following active} & \multirow{1}{4cm}{\centering(\IAOServiceLow, \IAOServiceHigh)} \\
            \hline
            \multirow{1}{*}{Inactive following treatment} & \multirow{1}{3cm}{\centering(\IATServiceLow, \IATServiceHigh)} \\
            \hline
            \multirow{1}{*}{Inactive following CJS} & \multirow{1}{3cm}{\centering(\IACServiceLow, \IACServiceHigh)} \\
            \hline
            \multirow{1}{*}{Inactive following hospital} & \multirow{1}{3cm}{\centering(\IAHServiceLow, \IAHServiceHigh)} \\
            \hline
        \end{tabular}
    \label{tab:output}
\end{table}

\begin{figure}[tp]
\centering
\begin{subfigure}[b]{.49\linewidth}
\includegraphics[width=\linewidth]{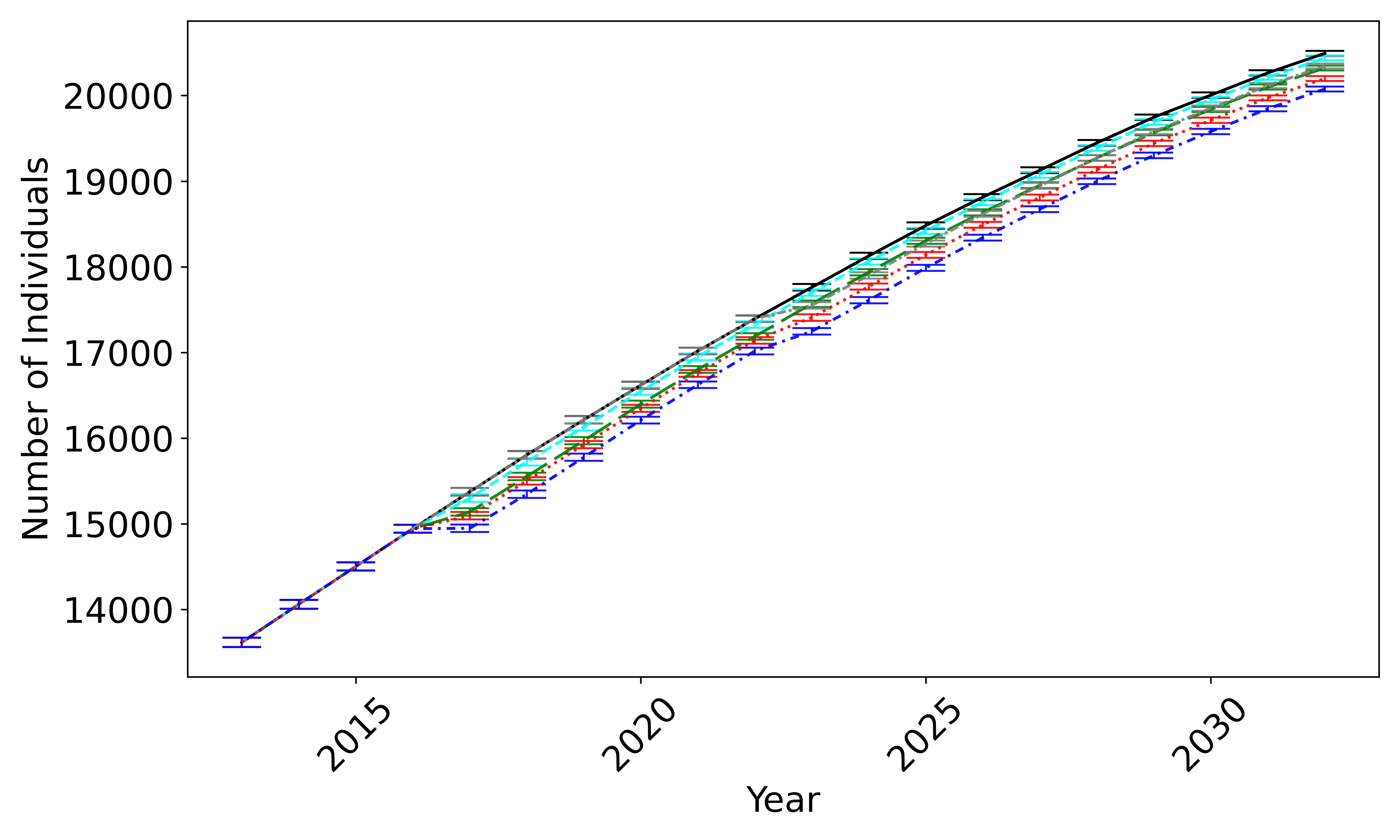}
\subcaption{Active Opioid Use}
\end{subfigure}
\begin{subfigure}[b]{.49\linewidth}
\includegraphics[width=\linewidth]{Figure_legend.png}
\caption{Legend}
\end{subfigure}
~
\begin{subfigure}[b]{.49\linewidth}
\includegraphics[width=\linewidth]{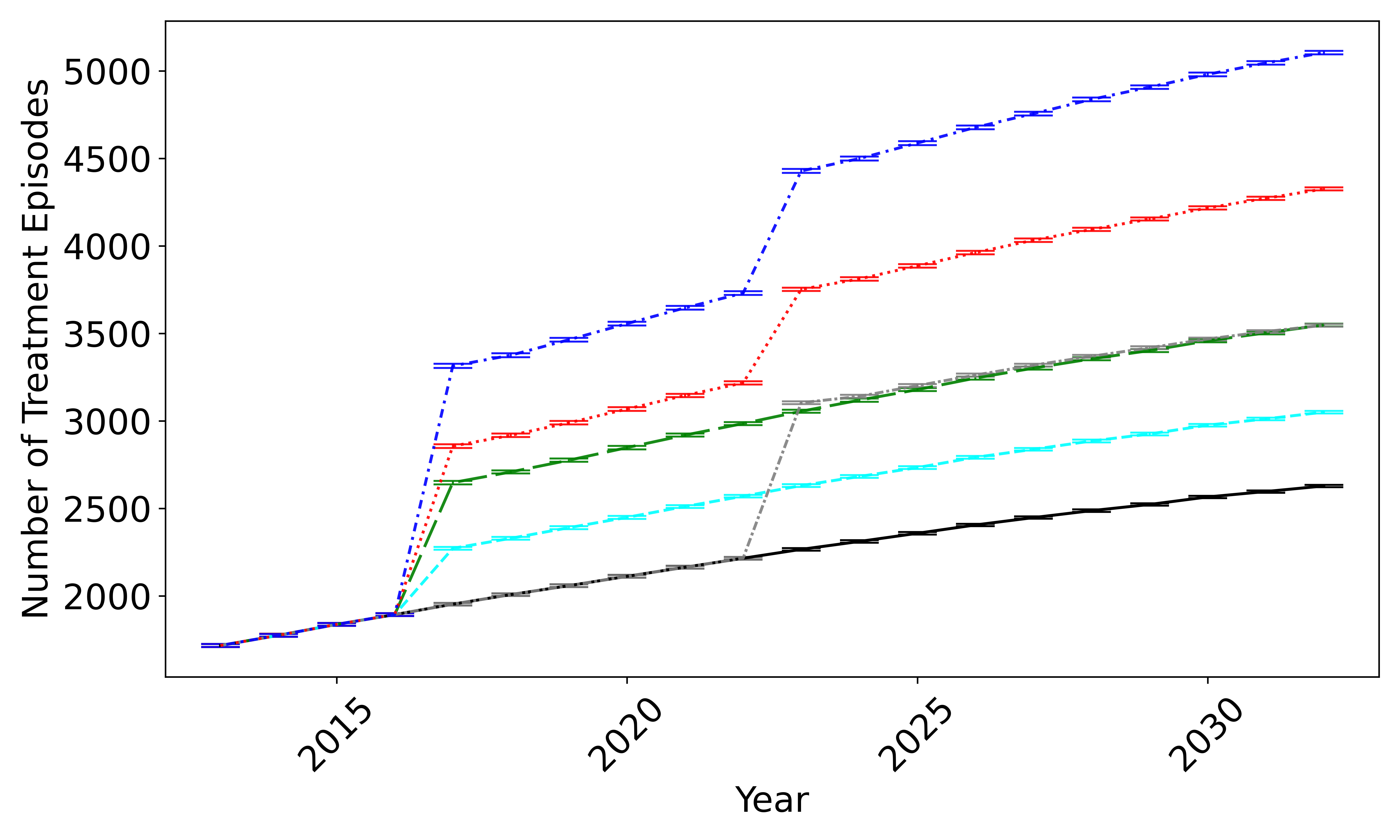}
\caption{Treatment Starts}
\end{subfigure}
\begin{subfigure}[b]{.49\linewidth}
\includegraphics[width=\linewidth]{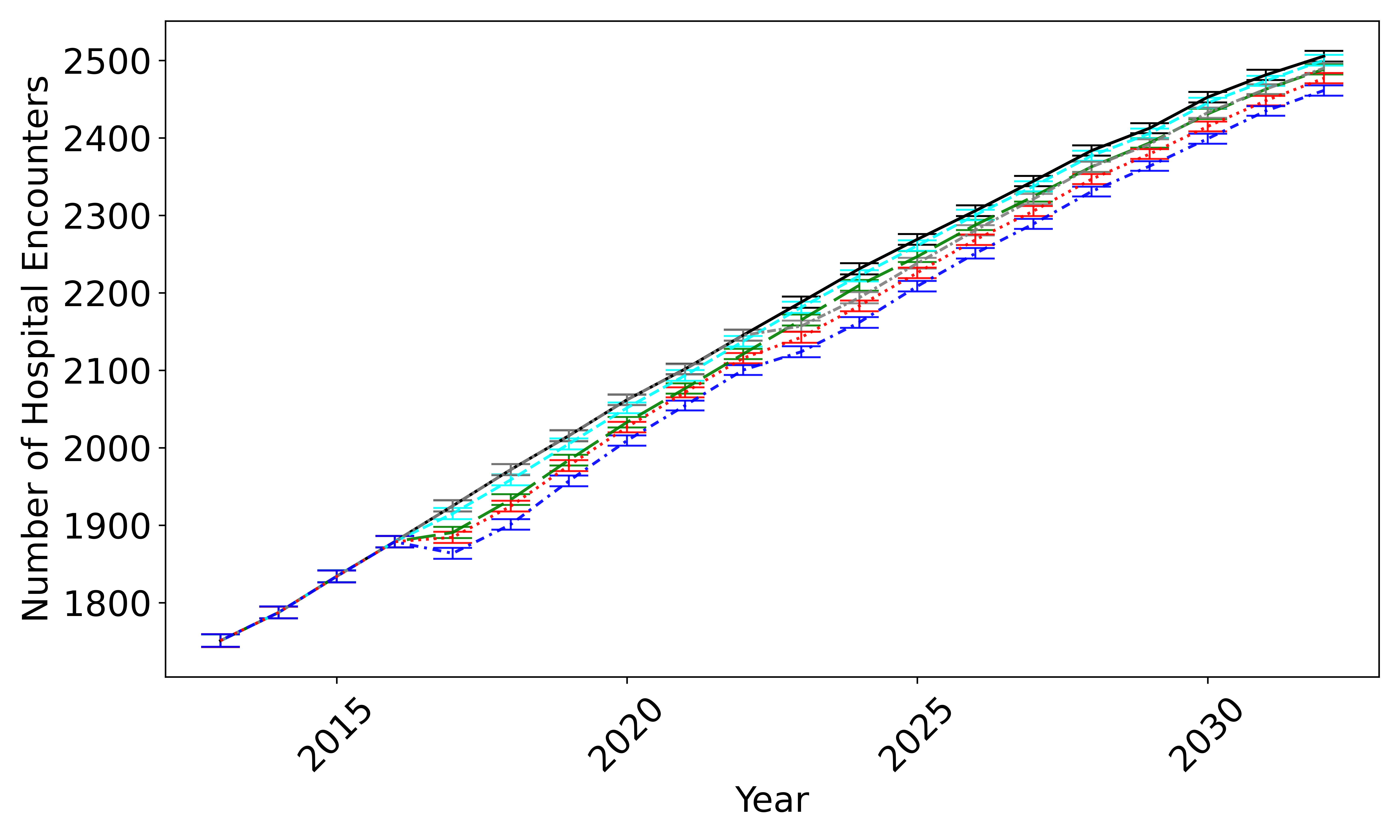}
\caption{Hospital Encounters}
\end{subfigure}
~
\begin{subfigure}[b]{.49\linewidth}
\includegraphics[width=\linewidth]{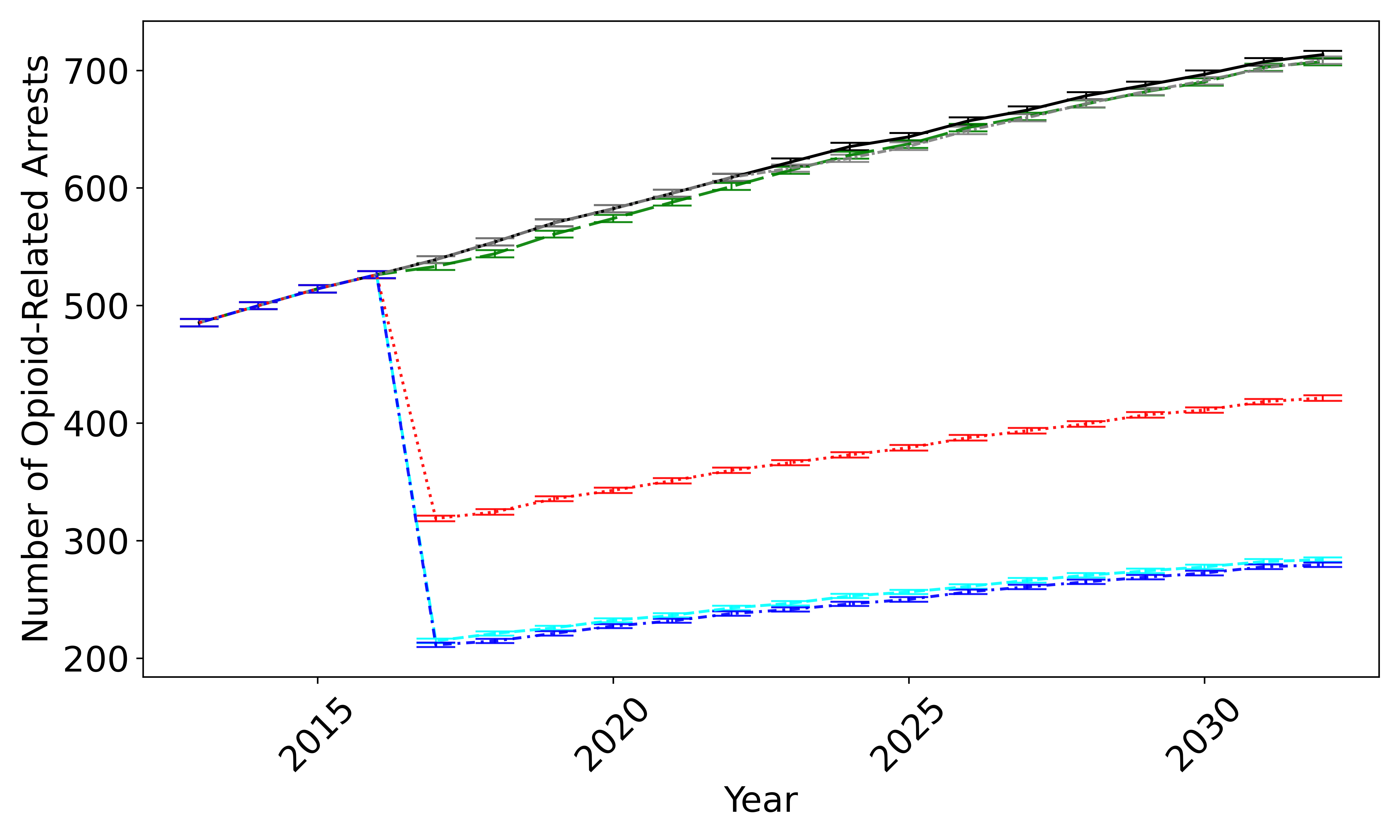}
\caption{Opioid-Related Arrests}
\end{subfigure}
\begin{subfigure}[b]{.49\linewidth}
\includegraphics[width=\linewidth]{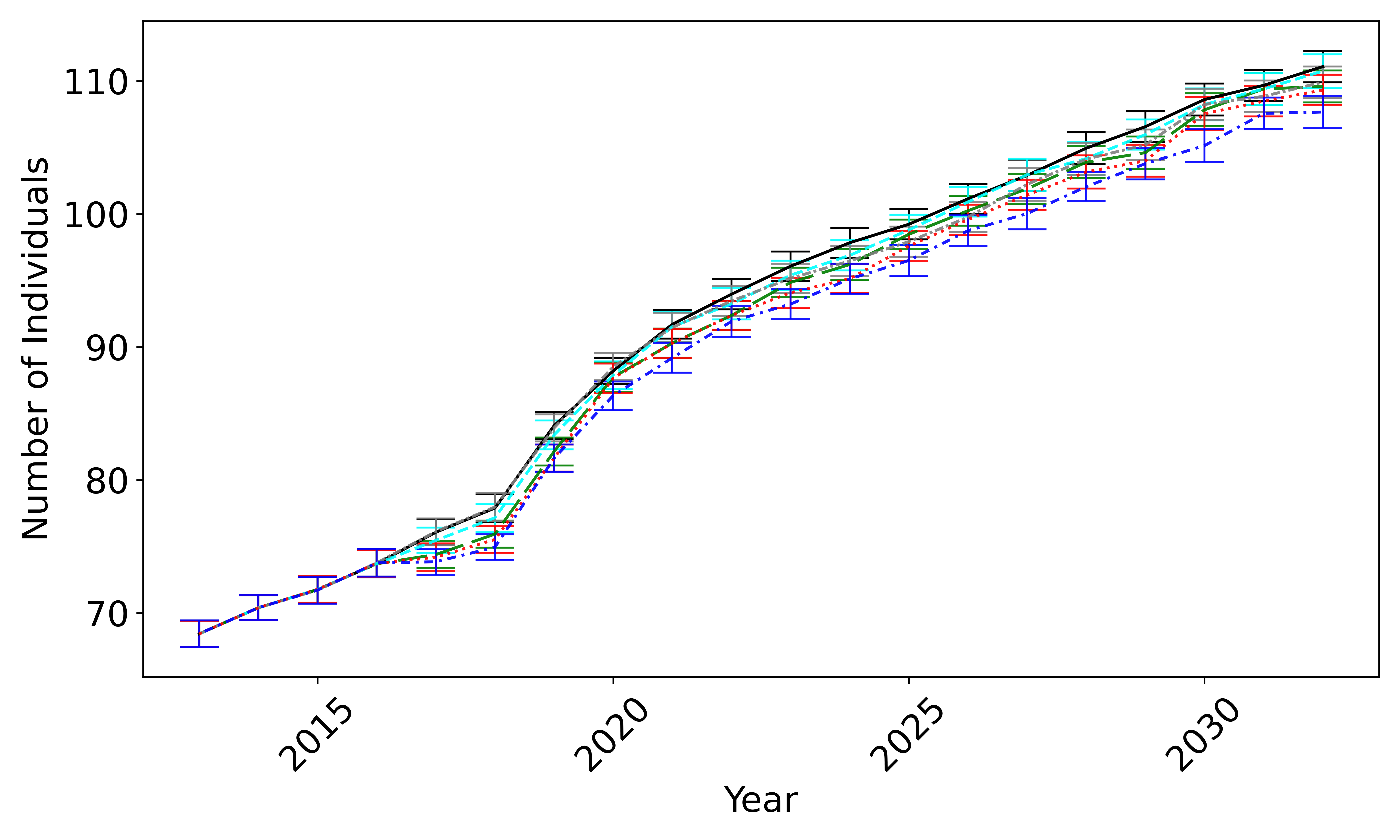}
\caption{Opioid-Related Deaths}
\end{subfigure}
\caption{Total mean simulated 95\% CI intervals }
\label{fig:Full_Results}
\end{figure}

Figure \ref{fig:Full_Results} shows the mean 95\% confidence intervals of the number of opioid-related deaths, hospital encounters, opioid-related arrests, and OUD treatment over time. Figure \ref{fig:Full_Results} shows similar results to Table \ref{tab:AD_All_Results}, which reports these values for the Year 2032. Additionally, \ref{fig:Full_Results} shows that statistically significant differences can be shown in the first year following program implementation (i.e., year 2017 for AD and OD, and year 2023 for CM) with diminishing mean differences from the baseline over time for arrests, hospital encounters, and opioid-related deaths. The reduced number of arrests and hospitalizations remain significantly different near the end of the simulation, as reported in Table \ref{tab:AD_All_Results}. However, as shown in Figure \ref{fig:Full_Results}(f), in most scenarios that implement OD policies, opioid-related deaths are significantly lower in the first three years following the intervention and not statistically significant in subsequent years. Therefore, continuing to scale up the OD policy while exploring other harm reduction and prevention strategies would be valuable to make a statistical difference in the number of overdose deaths. 

\begin{figure}[tp]
\centering
    \begin{subfigure}[b]{.49\linewidth}
        \includegraphics[width=\linewidth]{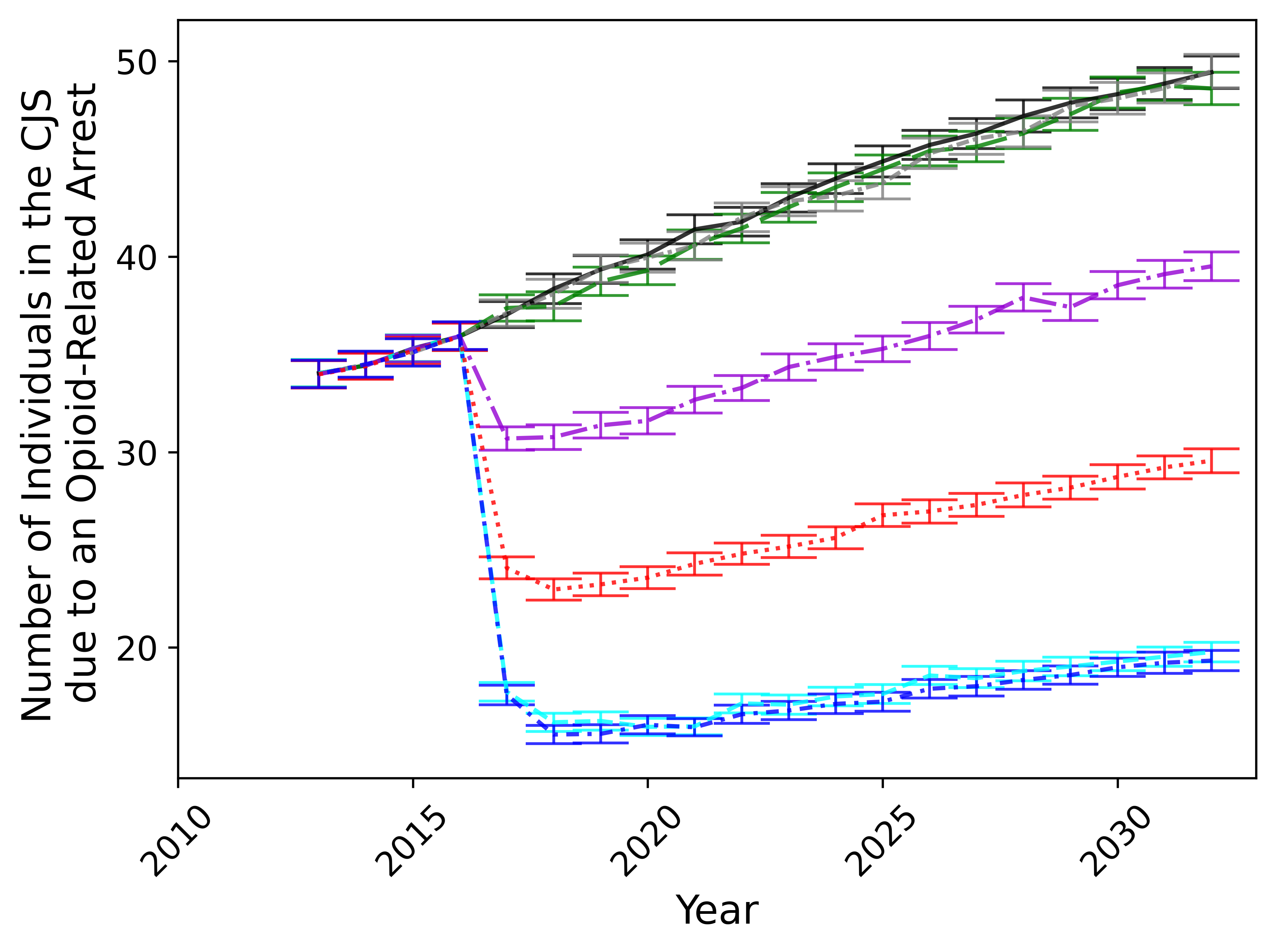}
        \caption{Opioid-Related Arrest 95\% CI}
    \end{subfigure}
    \begin{subfigure}[b]{.49\linewidth}
        \includegraphics[width=\linewidth]{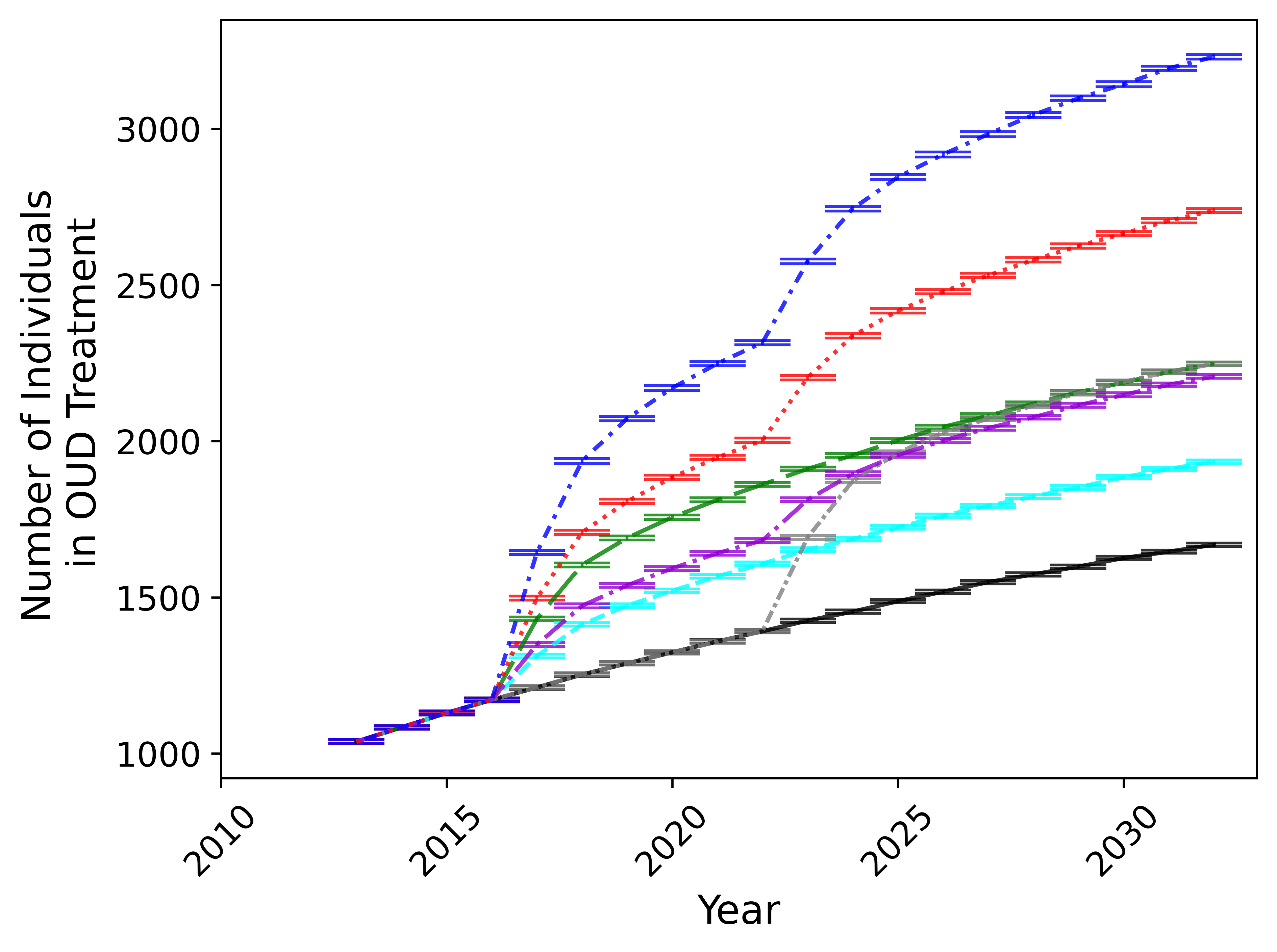}
        \caption{OUD Treatment 95\% CI}
    \end{subfigure}
    ~
    \begin{subfigure}[b]{.49\linewidth}
        \includegraphics[width=\linewidth]{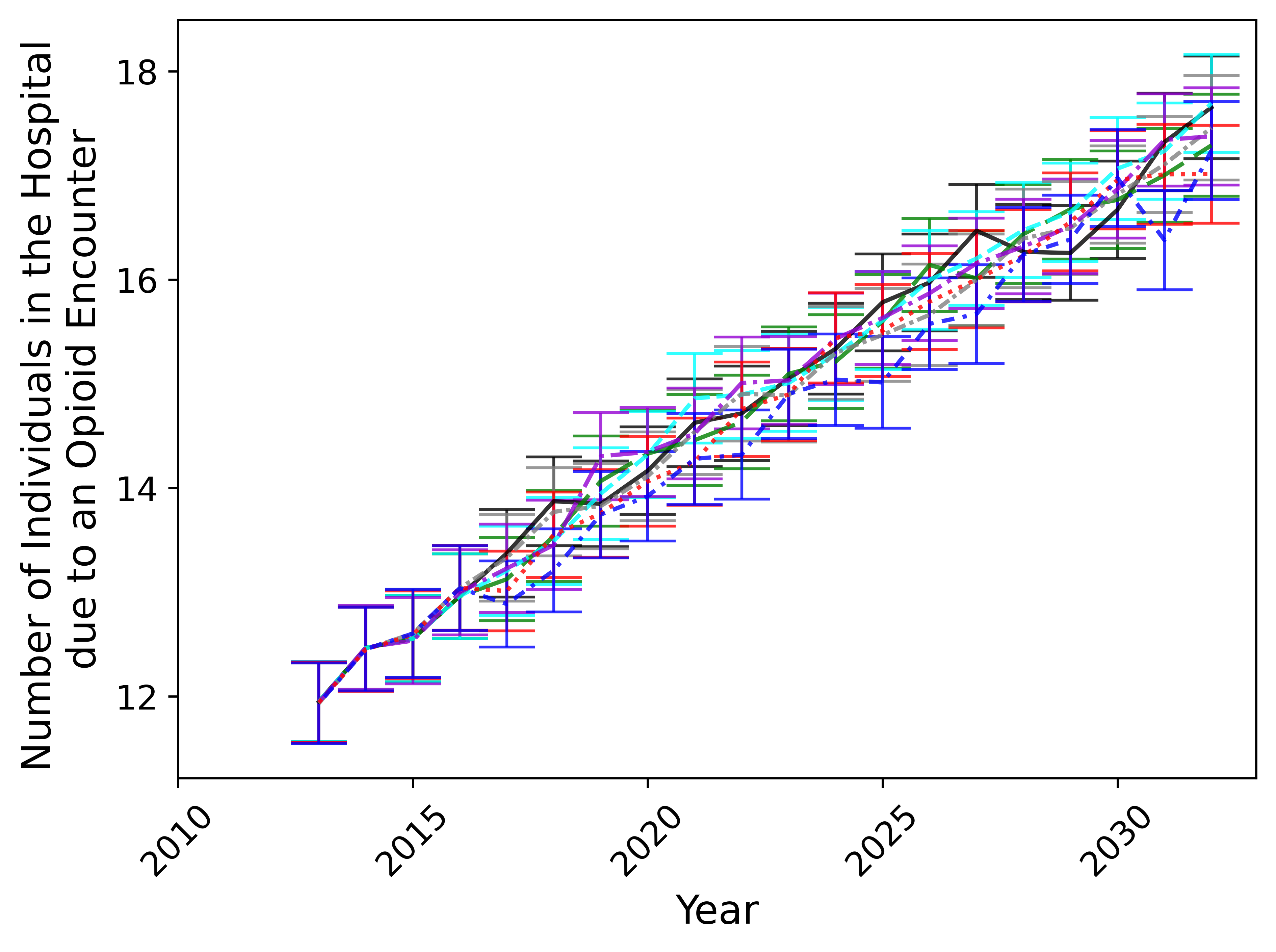}
        \caption{Opioid Hospital Encounters 95\% CI}
    \end{subfigure}
     \begin{subfigure}[b]{.49\linewidth}
     \centering
        \includegraphics[width=.9\textwidth]{Figure_legend.png}
        \\ $\qquad \quad$
        \caption{Legend}
    \end{subfigure}
    \caption{Mean 95\% CI intervals of resource usage }
\label{fig:MeanUsage_Results}
\end{figure}

Figure \ref{fig:MeanUsage_Results} shows the mean 95\% confidence intervals of the number of individuals currently in the \ac{CJS}, hospital, and treatment due to opioids at the midpoint of each year (i.e., June 30th). Due to increases in opioid use, the number of individuals in each system increases over time. Figure \ref{fig:MeanUsage_Results} shows similar results and interpretation of Figure  \ref{fig:Capacity}. The main difference is Figure \ref{fig:MeanUsage_Results} shows the mean demand of the \ac{CJS}, hospital, and OUD treatment systems rather than maximum usage over time. 

\begin{figure}[hp]
     \centering
\begin{subfigure}[b]{.49\linewidth}
\includegraphics[width=\linewidth]{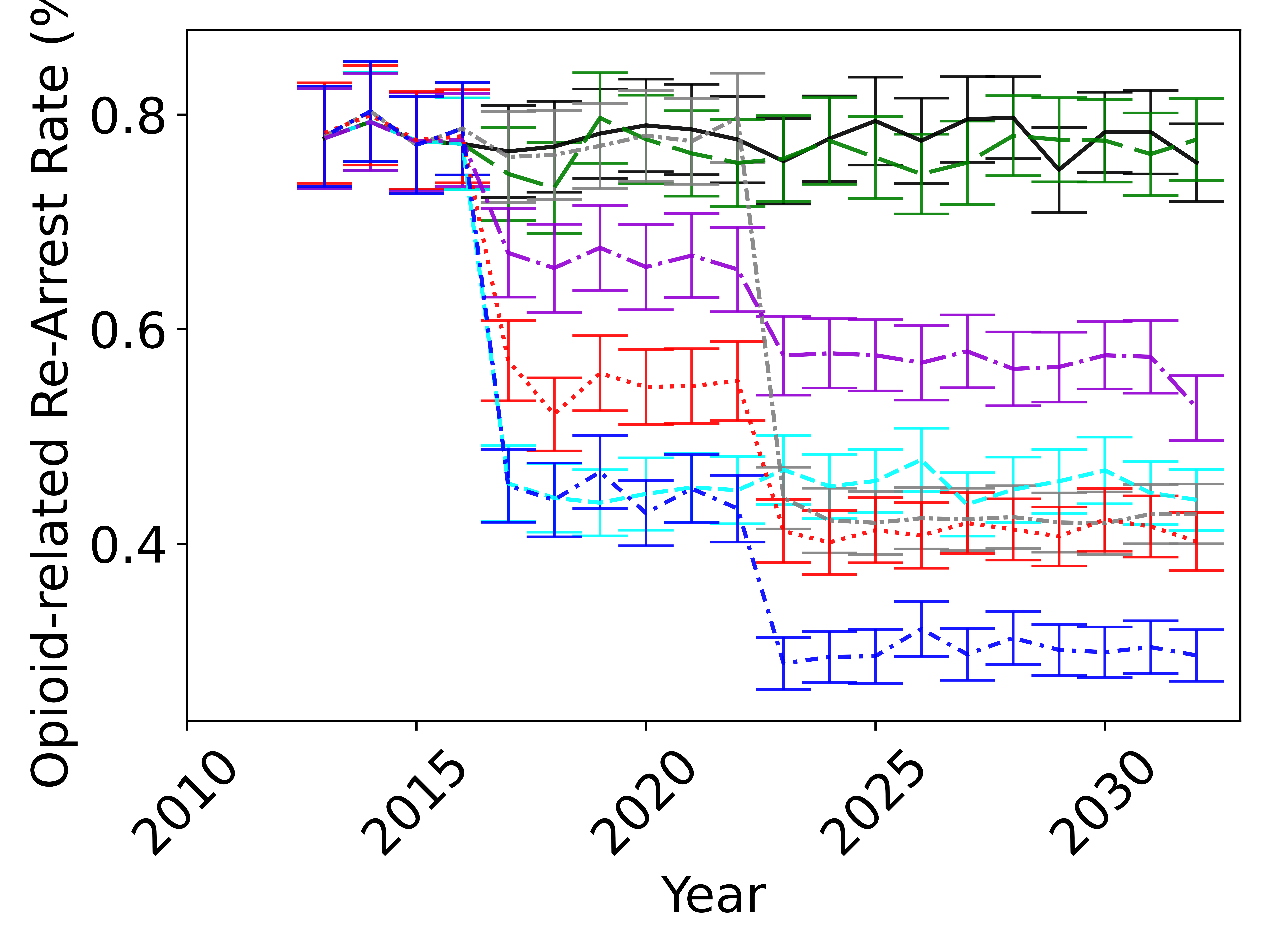 }
\caption{Opioid Related Arrests }
\end{subfigure}
\begin{subfigure}[b]{.49\linewidth}
\includegraphics[width=\linewidth]{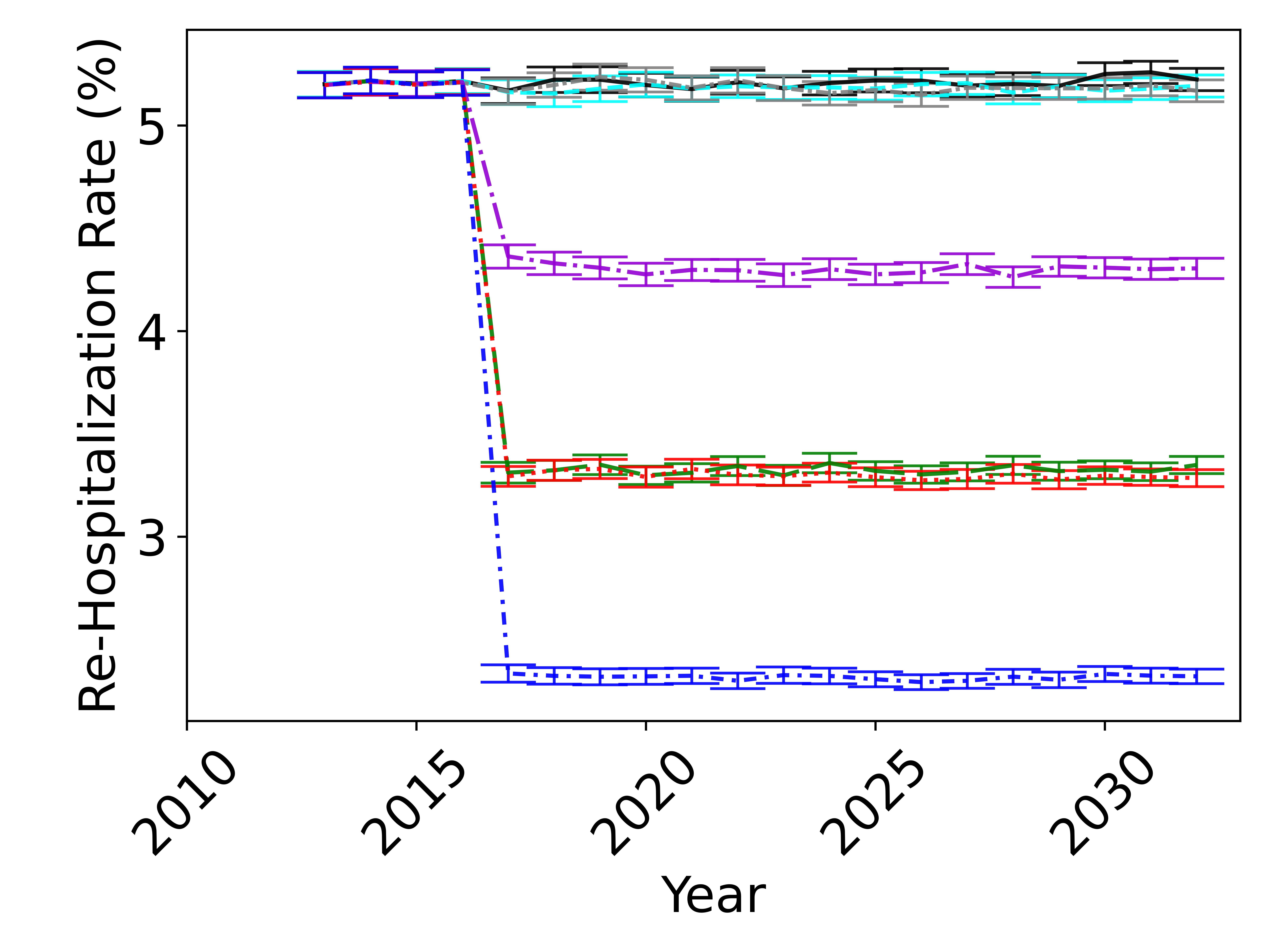 }
\caption{Hospitalizations}
\end{subfigure}
~
\centering
\begin{subfigure}[b]{.49\linewidth}
\includegraphics[width=\linewidth]{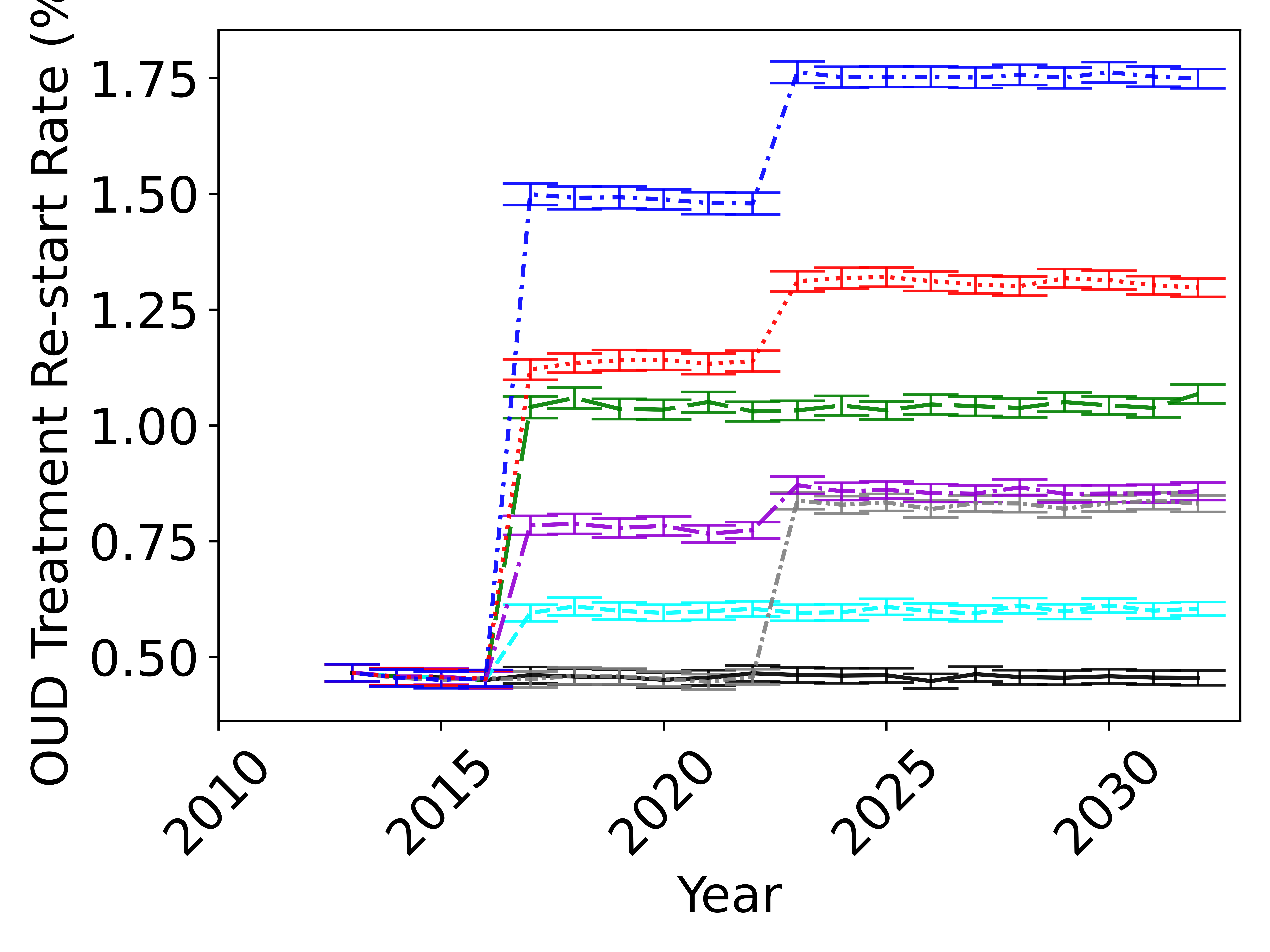 }
\caption{OUD Treatment}
\end{subfigure}
\begin{subfigure}[b]{.49\linewidth}
\includegraphics[width=\linewidth]{Figure_legend.png}
\\ $\qquad \quad$
\caption{Legend}
\end{subfigure}
\caption{Average number of events per person with 95\% CIs}
         \label{fig:PerPerson}
\end{figure}

Figure \ref{fig:PerPerson} illustrates the mean 95\% \ac{CI}s of arrests, hospitalizations, and treatment episodes per person over time. One is the lower bound of the re-arrest, re-hospitalization, and OUD treatment re-start rates. {A value of one can be interpreted as} everyone who used the given system only used it once in a given year. We expect the base model re-arrest and re-hospitalization rates to be a lower bound on the number of interactions for a given individual with these systems{, since, as stated earlier, the model assumes``memoryless” cumulative state durations, whereas in reality some state transitions may be history-dependent \citep{nosyk_characterizing_2014}}. In Figure \ref{fig:PerPerson}, we see statistically significant decreases in re-hospitalizations for policies with increases in OD, decreases in re-arrests for policies with AD or CM, and increases in OUD treatment for all policies in the first year or two after policy implementation. Figure \ref{fig:PerPerson} shows similar results to Table \ref{tab:perPerson_Results}, but over time. 

 Figures  \ref{fig:Full_Results} and \ref{fig:PerPerson} indicate statistically significant differences in opioid-related arrests, active opioid use, and hospital encounters between the policies from the base model after the first year of policy implementation. Therefore, statistically significant outcome changes can be seen in the first year of implementation. This suggests that an annual evaluation of an implemented policy is sufficient. However, Figure \ref{fig:Full_Results} shows narrowing differences between various policies and the baseline regarding active opioid use hospital encounters, opioid-related arrests, and deaths. This suggests stakeholders should expect diminishing returns over time in terms of the number of arrests, hospital encounters, and opioid-related deaths.

\section{Sensitivity Analysis}
\spacingset{1}
\label{Appendix:Sensitivity}
This section explores and describes the sensitivity of the DES model. We conducted the sensitivity analysis on the scenario with arrest diversion at 60\%, overdose diversion at 80\%, and case management at 60\%, referred to as scenario (60, 80, 60) for the remainder of this section. This scenario was selected for two main reasons. The first was due to the scenario's high percentages of individuals diverted by the interventions. Two, the scenario's scale of implementation is feasible for stakeholders to implement since we can expect some eligible individuals not to complete or participate in an available program \citep{white_impact_2021}. We evaluate changes to the following yearly outputs in 2016, 2018, and 2032: number of overdose deaths, individuals entering the \ac{CJS}, hospital encounters, treatment, and active use. These outputs were selected since they were the main focus of the results. The outputs in the year 2016 test input sensitivity to the base model when no interventions have been implemented yet. The year 2018 indicates the sensitivity of the model with increased overdose diversion and arrest diversion implemented, and the year 2032 indicates the effects from the results still hold and the model sensitivity to when all three interventions are implemented. 

The sensitivity analysis follows the approach taken by \citet{renardy_sobol_2021}, where we first assess the direction and monotonicity of input parameters' effect on outputs. Monotonicity for each input parameter is assessed via partial rank correlation coefficients (PRCC) and is evaluated using t-tests. A parameter is deemed significant with a Bonferroni corrected p-value of 0.05. Since PRCC only assesses the direction of a parameter\textquotesingle s effect on output, we also estimate effect sizes using linear regression for each of the {41} varied parameters. As in \citet{renardy_sobol_2021}, we map each parameter range to [0,1] to account for differences in parameter magnitudes and use a log-normal transformation of both the inputs and outputs to make it easier to assess the relative effect between input variables on a given output. 

We vary {41} parameters in an uncertainty analysis where the ranges are $X \pm 5\%$ of the base model values. Table \ref{tab:SensParams} shows the ranges considered for each of the {41} parameters. We use Sobol Sequences to generate  {$2^{10}=1024$} parameter sets. To account for aleatory uncertainty (i.e., randomness in the model dynamics), we conduct three replications for each parameter set for a total of {3,072} simulation runs.  
\spacingset{1}
\begin{table}[htbp]
\caption{Range of parameters for sensitivity analysis  }
\label{tab:SensParams}
\resizebox{\textwidth}{!}{
\begin{tabular}{|l|l|l|l|l|}
\hline
\textbf{\#} & \textbf{Arc} & \textbf{Parameter Type} & \textbf{Base Model Value} & \textbf{Range Considered} \\ \hline \hline
1& Warmup & Initiation Age, $\mu$&{2.08} & ({1.98, 2.18}) \\ \hline
2 & Warmup &  Initiation Age, $\sigma$  &{0.76} & ({0.72, 0.80}) \\ \hline
3&Warmup &Prevalence Age, $\mu$ &{3.74} & ({3.55, 3.93}) \\ \hline
4&Warmup &Prevalence Age, $\sigma$ &{0.49} & ({0.47, 0.51}) \\ \hline \hline
5$^\alpha$&Warmup & Total Starting Population   & Tri( 27299, 34224, 44087)   & (25934.05, 46291.35) \\ \hline \hline
6$^\alpha$&Warmup & Expected CJS Starting Population   & Tri(   5, 11, 15) & (4.75, 15.75) \\ \hline
7$^\alpha$&Warmup &  Expected HE Starting Population   & Tri( 15, 25, 50) & (14.25, 52.50) \\ \hline
8$^\alpha$&Warmup &  Expected Treat Starting Population   & Tri(   300, 450, 500)  & (285, 525) \\ \hline
9$^\alpha$&Warmup &  Prob. Individual Starts in Inactive State  & Tri( 0.2, 0.4, 0.8)  & (0.19, 0.84) \\ \hline
NA$^\beta$ &Warmup & Prob. Individual Starts in Active State  & See Section \ref{sec:InitiatingModel} & NA$^\beta$ \\ \hline 
10&Arc (1) & $\lambda$ & 10.87 & (10.32, 11.41) \\ \hline \hline
11&Arc (2) & $\mu$, pre 2019 & 17.60 & (16.72, 18.47) \\ \hline
12&Arc (2) & $\sigma$, pre 2019 & 4.19 & (3.98, 4.40) \\ \hline
13&Arc (2) & $\mu$, 2019 to 2032 & 17.13 & (16.28, 17.99) \\ \hline
14&Arc (2) & $\sigma$, 2019 to 2032 & 4.14 & (3.93, 4.35) \\ \hline
15&Arc (3) & $\mu$& 9.07 & (8.62, 9.53) \\ \hline \hline
16&Arc (3) & $\sigma$ & 3.01 & (2.86, 3.16) \\ \hline
17&Arc (4) & $\mu$& 10.04 & (9.54, 10.55) \\ \hline
18&Arc (4) & $\sigma$ & 2.35 & (2.24, 2.47) \\ \hline
19&Arc (5) & $\mu$& 7.48 & (7.10, 7.85) \\ \hline
20&Arc (5) & $\sigma$ & 1.03 & (0.98, 1.08) \\ \hline \hline
21&Arc (6) & $\mu$& 4.82 & (4.58, 5.06) \\ \hline
22 & Arc (6) & $\sigma$ & 2.20 & (2.09, 2.31) \\ \hline
{23}&{Arc (8)} &{ $\mu$}& {7.88 }&{ (7.49, 8.27)} \\ \hline
{24} &{ Arc (8)} & {$\sigma$ }& {2.38 }&{ (2.26, 2.50)} \\ \hline
25&\multicolumn{2}{|l|}{$p_A$} & 0.0100 & (0.0095, 0.0105) \\ \hline
26&\multicolumn{2}{|l|}{$p_{OD}$} & 0.2227 & (0.2116, 0.2338) \\ \hline \hline
27&\multicolumn{2}{|l|}{$p_D$} & 0.0218 & (0.0207, 0.0229) \\ \hline

28&Arc (A) & $\mu$& 0.82 & (0.78, 0.86) \\ \hline
29&Arc (A) & $\sigma$ & 0.48 & (0.45, 0.50) \\ \hline
30&Arc (B) & $\mu$& 2.16 & (2.05, 2.27) \\ \hline
31&Arc (B) & $\sigma$ & 1.47 & (1.40, 1.54) \\ \hline
32&Arc (C) & $\mu$& 4.78 & (4.54, 5.02) \\ \hline \hline
33&Arc (C) & $\sigma$ & 1.18 & (1.12, 1.23) \\ \hline
34&Arc (D) & $\mu$& 3.29 & (3.12, 3.45) \\ \hline
35&Arc (D) & $\sigma$ & 1.61 & (1.53, 1.69) \\ \hline
36&Arc (E) & $\mu$& 4.52 & (4.30, 4.75) \\ \hline
37&Arc (E) & $\sigma$ & 1.09 & (1.04, 1.15) \\ \hline \hline
38&Arc (F) & $\mu$& 1.95 & (1.86, 2.05) \\ \hline
39&Arc (F) & $\sigma$ & 1.40 & (1.33, 1.47) \\ \hline
40 &Arc (G) & $\mu$& 6.29 & (5.98, 6.61) \\ \hline
41 &Arc (G) & $\sigma$ & 2.51 & (2.38, 2.63) \\ \hline
\multicolumn{5}{l}{\multirow{3}{18cm}{$^\alpha$ Parameters 5-9 use a triangular distribution. These distributions are only sampled once during the warm-up period. Therefore, we choose to sample the range from min $-5\%$ to max $+5\%$ of the stated triangular distributions to reduce our parameter space }} \\ \multicolumn{5}{l}{} \\ \multicolumn{5}{l}{}\\
\multicolumn{5}{l}{\multirow{3}{18cm}{$^\beta$ The probability an individual starts in the active state is dependent on parameters 5-9. Therefore, it does not require additional testing since the sensitivities attributed to parameters 5-9 can also be applied to this parameter.}}
\end{tabular}}
\end{table}
\spacingset{1}

Table \ref{tab:PRCC_Results} reports the PRCC coefficients and associated t-tests for all parameters and each of the 2016, 2018, and 2032 outputs under scenarios (60, 80, 60). We categorize monotonic relationships with a value of 0.00-0.19 as ``very weak'', 0.20-0.39 as ``weak'', 0.40-0.59 as ``moderate'', 0.60-0.79 as ``strong'', and 0.80-1.0 as ``very strong'' monotonicity. Linearity assumptions, required by linear regression to estimate effect sizes, are assessed visually through scatter plots. We found the linear assumption is not unreasonable for all parameters and outputs. Figures were left out for simplicity. Table \ref{tab:OLS_Results} shows each parameter's estimated effect sizes on the main outputs of scenario (60, 80, 60) in years 2016, 2018, and 2032.
{\spacingset{1}
\begin{sidewaystable}
\caption{ Partial rank correlation coefficient (p-value) of input parameters vs main outputs in 2016, 2018, and 2032 for (60,80,60) scenario }
\label{tab:PRCC_Results}
\resizebox{.95\textwidth}{!}{
\begin{tabular}{|c||c|c|c||c|c|c||c|c|c||c|c|c||c|c|c|}
\hline
 \multicolumn{1}{|c||}{\multirow{3}{2cm}{Parameter Number}} & \multicolumn{3}{c||}{Opioid-Related Deaths} & \multicolumn{3}{c||}{Opioid-Related Arrests} & \multicolumn{3}{c||}{Opioid-Related Hospital Encounters} & \multicolumn{3}{c||}{OUD Treatment Starts} & \multicolumn{3}{|c|}{Active Use Starts}\\ \cline{2-16}
 & \multicolumn{3}{c||}{ coefficient (p-value)}  & \multicolumn{3}{c||}{ coefficient (p-value)}  & \multicolumn{3}{c||}{ coefficient (p-value)}  & \multicolumn{3}{c||}{ coefficient (p-value)}  & \multicolumn{3}{|c|}{coefficient  (p-value)}  \\ \cline{2-16} \cline{2-16}
 & Year 2016 & Year 2018 & Year 3032 & Year 2016 & Year 2018 & Year 3032  & Year 2016 & Year 2018 & Year 3032   &Year 2016 & Year 2018 & Year 3032   & Year 2016 & Year 2018 & Year 3032   \\ \hline \hline 

\multirow{1}{*}{1} & 0.02(1.0) & 0.02(1.0) & 0.02(1.0) & 0.01(1.0) & 0.0(1.0) & -0.0(1.0) & 0.02(1.0) & 0.02(1.0) & 0.02(1.0) & 0.01(1.0) & 0.01(1.0) & 0.01(1.0) & -0.01(1.0) & -0.01(1.0) & -0.01(1.0)\\ \hline 

 \multirow{1}{*}{2} & 0.0(1.0) & 0.01(1.0) & 0.01(1.0) & -0.02(1.0) & -0.02(1.0) & -0.02(1.0) & 0.01(1.0) & 0.01(1.0) & 0.01(1.0) & -0.02(1.0) & -0.02(1.0) & -0.02(1.0) & -0.03(1.0) & -0.03(1.0) & -0.03(1.0)\\ \hline 

 \multirow{1}{*}{3} & -0.01(1.0) & -0.01(1.0) & -0.02(1.0) & -0.01(1.0) & -0.02(1.0) & -0.02(1.0) & -0.02(1.0) & -0.02(1.0) & -0.02(1.0) & -0.02(1.0) & -0.02(1.0) & -0.02(1.0) & -0.02(1.0) & -0.02(1.0) & -0.02(1.0)\\ \hline 

 \multirow{1}{*}{4} & -0.02(1.0) & -0.02(1.0) & -0.01(1.0) & -0.08(1.0) & -0.08(1.0) & -0.08(1.0) & -0.02(1.0) & -0.02(1.0) & -0.01(1.0) & -0.01(1.0) & -0.03(1.0) & -0.03(1.0) & -0.03(1.0) & -0.03(1.0) & -0.03(1.0)\\ \hline 

 \multirow{1}{*}{5} & 0.1(1.0) & 0.09(1.0) & 0.02(1.0) & 0.08(1.0) & 0.07(1.0) & 0.04(1.0) & 0.08(1.0) & 0.08(1.0) & 0.02(1.0) & 0.12(0.159) & 0.12(0.1) & 0.07(1.0) & \bf{0.17$^*$  (0.0) } & \bf{0.15$^*$  (0.001) } & 0.08(1.0)\\ \hline \hline

 \multirow{1}{*}{6} & -0.04(1.0) & -0.03(1.0) & -0.05(1.0) & 0.02(1.0) & 0.02(1.0) & 0.02(1.0) & -0.03(1.0) & -0.03(1.0) & -0.04(1.0) & -0.04(1.0) & -0.04(1.0) & -0.05(1.0) & -0.03(1.0) & -0.03(1.0) & -0.03(1.0)\\ \hline 

 \multirow{1}{*}{7} & -0.01(1.0) & -0.02(1.0) & -0.02(1.0) & -0.0(1.0) & 0.0(1.0) & -0.01(1.0) & -0.03(1.0) & -0.03(1.0) & -0.02(1.0) & -0.01(1.0) & -0.01(1.0) & -0.01(1.0) & 0.0(1.0) & 0.0(1.0) & -0.0(1.0)\\ \hline 

 \multirow{1}{*}{8} & -0.02(1.0) & -0.03(1.0) & -0.03(1.0) & -0.02(1.0) & -0.02(1.0) & -0.02(1.0) & -0.03(1.0) & -0.03(1.0) & -0.03(1.0) & -0.04(1.0) & -0.05(1.0) & -0.05(1.0) & 0.0(1.0) & 0.01(1.0) & 0.01(1.0)\\ \hline 

 \multirow{1}{*}{9} & 0.01(1.0) & 0.06(1.0) & 0.04(1.0) & -0.03(1.0) & -0.05(1.0) & -0.04(1.0) & 0.07(1.0) & 0.07(1.0) & 0.07(1.0) & -0.01(1.0) & 0.0(1.0) & -0.02(1.0) & -0.03(1.0) & -0.04(1.0) & -0.05(1.0)\\ \hline 

 \multirow{1}{*}{10} & 0.01(1.0) & 0.01(1.0) & 0.01(1.0) & 0.04(1.0) & 0.04(1.0) & 0.04(1.0) & 0.0(1.0) & 0.01(1.0) & 0.01(1.0) & -0.0(1.0) & 0.01(1.0) & 0.02(1.0) & 0.03(1.0) & 0.03(1.0) & 0.03(1.0)\\ \hline \hline

 \multirow{1}{*}{11} & \bf{-0.18$^*$  (0.0) } & -0.04(1.0) & 0.03(1.0) & 0.05(1.0) & 0.05(1.0) & 0.06(1.0) & 0.01(1.0) & 0.01(1.0) & 0.01(1.0) & 0.03(1.0) & 0.04(1.0) & 0.05(1.0) & 0.03(1.0) & 0.04(1.0) & 0.04(1.0)\\ \hline 

 \multirow{1}{*}{12} & \bf{0.15$^*$  (0.001) } & 0.03(1.0) & -0.01(1.0) & -0.02(1.0) & -0.02(1.0) & -0.01(1.0) & -0.02(1.0) & -0.02(1.0) & -0.01(1.0) & -0.0(1.0) & -0.01(1.0) & 0.0(1.0) & 0.0(1.0) & 0.01(1.0) & 0.02(1.0)\\ \hline 

 \multirow{1}{*}{13} & -0.01(1.0) & -0.02(1.0) & -0.08(1.0) & 0.03(1.0) & 0.03(1.0) & 0.04(1.0) & -0.03(1.0) & -0.03(1.0) & -0.03(1.0) & 0.01(1.0) & -0.0(1.0) & 0.01(1.0) & -0.01(1.0) & -0.01(1.0) & -0.0(1.0)\\ \hline 

 \multirow{1}{*}{14} & -0.02(1.0) & -0.01(1.0) & 0.02(1.0) & 0.0(1.0) & 0.01(1.0) & 0.0(1.0) & -0.02(1.0) & -0.01(1.0) & -0.03(1.0) & -0.03(1.0) & -0.02(1.0) & -0.04(1.0) & -0.03(1.0) & -0.03(1.0) & -0.04(1.0)\\ \hline 

 \multirow{1}{*}{15} & -0.08(1.0) & \bf{-0.2$^*$  (0.0) } & \bf{-0.19$^*$  (0.0) } & -0.0(1.0) & 0.01(1.0) & 0.01(1.0) & \bf{-0.22$^*$  (0.0) } & \bf{-0.21$^*$  (0.0) } & \bf{-0.22$^*$  (0.0) } & -0.02(1.0) & -0.09(1.0) & -0.09(1.0) & 0.02(1.0) & 0.04(1.0) & 0.06(1.0)\\ \hline \hline

 \multirow{1}{*}{16} & 0.07(1.0) & \bf{0.14$^*$  (0.014) } & \bf{0.13$^*$  (0.046) } & 0.01(1.0) & 0.02(1.0) & 0.01(1.0) & \bf{0.15$^*$  (0.003) } & \bf{0.14$^*$  (0.004) } & \bf{0.15$^*$  (0.001) } & 0.01(1.0) & 0.06(1.0) & 0.06(1.0) & -0.01(1.0) & -0.02(1.0) & -0.02(1.0)\\ \hline 

 \multirow{1}{*}{17} & -0.05(1.0) & 0.0(1.0) & 0.02(1.0) & \bf{-0.29$^*$  (0.0) } & \bf{-0.29$^*$  (0.0) } & \bf{-0.29$^*$  (0.0) } & 0.02(1.0) & 0.02(1.0) & 0.02(1.0) & -0.01(1.0) & -0.05(1.0) & -0.08(1.0) & -0.01(1.0) & -0.01(1.0) & -0.02(1.0)\\ \hline 

 \multirow{1}{*}{18} & -0.0(1.0) & -0.01(1.0) & -0.01(1.0) & 0.11(0.386) & 0.11(0.226) & 0.11(0.253) & -0.01(1.0) & -0.01(1.0) & -0.0(1.0) & 0.0(1.0) & 0.02(1.0) & 0.03(1.0) & -0.01(1.0) & -0.0(1.0) & 0.0(1.0)\\ \hline 

 \multirow{1}{*}{19} & 0.03(1.0) & 0.0(1.0) & 0.02(1.0) & 0.03(1.0) & 0.03(1.0) & 0.04(1.0) & 0.01(1.0) & 0.01(1.0) & 0.01(1.0) & \bf{-0.31$^*$  (0.0) } & \bf{-0.22$^*$  (0.0) } & \bf{-0.22$^*$  (0.0) } & 0.12(0.096) & 0.12(0.064) & \bf{0.13$^*$  (0.035) }\\ \hline 

 \multirow{1}{*}{20} & -0.05(1.0) & -0.06(1.0) & -0.06(1.0) & -0.01(1.0) & -0.01(1.0) & -0.01(1.0) & -0.06(1.0) & -0.05(1.0) & -0.05(1.0) & -0.01(1.0) & -0.03(1.0) & -0.02(1.0) & -0.05(1.0) & -0.05(1.0) & -0.04(1.0)\\ \hline \hline
  \multirow{1}{*}{21} & 0.03(1.0) & 0.08(1.0) & 0.06(1.0) & 0.09(1.0) & 0.08(1.0) & 0.07(1.0) & 0.08(1.0) & 0.09(1.0) & 0.08(1.0) & 0.12(0.215) & \bf{0.14$^*$  (0.007) } & \bf{0.17$^*$  (0.0) } & \bf{0.16$^*$  (0.0) } & \bf{0.16$^*$  (0.0) } & \bf{0.15$^*$  (0.001) }\\ \hline 

 \multirow{1}{*}{22} & -0.01(1.0) & 0.02(1.0) & 0.03(1.0) & 0.01(1.0) & 0.01(1.0) & 0.01(1.0) & 0.02(1.0) & 0.02(1.0) & 0.02(1.0) & 0.04(1.0) & 0.05(1.0) & 0.05(1.0) & 0.05(1.0) & 0.05(1.0) & 0.05(1.0)\\ \hline

 \multirow{1}{*}{23} & -0.01(1.0) & -0.03(1.0) & -0.06(1.0) & -0.01(1.0) & -0.02(1.0) & -0.02(1.0) & -0.04(1.0) & -0.04(1.0) & -0.04(1.0) & -0.03(1.0) & -0.05(1.0) & -0.1(0.832) & -0.06(1.0) & -0.06(1.0) & -0.07(1.0)\\ \hline 

 \multirow{1}{*}{24} & -0.03(1.0) & -0.02(1.0) & -0.02(1.0) & -0.03(1.0) & -0.04(1.0) & -0.04(1.0) & -0.01(1.0) & -0.01(1.0) & -0.01(1.0) & -0.04(1.0) & -0.04(1.0) & -0.04(1.0) & -0.01(1.0) & -0.01(1.0) & -0.02(1.0)\\ \hline

 \multirow{1}{*}{25} & -0.01(1.0) & -0.01(1.0) & 0.02(1.0) & 0.02(1.0) & 0.02(1.0) & 0.01(1.0) & 0.01(1.0) & 0.01(1.0) & 0.01(1.0) & -0.01(1.0) & -0.0(1.0) & 0.01(1.0) & -0.01(1.0) & -0.01(1.0) & -0.0(1.0)\\ \hline \hline

 \multirow{1}{*}{26} & -0.03(1.0) & -0.03(1.0) & -0.01(1.0) & -0.02(1.0) & -0.02(1.0) & -0.02(1.0) & -0.01(1.0) & -0.02(1.0) & -0.02(1.0) & 0.01(1.0) & -0.0(1.0) & -0.01(1.0) & -0.01(1.0) & -0.01(1.0) & -0.01(1.0)\\ \hline 

 \multirow{1}{*}{27} & -0.0(1.0) & -0.03(1.0) & -0.02(1.0) & 0.02(1.0) & 0.02(1.0) & 0.02(1.0) & -0.04(1.0) & -0.03(1.0) & -0.04(1.0) & -0.01(1.0) & -0.01(1.0) & -0.01(1.0) & 0.03(1.0) & 0.03(1.0) & 0.04(1.0)\\ \hline 
 \multirow{1}{*}{28} & 0.0(1.0) & -0.01(1.0) & 0.01(1.0) & 0.04(1.0) & 0.04(1.0) & 0.04(1.0) & -0.0(1.0) & 0.0(1.0) & -0.01(1.0) & -0.02(1.0) & -0.01(1.0) & -0.01(1.0) & 0.02(1.0) & 0.02(1.0) & 0.01(1.0)\\ \hline 

 \multirow{1}{*}{29} & -0.03(1.0) & 0.01(1.0) & 0.01(1.0) & -0.01(1.0) & -0.02(1.0) & -0.02(1.0) & 0.01(1.0) & 0.02(1.0) & 0.01(1.0) & -0.0(1.0) & 0.0(1.0) & -0.01(1.0) & -0.03(1.0) & -0.04(1.0) & -0.04(1.0)\\ \hline 

 \multirow{1}{*}{30} & -0.02(1.0) & -0.04(1.0) & -0.04(1.0) & 0.01(1.0) & 0.01(1.0) & 0.01(1.0) & -0.04(1.0) & -0.03(1.0) & -0.04(1.0) & 0.01(1.0) & -0.0(1.0) & -0.02(1.0) & 0.0(1.0) & 0.0(1.0) & -0.0(1.0)\\ \hline \hline

 \multirow{1}{*}{31} & -0.03(1.0) & -0.05(1.0) & -0.02(1.0) & -0.02(1.0) & -0.03(1.0) & -0.01(1.0) & -0.05(1.0) & -0.05(1.0) & -0.04(1.0) & -0.01(1.0) & -0.03(1.0) & -0.03(1.0) & -0.05(1.0) & -0.04(1.0) & -0.03(1.0)\\ \hline 

 \multirow{1}{*}{32} & 0.01(1.0) & -0.04(1.0) & -0.04(1.0) & -0.04(1.0) & -0.04(1.0) & -0.03(1.0) & -0.04(1.0) & -0.04(1.0) & -0.04(1.0) & 0.0(1.0) & -0.02(1.0) & -0.02(1.0) & -0.08(1.0) & -0.09(1.0) & -0.08(1.0)\\ \hline 

 \multirow{1}{*}{33} & 0.03(1.0) & 0.02(1.0) & 0.03(1.0) & -0.02(1.0) & -0.02(1.0) & -0.02(1.0) & 0.01(1.0) & 0.02(1.0) & 0.02(1.0) & -0.02(1.0) & -0.01(1.0) & -0.0(1.0) & -0.0(1.0) & -0.0(1.0) & 0.0(1.0)\\ \hline 

 \multirow{1}{*}{34} & 0.03(1.0) & 0.04(1.0) & 0.02(1.0) & 0.03(1.0) & 0.03(1.0) & 0.02(1.0) & 0.03(1.0) & 0.04(1.0) & 0.03(1.0) & 0.02(1.0) & 0.04(1.0) & 0.03(1.0) & 0.05(1.0) & 0.05(1.0) & 0.04(1.0)\\ \hline 

 \multirow{1}{*}{35} & 0.01(1.0) & 0.0(1.0) & 0.0(1.0) & 0.0(1.0) & 0.01(1.0) & 0.0(1.0) & 0.01(1.0) & 0.0(1.0) & 0.01(1.0) & -0.04(1.0) & -0.03(1.0) & -0.03(1.0) & -0.02(1.0) & -0.02(1.0) & -0.02(1.0)\\ \hline \hline

 \multirow{1}{*}{36} & -0.07(1.0) & 0.02(1.0) & 0.03(1.0) & 0.02(1.0) & 0.02(1.0) & 0.01(1.0) & 0.03(1.0) & 0.03(1.0) & 0.03(1.0) & -0.01(1.0) & 0.0(1.0) & 0.0(1.0) & -0.02(1.0) & -0.02(1.0) & -0.02(1.0)\\ \hline 

 \multirow{1}{*}{37} & -0.05(1.0) & -0.04(1.0) & -0.04(1.0) & 0.01(1.0) & 0.02(1.0) & 0.02(1.0) & -0.04(1.0) & -0.04(1.0) & -0.05(1.0) & -0.03(1.0) & -0.04(1.0) & -0.03(1.0) & -0.02(1.0) & -0.02(1.0) & -0.02(1.0)\\ \hline 

 \multirow{1}{*}{38} & -0.01(1.0) & -0.01(1.0) & 0.0(1.0) & 0.02(1.0) & 0.03(1.0) & 0.03(1.0) & -0.01(1.0) & -0.01(1.0) & -0.01(1.0) & 0.02(1.0) & 0.01(1.0) & 0.03(1.0) & -0.01(1.0) & -0.01(1.0) & -0.01(1.0)\\ \hline 

 \multirow{1}{*}{39} & -0.01(1.0) & -0.02(1.0) & -0.0(1.0) & 0.0(1.0) & 0.01(1.0) & 0.01(1.0) & -0.02(1.0) & -0.02(1.0) & -0.01(1.0) & 0.03(1.0) & 0.02(1.0) & 0.03(1.0) & -0.05(1.0) & -0.05(1.0) & -0.03(1.0)\\ \hline 

 \multirow{1}{*}{40} & -0.04(1.0) & -0.12(0.12) & \bf{-0.14$^*$  (0.006) } & -0.07(1.0) & -0.07(1.0) & -0.06(1.0) & \bf{-0.14$^*$  (0.011) } & \bf{-0.14$^*$  (0.01) } & \bf{-0.13$^*$  (0.024) } & -0.09(1.0) & \bf{-0.14$^*$  (0.011) } & \bf{-0.13$^*$  (0.042) } & \bf{-0.18$^*$  (0.0) } & \bf{-0.18$^*$  (0.0) } & \bf{-0.17$^*$  (0.0) }\\ \hline \hline

 \multirow{1}{*}{41} & -0.05(1.0) & -0.07(1.0) & -0.03(1.0) & -0.0(1.0) & -0.0(1.0) & 0.0(1.0) & -0.05(1.0) & -0.06(1.0) & -0.04(1.0) & -0.06(1.0) & -0.06(1.0) & -0.06(1.0) & -0.06(1.0) & -0.06(1.0) & -0.06(1.0)\\ \hline
\multicolumn{16}{l}{*{Statistically significant PRCC t-test with Bonforroni Corrected p-val at 0.05 }}\\
\end{tabular} 
}
\end{sidewaystable}
}


{\spacingset{1}
\begin{table}
\caption{ OLS effect size of input parameters vs main outputs in 2016, 2018, and 2032 for (60,80,60) scenario }
\label{tab:OLS_Results}
\resizebox{\textwidth}{!}{
\begin{tabular}{|c||c|c|c||c|c|c||c|c|c||c|c|c||c|c|c|}
\hline
\multicolumn{1}{|c||}{\multirow{3}{2cm}{Parameter Number \centering}} & \multicolumn{3}{c||}{Opioid-Related Deaths} & \multicolumn{3}{c||}{Opioid-Related Arrests} & \multicolumn{3}{c||}{Opioid-Related Hospital Encounters} & \multicolumn{3}{c||}{OUD Treatment Starts} & \multicolumn{3}{c||}{Active Use Starts}\\ \cline{2-16}
 & \multicolumn{3}{c||}{ effect size}  & \multicolumn{3}{c||}{ effect size}  & \multicolumn{3}{c||}{ effect size}  & \multicolumn{3}{c||}{ effect size}  & \multicolumn{3}{c||}{effect size}  \\\cline{2-16} \cline{2-16}
 & Year 2016 & Year 2018 & Year 3032 & Year 2016 & Year 2018 & Year 3032  & Year 2016 & Year 2018 & Year 3032   &Year 2016 & Year 2018 & Year 3032   & Year 2016 & Year 2018 & Year 3032   \\ \hline \hline 
 
 \multirow{1}{*}{1} & 0.174 & 0.157 & 0.102 & 0.066 & 0.065 & -0.007 & 0.116 & 0.112 & 0.114 & 0.031 & 0.048 & 0.023 & -0.031 & -0.04 & -0.046\\ \hline 

 \multirow{1}{*}{2} & 0.036 & 0.033 & 0.065 & -0.121 & -0.179 & -0.134 & 0.053 & 0.042 & 0.053 & -0.152 & -0.095 & -0.092 & -0.116 & -0.121 & -0.126\\ \hline 

 \multirow{1}{*}{3} & -0.039 & -0.019 & -0.103 & -0.106 & -0.133 & -0.108 & -0.092 & -0.089 & -0.1 & -0.127 & -0.12 & -0.109 & -0.098 & -0.109 & -0.103\\ \hline 

 \multirow{1}{*}{4} & -0.287 & -0.195 & -0.106 & -0.968 & -0.911 & -0.955 & -0.189 & -0.206 & -0.125 & -0.134 & -0.209 & -0.176 & -0.182 & -0.161 & -0.127\\ \hline 

 \multirow{1}{*}{5} & 0.167 & 0.097 & 0.016 & 0.147 & 0.122 & 0.062 & 0.1 & 0.089 & 0.022 & 0.158 & 0.119 & 0.053 & \bf{0.126$^*$  }  & \bf{0.111$^*$  }  & 0.052\\ \hline \hline

 \multirow{1}{*}{6} & -0.009 & -0.015 & -0.008 & 0.005 & 0.008 & 0.001 & -0.014 & -0.013 & -0.013 & -0.01 & -0.008 & -0.007 & 0.002 & 0.002 & 0.0\\ \hline 

 \multirow{1}{*}{7} & -0.022 & -0.012 & -0.015 & 0.008 & 0.007 & 0.009 & -0.008 & -0.01 & -0.011 & -0.019 & -0.015 & -0.014 & -0.007 & -0.007 & -0.008\\ \hline 

 \multirow{1}{*}{8} & -0.029 & -0.031 & -0.021 & -0.038 & -0.023 & -0.032 & -0.027 & -0.027 & -0.029 & -0.045 & -0.038 & -0.033 & 0.007 & 0.01 & 0.008\\ \hline 

 \multirow{1}{*}{9} & 0.016 & 0.028 & 0.021 & -0.023 & -0.031 & -0.031 & 0.033 & 0.03 & 0.027 & 0.002 & 0.006 & -0.001 & -0.005 & -0.008 & -0.012\\ \hline 

 \multirow{1}{*}{10} & 0.058 & 0.058 & 0.096 & 0.308 & 0.317 & 0.352 & 0.02 & 0.053 & 0.074 & -0.015 & 0.064 & 0.12 & 0.093 & 0.099 & 0.142\\ \hline \hline

 \multirow{1}{*}{11} & \bf{-1.527$^*$}    & -0.23 & 0.159 & 0.516 & 0.51 & 0.535 & 0.057 & 0.053 & 0.053 & 0.255 & 0.217 & 0.217 & 0.192 & 0.207 & 0.218\\ \hline 

 \multirow{1}{*}{12} & \bf{1.365$^*$}    & 0.184 & -0.102 & -0.19 & -0.137 & -0.119 & -0.118 & -0.114 & -0.05 & -0.038 & -0.075 & -0.024 & 0.005 & 0.016 & 0.081\\ \hline 

 \multirow{1}{*}{13} & -0.123 & -0.134 & -0.473 & 0.269 & 0.225 & 0.352 & -0.248 & -0.195 & -0.171 & -0.001 & -0.019 & 0.063 & -0.05 & -0.039 & 0.024\\ \hline 

 \multirow{1}{*}{14} & -0.158 & -0.051 & 0.101 & 0.024 & 0.074 & 0.021 & -0.125 & -0.058 & -0.137 & -0.289 & -0.19 & -0.193 & -0.125 & -0.137 & -0.186\\ \hline 

 \multirow{1}{*}{15} & -0.666 & \bf{-1.196$^*$}    & \bf{-1.019$^*$}    & 0.014 & 0.103 & 0.101 & \bf{-1.385$^*$ }   & \bf{-1.307$^*$ }   & \bf{-1.25$^*$ }   & -0.193 & -0.502 & -0.381 & 0.099 & 0.168 & 0.222\\ \hline \hline

 \multirow{1}{*}{16} & 0.68 & \bf{0.856$^*$}    & \bf{0.699$^*$}    & 0.099 & 0.135 & 0.054 & \bf{0.954$^*$ }   & \bf{0.909$^*$ }   & \bf{0.892$^*$ }   & 0.146 & 0.349 & 0.278 & -0.077 & -0.111 & -0.128\\ \hline 

 \multirow{1}{*}{17} & -0.461 & 0.031 & 0.137 & \bf{-2.813$^*$ }   & \bf{-2.735$^*$ }   & \bf{-2.708$^*$ }   & 0.125 & 0.116 & 0.146 & -0.009 & -0.225 & -0.312 & -0.021 & -0.015 & -0.027\\ \hline 

 \multirow{1}{*}{18} & -0.02 & -0.089 & -0.016 & 1.154 & 1.158 & 1.149 & -0.048 & -0.082 & -0.018 & 0.033 & 0.107 & 0.17 & -0.028 & -0.021 & 0.027\\ \hline 

 \multirow{1}{*}{19} & 0.252 & 0.017 & 0.114 & 0.397 & 0.318 & 0.442 & 0.081 & 0.058 & 0.064 & \bf{-2.409$^*$  }  & \bf{-1.272$^*$  }  & \bf{-1.02$^*$  }  & 0.502 & 0.518 & \bf{0.542$^*$  } \\ \hline 

 \multirow{1}{*}{20} & -0.387 & -0.391 & -0.367 & -0.074 & -0.06 & -0.054 & -0.376 & -0.365 & -0.34 & -0.123 & -0.212 & -0.147 & -0.254 & -0.238 & -0.205\\ \hline \hline

 \multirow{1}{*}{21} & 0.256 & 0.509 & 0.391 & 0.902 & 0.819 & 0.728 & 0.549 & 0.561 & 0.504 & 0.916 & \bf{0.794$^*$  }  & \bf{0.793$^*$  }  & \bf{0.675$^*$  }  & \bf{0.662$^*$  }  & \bf{0.634$^*$  } \\ \hline 

 \multirow{1}{*}{22} & -0.054 & 0.113 & 0.136 & 0.12 & 0.133 & 0.06 & 0.147 & 0.135 & 0.136 & 0.364 & 0.24 & 0.213 & 0.185 & 0.189 & 0.181\\ \hline 

 \multirow{1}{*}{23} & -0.157 & -0.189 & -0.343 & -0.191 & -0.226 & -0.217 & -0.274 & -0.255 & -0.267 & -0.231 & -0.262 & -0.46 & -0.216 & -0.214 & -0.26\\ \hline 

 \multirow{1}{*}{24} & -0.237 & -0.128 & -0.114 & -0.308 & -0.345 & -0.374 & -0.049 & -0.058 & -0.066 & -0.327 & -0.224 & -0.181 & -0.046 & -0.039 & -0.059\\ \hline 

 \multirow{1}{*}{25} & -0.005 & -0.001 & 0.004 & 0.001 & 0.002 & 0.001 & 0.001 & 0.002 & 0.002 & -0.001 & 0.001 & 0.001 & -0.001 & -0.0 & 0.0\\ \hline \hline

 \multirow{1}{*}{26} & -0.2 & -0.222 & -0.108 & -0.19 & -0.208 & -0.145 & -0.121 & -0.144 & -0.119 & 0.062 & -0.024 & -0.026 & -0.033 & -0.021 & -0.016\\ \hline 

 \multirow{1}{*}{27} & -0.031 & -0.102 & -0.042 & 0.097 & 0.133 & 0.104 & -0.122 & -0.105 & -0.104 & -0.004 & -0.025 & 0.001 & 0.065 & 0.072 & 0.085\\ \hline 

 \multirow{1}{*}{28} & -0.009 & -0.09 & 0.024 & 0.381 & 0.409 & 0.396 & -0.027 & -0.017 & -0.043 & -0.152 & -0.06 & -0.063 & 0.08 & 0.083 & 0.07\\ \hline 

 \multirow{1}{*}{29} & -0.18 & 0.061 & 0.056 & -0.122 & -0.184 & -0.153 & 0.086 & 0.083 & 0.063 & -0.02 & -0.012 & -0.045 & -0.1 & -0.109 & -0.14\\ \hline 

 \multirow{1}{*}{30} & -0.079 & -0.251 & -0.216 & 0.095 & 0.072 & 0.093 & -0.25 & -0.231 & -0.255 & 0.083 & -0.02 & -0.034 & 0.042 & 0.047 & 0.032\\ \hline \hline

 \multirow{1}{*}{31} & -0.347 & -0.331 & -0.195 & -0.25 & -0.259 & -0.116 & -0.346 & -0.334 & -0.242 & -0.126 & -0.178 & -0.158 & -0.236 & -0.213 & -0.162\\ \hline 

 \multirow{1}{*}{32} & 0.114 & -0.242 & -0.227 & -0.42 & -0.411 & -0.358 & -0.284 & -0.281 & -0.226 & 0.048 & -0.102 & -0.082 & -0.338 & -0.352 & -0.327\\ \hline 

 \multirow{1}{*}{33} & 0.18 & 0.127 & 0.112 & -0.214 & -0.175 & -0.189 & 0.074 & 0.08 & 0.1 & -0.18 & -0.083 & -0.041 & 0.007 & 0.007 & 0.044\\ \hline 

 \multirow{1}{*}{34} & 0.243 & 0.244 & 0.109 & 0.296 & 0.244 & 0.188 & 0.188 & 0.209 & 0.135 & 0.2 & 0.202 & 0.144 & 0.201 & 0.191 & 0.137\\ \hline 

 \multirow{1}{*}{35} & 0.18 & 0.002 & 0.018 & 0.015 & 0.044 & 0.04 & 0.039 & 0.025 & 0.067 & -0.304 & -0.164 & -0.119 & -0.05 & -0.053 & -0.04\\ \hline \hline

 \multirow{1}{*}{36} & -0.54 & 0.142 & 0.165 & 0.148 & 0.144 & 0.08 & 0.185 & 0.18 & 0.194 & -0.059 & 0.031 & 0.008 & -0.087 & -0.116 & -0.123\\ \hline 

 \multirow{1}{*}{37} & -0.471 & -0.276 & -0.239 & 0.13 & 0.176 & 0.138 & -0.29 & -0.293 & -0.275 & -0.254 & -0.218 & -0.143 & -0.097 & -0.095 & -0.086\\ \hline 

 \multirow{1}{*}{38} & -0.173 & -0.085 & 0.01 & 0.202 & 0.267 & 0.277 & -0.064 & -0.079 & -0.059 & 0.153 & 0.088 & 0.128 & -0.019 & -0.014 & -0.016\\ \hline 

 \multirow{1}{*}{39} & -0.119 & -0.164 & -0.086 & -0.007 & 0.045 & 0.08 & -0.184 & -0.163 & -0.093 & 0.211 & 0.084 & 0.111 & -0.172 & -0.158 & -0.097\\ \hline 

 \multirow{1}{*}{40} & -0.335 & -0.729 & \bf{-0.817$^*$}    & -0.721 & -0.712 & -0.632 & \bf{-0.923$^*$ }   & \bf{-0.889$^*$ }   & \bf{-0.806$^*$ }   & -0.704 & \bf{-0.745$^*$  }  & \bf{-0.598$^*$  }  & \bf{-0.754$^*$  }  & \bf{-0.752$^*$  }  & \bf{-0.722$^*$  } \\ \hline \hline

 \multirow{1}{*}{41} & -0.379 & -0.403 & -0.236 & -0.032 & -0.016 & 0.025 & -0.347 & -0.369 & -0.301 & -0.413 & -0.361 & -0.292 & -0.27 & -0.263 & -0.256\\ \hline
\multicolumn{16}{l}{*{Statistically significant PRCC t-test with Bonforroni Corrected p-val at 0.05 }}\\
\end{tabular} 
}
\end{table}
}
Table \ref{tab:PRCC_Results} indicates three {weak} negative monotonic relationships between the inputs and outputs. The first is the relationship between the number of arrests and parameter 17, $\mu$ of the distribution for the time in the active state given the next event is an opioid-related arrest. The second is the relationship between the number of hospital encounters and parameter 15, $\mu$  of the distribution for the time in the active state given that the next event is an opioid-related hospital encounter. The third is the relationship between the number of OUD treatment starts and parameter 19, $\mu$ of the distribution for the time in the active state given the next event is OUD treatment. This relationship between OUD treatment and parameter 19 is less monotonic after implementing the OUD policies. As shown in Table \ref{tab:OLS_Results}, all three {weak} monotonicity relationships have effect sizes of {-2.813, -1.385, and -2.409}, respectively, the greatest of which is that of the relationship of parameter 17 and the number of opioid-related arrests. It is also worth noting the corresponding $\sigma$ parameters of these LN distributions had notable OLS effects of {1.154, 0.954, and -0.123}, respectively, even though the relationships were {very} weakly monotone or determined non-monotone. This suggests that the corresponding model outputs are more sensitive to small changes to the $\mu$ parameters over the $\sigma$ parameters. Additionally, the adjustments are more predictable (i.e., linear) to the corresponding output with changes to the $\mu$ parameters. Therefore, parameters 15 through 20 should be carefully selected as they directly impact the magnitude of outputs over time. We do note that the effect size of parameter 15 on hospital encounters and parameter 18 on OUD treatment starts to decrease after the sustained implementation of the policies. This means that while they still impact model outputs, they seem to affect the outputs less after the policies are implemented.

{There are several other very weak} monotone relationships in the model{, such as} between the number of opioid-related deaths and parameters {11, 12, 15, and 16}. Parameter 11 is the $\mu$ of the distribution for the active state time given that the next event is an opioid-related death pre-2019. The corresponding $\sigma$ parameter, 12, is also {very weakly} monotone for this distribution. We note that the relationship between opioid-related deaths and parameter 13, the $\mu$ of the distribution for the active state time given that the next event is an opioid-related death post-2019, is {non-}monotone with a smaller effect size than that of parameter 11. Since the model is sensitive to parameters 11 and 12, care should be given to parameterizing opioid-related deaths. We also note that after the implementation of AD and OD, the absolute value of the OLS effects of parameters 11 and 12 reduces the magnitude from {-1.527 to -0.23 and 1.365 to 0.184}, respectively. This suggests that the AD and OD interventions make the model less sensitive to parameter 11. Parameters 15 {and 16 are the $\mu$ and $\sigma$} of the distribution for the active state time given that the next event is a hospital encounter. Parameter 15 becomes more monotone, and its effect size increases after the AD and OD policies are implemented in 2018, but is decreased for 2032. This suggests that as parameter 15 increases, an OD policy would look more competitive for opioid-related deaths. The $\mu$ parameters 11, 13, and 15 have larger effect sizes than their corresponding distribution's $\sigma$ parameters, meaning the model is more sensitive to the $\mu$ parameters of the distribution than their $\sigma$ parameters. 

Several additional {very} weakly monotone relationships exist for parameters 5, {19,} 21, and 40. As the size of the total starting population (i.e., parameter 5) increases, the number of individuals in the active use state at the end of {2016 and 2018 increases. The effect reduces from 0.126 to 0.111 from 2016 to 2018 and is non-monotone with effect size 0.052 in 2032.} This suggests that the starting population slightly affects the active use state, but its influence reduces the longer the simulation runs. {Parameter 19, $\mu$ of the distribution for the time in the active state given the next event is OUD treatment, also has a very weak monotone relationship with the number in the active use state in 2032 with an effect size of 0.542. This means parameter 19 becomes more sensitive after implementing OUD treatment policies. }
Parameter 21, $\mu$ of the distribution for the time in the active state given the next event is to stop opioid use, is {very weakly} monotone, with the number of active use with an effect size of {0.675} in 2016, and OUD treatments with an effect size of {0.794} in 2018. Lastly, the $\mu$ of the LN distribution for the time in the inactive state following the active state (i.e., parameter 40) has a {very weak} monotone relationship with opioid-related deaths in 2032, opioid-related hospital encounters, the number of OUD treatments in 2018 and 2032 and the number of active use starts. This suggests that the model is {most sensitive} to parameter 40 as it impacts the most model outputs. 

The monotonic relationships between the input and output parameters described in this section are expected based on the model structure. Additionally, the effect size of parameters on outputs does not change as policies are implemented. This suggests that changes to these parameters would not affect the conclusions of this paper between the base model and treatment policy scenarios regarding opioid-related deaths, opioid-related arrests, hospital encounters, treatments, and active use. 

\end{document}